\documentclass[a4paper,11pt]{article}
\pdfoutput=1 % to allow compilation in JCAP
\usepackage{jcappub}
\usepackage{amsmath}
\usepackage{bm}
\usepackage{graphicx}
\usepackage{enumitem}
\usepackage{geometry}
\usepackage{xspace}
\usepackage{pdflscape}
\usepackage{dsfont}
\usepackage[normalem]{ulem}

\usepackage[compat=1.1.0]{tikz-feynman}

\usepackage{mathtools}

\DeclarePairedDelimiter\floor{\lfloor}{\rfloor}

\usepackage{relsize}

\allowdisplaybreaks % allows long equations to be displayed over several pages

\makeatletter
\gdef\@fpheader{}
\g@addto@macro\bfseries{\boldmath}
\makeatother

\newcommand{\ie}{{i.e.~}}

\newcommand{\eg}{e.g.~}

\newcommand{\etc}{etc.}

\let\oldsqrt\sqrt
\def\sqrt{\mathpalette\DHLhksqrt}
\def\DHLhksqrt#1#2{%
\setbox0=\hbox{$#1\oldsqrt{#2\,}$}\dimen0=\ht0
\advance\dimen0-0.2\ht0
\setbox2=\hbox{\vrule height\ht0 depth -\dimen0}%
{\box0\lower0.4pt\box2}}

\newcommand{\dd}{\mathrm{d}}
\newcommand{\ee}{e}

\newcommand{\lE}{\ell_{\mathrm{E}} }

\newcommand{\sss}[1]{{\scriptscriptstyle{#1}}}
\newcommand{\boldmathsymbol}[1]{{\ensuremath{\boldsymbol{#1}}}}

\newcommand{\uPl}{\mathrm{Pl}}
\newcommand{\uin}{\mathrm{in}}

\newcommand{\uend}{\mathrm{end}}

\newcommand{\ueff}{\mathrm{eff}}

\newcommand{\uc}{\mathrm{c}}

\newcommand{\uS}{\mathrm{S}}

\newcommand{\usssS}{\sss{\uS}}

\newcommand{\usssPl}{\sss{\uPl}}

\newcommand{\nS}{n_\usssS}

\newcommand{\bmk}{\boldmathsymbol{k}}

\newcommand{\calP}{\mathcal{P}}

\newcommand{\setR}{\mathbb{R}}

\newcommand{\setI}{\mathbb{I}}

\newcommand{\Mp}{M_\usssPl}

\newcommand{\kgamma}{k_{\gamma}}

\newcommand{\efolds}{$e$-folds}

\newcommand{\beq}{\begin{equation}}
\newcommand{\eeq}{\end{equation}}
\newcommand{\bea}{\begin{equation}\begin{aligned}}
\newcommand{\eea}{\end{aligned}\end{equation}}

\newlength{\wsingfig}
\newlength{\wdblefig}
\newlength{\wquadfig}
\newlength{\wtriplefig}
\newcommand{\Eq}[1]{Eq.~(\ref{#1})}
\newcommand{\Eqs}[1]{Eqs.~(\ref{#1})}
\newcommand{\Fig}[1]{Fig.~{\ref{#1}}}
\newcommand{\Figs}[1]{Figs.~{\ref{#1}}}
\newcommand{\Refa}[1]{Ref.~{\cite{#1}}}
\newcommand{\Refs}[1]{Refs.~{\cite{#1}}}
\newcommand{\Sec}[1]{Sec.~\ref{#1}}

\newcommand{\App}[1]{Appendix~\ref{#1}}

\newcommand{\deflen}[2]{%      
    \expandafter\newlength\csname #1\endcsname
    \expandafter\setlength\csname #1\endcsname{#2}%
}

\subheader{}

\title{Observational constraints on quantum decoherence \\during inflation}

\author[a]{J\'er\^ome Martin,}
\author[b]{Vincent Vennin}

\affiliation[a]{Institut d'Astrophysique de Paris, UMR
7095-CNRS, Universit\'e Pierre et Marie Curie, 98bis boulevard Arago,
75014 Paris, France}

\affiliation[b]{Laboratoire Astroparticule et Cosmologie, Universit\'e
  Denis Diderot Paris 7, 10 rue Alice Domon et L\'eonie Duquet, 75013 Paris, France}

\emailAdd{jmartin@iap.fr}
\emailAdd{vincent.vennin@apc.univ-paris7.fr}

\date{today}

\begin{document}
\sloppy

\abstract{
Since inflationary perturbations must generically couple to all degrees of freedom present in the early Universe, it is more realistic to view these fluctuations as an open quantum system interacting with an environment. Then, on very general grounds, their evolution can be modelled with a Lindblad equation. This modified evolution leads to quantum decoherence of the system, as well as to corrections to observables such as the power spectrum of curvature fluctuations. On one hand, current cosmological observations constrain the properties of possible environments and place upper bounds on the interaction strengths. On the other hand, imposing that decoherence completes by the end of inflation implies lower bounds on the interaction strengths. Therefore, the question arises of whether successful decoherence can occur without altering the power spectrum. In this paper, we systematically identify all scenarios in which this is possible. As an illustration, we discuss the case in which the environment consists of a heavy test scalar field. We show that this realises the very peculiar configuration where the correction to the power spectrum is quasi scale invariant. In that case, the presence of the environment improves the fit to the data for some inflationary models but deteriorates it for others. This clearly demonstrates that decoherence is not only of theoretical importance but can also be crucial for astrophysical observations.}

%\arxivnumber{18XX.XXXXX}
\maketitle
\section{Introduction}
\label{sec:intro}
One of the deepest insights of modern cosmology is that all structures
in our Universe (galaxies, clusters of galaxies, Cosmic Microwave
Background - CMB - anisotropies \etc) originate from vacuum quantum
fluctuations stretched by the expansion and amplified by gravitational
instability~\cite{Mukhanov:1981xt, Mukhanov:1982nu,
  Starobinsky:1982ee, Guth:1982ec, Hawking:1982cz, Bardeen:1983qw}
during an early epoch of accelerated expansion named
inflation~\cite{Starobinsky:1980te, Sato:1980yn, Guth:1980zm,
  Linde:1981mu, Albrecht:1982wi, Linde:1983gd}. This idea is strongly
supported by the data~\cite{Adam:2015rua, Ade:2015tva}, in particular
by the fact that we observe an almost scale invariant power spectrum
of curvature perturbations.

However, the quantum origins of the perturbations also raise new
issues. Clearly, the structures observed today are classical objects
and the question of how the quantum-to-classical transition occurred
in a cosmological context remains unanswered and has been the subject
of many investigations~\cite{Starobinsky:1986fx, Lesgourgues:1996jc,
  Kiefer:1998qe, Perez:2005gh, Campo:2005sv, Ellis:2006fy,
  Kiefer:2008ku, Valentini:2008dq, Sudarsky:2009za, Bassi:2010ss,
  PintoNeto:2011ui, Martin:2012pea, Canate:2012ua, Lochan:2012di,
  Das:2013qwa, Markkanen:2014dba, Maldacena:2015bha,
  Goldstein:2015mha, Leon:2015hwa, Valentini:2015sna, Martin:2015qta,
  Martin:2017zxs}. It is widely believed that
decoherence~\cite{Zurek:1981xq, Zurek:1982ii, Joos:1984uk} could have
played an important role in that process~\cite{Brandenberger:1992sr, Polarski:1995jg,
  Barvinsky:1998cq, Bellini:2001jm, Lombardo:2005iz, Kiefer:2006je,
  Martineau:2006ki, Burgess:2006jn, Prokopec:2006fc, Sharman:2007gi,
  Koksma:2010dt, Weenink:2011dd, Boyanovsky:2015jen, Nelson:2016kjm, Rostami:2017akw}. On
more general grounds, cosmological fluctuations have to couple (at
least gravitationally) to the other degrees of freedom present in the
Universe. They should thus be treated as an open quantum system rather
than an isolated one, and studying decoherence is an effective way to
investigate the role played by these additional degrees of freedom. As a consequence, decoherence is important not only for theoretical considerations but also for observational reasons. In other words, even if one denies its relevance in foundational issues of quantum mechanics, on the practical side, it must be taken into account since the presence of these interacting extra degrees of freedom appears to be unavoidable.

Under some very general conditions (to be discussed in this article),
the evolution of an open quantum system can be modelled through a
Lindblad equation, which describes how the interaction with the
environment modifies the evolution of the system. This modification is
such that the off-diagonal terms of the system density matrix go to
zero in a preferred basis selected by the form of the
interaction. Although this does not solve the quantum measurement
problem~\cite{Adler:2001us, Schlosshauer:2003zy} since it does not
explain how a definite outcome is obtained, it explains how a
preferred basis can be selected and why we do not see superpositions
of macroscopic objects.

Decoherence is not an instantaneous process but proceeds over a finite
time scale, controlled by the strength of the interaction between the
system and its environment. Over the duration of that interaction, the
environment can also change the diagonal elements of the density
operator, \ie the probabilities associated to the possible final
results. This is why in general, the environment does not only
suppress interferences between possible outcomes of a measurement, it
also changes their predicted probabilities of occurrence.

In a cosmological setting, this means that if decoherence occurs in
the early Universe, the statistical properties of cosmological
perturbations may be modified. Since the later are very well
constrained, in particular by measurements of the CMB temperature and
polarisation anisotropies~\cite{Adam:2015rua, Ade:2015tva}, this opens
up the possibility to observationally constrain cosmic
decoherence. The question we investigate in the present work is
therefore the following: Which interactions with which environments
allow sufficient decoherence to take place in the early Universe while
preserving the standard statistical properties of primordial
cosmological fluctuations as predicted from inflation and confirmed by
observations?

This article is organised as follows. In \Sec{sec:lindblad}, we
present the general Lindblad equation formalism for cosmological
perturbations during inflation. We pay special attention to how
correlation functions of the environment enter this equation, and to
the conditions under which it is valid. We also relate its parameters
to microphysical quantities in the case where the environment is made
of a heavy scalar field. In \Sec{sec:powerspectrum}, we calculate
environment induced corrections to the power spectrum of cosmological
perturbations in the case where the interaction is linear in the
Mukhanov-Sasaki variable (see \Sec{sec:PowerSpectrum:Linear}, where an
exact solution for the full density matrix is obtained), and in the
case where the interaction is quadratic (see
\Sec{sec:PowerSpectrum:Quadratic}, where mode coupling renders the
analysis more involved but the modified power spectrum can still be calculated exactly). Requiring that quasi scale invariance
is preserved, we derive an upper bound on the effective interaction
strength. In \Sec{sec:decoherencetime}, we calculate the level of
decoherence that cosmological perturbations undergo during inflation, in the case of linear interactions in \Sec{subsec:decotimelinear} and for quadratic interactions in \Sec{subsec:decotimequadratic}. Requiring that
decoherence has occurred by the end of inflation, we derive a lower
bound on the effective interaction strength that we compare with the
above-mentioned upper bound. This allows us to identify the class of
viable scenarios. In \Sec{sec:generalisation}, we generalise our approach to  arbitrary order interactions (\ie cubic and beyond), and we show how the power spectrum and the decoherence rate can still be calculated exactly. In \Sec{sec:conclusions}, we summarise our results and conclude by mentioning a few possible extensions. Finally, the
paper ends with a series of technical
appendices. \App{sec:DerivingLindblad} presents a detailed derivation
of the Lindblad equation, focusing on the assumptions needed to obtain
it, in order to determine whether and when they are satisfied in a
cosmological context. In \App{sec:massivescalarfield}, we discuss the
case where the environment is made of a heavy scalar field, which
allows us to relate the phenomenological parameters appearing in the
Lindblad equation to microphysical quantities. In
\App{sec:dmlinearinteraction}, we explain how an exact solution to the
Lindblad equation can be found if the system is linearly coupled to
the environment, and \App{sec:Cr} provides details for the calculation
of the power spectrum if the coupling is quadratic.
\section{Lindblad equation for inflationary perturbations in interaction with an environment}
\label{sec:lindblad}
During inflation curvature perturbations are described by the
Mukhanov-Sasaki variable~\cite{Mukhanov:1981xt,Kodama:1985bj}
$v\left(\eta,\bm{x}\right)$, where $\eta$ denotes conformal time and
$\bm{x}$ is the conformal spatial coordinate. This variable is a
combination of the perturbed inflaton field and of the Bardeen
potential, the latter being a generalisation of the gravitational
Newtonian potential~\cite{Bardeen:1980kt}. The free evolution
Hamiltonian for cosmological perturbations, $\hat{H}_{v}$, can be
expressed as~\cite{Mukhanov:1990me} 
\bea
\label{eq:Hsystem}
\hat{H}_v=\int _{\setR^3}\dd ^3 \bm{k} \, \hat{\mathcal{H}}_{\bm{k}}
=\frac 12\int_{\setR^3} \dd ^3 \bm{k}\left[
 \hat{p}_{\bm{k}}\hat{p}_{\bm{k}}^{\dagger}+\omega^2\left(\eta,\bm{k}\right)
\hat{v}_{\bm{k}}\hat{v}_{\bm{k}}^{\dagger}\right],
\eea
where $\hat{v}_{\bm k}(\eta)$ is the Fourier transform of the
Mukhanov-Sasaki variable, namely
\bea
\label{eq:v:Fourier}
\hat{v}\left(\eta,\bm{x}\right)=\frac{1}{\left(2\pi\right)^{3/2}}
\int _{\setR^3}\dd ^ 3\bm{k}
\, \hat{v}_{\bm{k}}(\eta)\ee^{i\bm{k}.\bm{x}}\, ,
\eea
and $\hat{p}_{\bm k}=\hat{v}_{\bm k}'$ is the Fourier transform of its conjugate momentum. Here a prime denotes a derivative with respect to conformal time $\eta $. The Hamiltonian~(\ref{eq:Hsystem}) describes a collection of parametric oscillators, the time-dependent frequency of which is given by
\bea
\label{eq:defomega}
\omega^2\left(\eta,\bm{k}\right)=
k^2-\frac{\left(a\sqrt{\epsilon_1}\right)^{\prime\prime}}
{a\sqrt{\epsilon_1}}\, , \eea where $a$ is the Friedman-Lema\^ \i
tre-Robertson-Walker scale factor, $\epsilon_1=1-{\cal H}'/{\cal H}^2$
the first slow-roll parameter and ${\cal H}=a'/a=aH$, $H$ being the
Hubble parameter (not to be confused with the Hamiltonian).
\subsection{Quantising inflationary perturbations}
\label{sec:quantising}
Since $\hat{v}(\eta,{\bm x})$ is real, one has
$\hat{v}_{-\bm k}=\hat{v}_{\bm k}^{\dagger}$. Decomposing
$\hat{v}_{\bm k}$ and $\hat{p}_{\bm k}$ into real and imaginary parts
according to
$\hat{v}_{\bm k}= \left(\hat{v}_{\bm k}^{\mathrm{R}}+i\hat{v}_{\bm
    k}^{\mathrm{I}}\right)/\sqrt{2}$
and
$\hat{p}_{\bm k}= \left(\hat{p}_{\bm k}^{\mathrm{R}}+i\hat{p}_{\bm
    k}^{\mathrm{I}}\right)/\sqrt{2}$,
this gives rise to
$\hat{v}_{-\bm k}^{\mathrm R}=\hat{v}_{\bm k}^{\mathrm R}$ and
$\hat{v}_{-\bm k}^{\mathrm I}=-\hat{v}_{\bm k}^{\mathrm I}$, and similar
expressions for $\hat{p}_{\pm\bm k}^{\mathrm{R,I}}$. This shows that not all
$v_{\bm k}$'s are independent degrees of freedom and that only the
variables $v_{\bm k}^{\mathrm{R}}$ and $v_{\bm k}^{\mathrm{I}}$ for
$\bm{k}\in\setR^{3+}$ must be quantised, \ie $\bm{k}$ runs on half of
the Fourier space. This is done through the canonical commutation
relations \bea \left[\hat{v}_{\bm k},\hat{p}_{\bm
    q}\right]=i\delta\left({\bm k}+{\bm q}\right)\, , \eea which also
imply that
$\left[\hat{v}_{\bm k}^{\dagger},\hat{p}_{\bm
    q}\right]=\left[\hat{v}_{\bm k},\hat{p}_{\bm q}^{\dagger}\right]
=i\delta\left({\bm k}-{\bm q}\right)$.

The quantum state of the perturbations is represented by a wavefunctional $\Psi[v(\eta,{\bm x})]$. In Fourier space, it reads 
\bea
\label{eq:psifouriergeneral}
\Psi\left[v(\eta,{\bm x})\right]=\Psi\left[\left\lbrace v_{\bm k}^{s}\left(\eta\right)\right\rbrace_{\bm{k}\in\setR^{3+},\,s\in\left\lbrace  \mathrm{R},\mathrm{I}\right\rbrace}\right]
\eea
\ie it is a function of the infinite number of Fourier components in $\setR^{3+}$. Since the free Hamiltonian~(\ref{eq:Hsystem}) can be written as a sum of independent Hamiltonians on each component of the Fock space, $\hat{H}_v=\int_{\setR^{3+}} \dd ^3\bm k\sum_{s=\mathrm{R,I}}\hat{\mathcal{H}}_{\bm k}^s$, with $\hat{\mathcal{H}}_{\bm k}^s=(\hat{p}_{\bm k}^s)^2/2+\omega^2(\hat{v}_{\bm k}^s)^2/2$, if the wavefunctional can be factorised initially it remains so at later times, 
\bea
\Psi\left[v\left(\eta,{\bm x}\right)\right]=\prod _{\bm k \in \setR^{3+}}
\Psi_{\bm k}\left(v_{\bm{k}}^\mathrm{R},
v_{\bm{k}}^\mathrm{I}\right)
=\prod_{\bm k \in \setR^{3+}}\Psi^{\mathrm R}_{\bm k}\left(v_{\bm{k}}^\mathrm{R}\right)
\Psi ^{\mathrm I}_{\bm k}\left(v_{\bm{k}}^\mathrm{I}\right)\, .
\label{eq:Psi:factorised}
\eea
Here, the dependence in $\eta$ has been dropped in the two latest expressions for display convenience, and the product has to be understood as a tensorial one (sometimes noted $\otimes$). As will be shown below, in the presence of non-linear interaction (\ie in the presence of mode coupling), this factorisation is no longer possible and one has to work with \Eq{eq:psifouriergeneral}. 

In order to include non-pure states in the analysis, one has to work in terms of the density matrix $\hat{\rho}_v=\left\vert\Psi\left[v\right]\right\rangle\left\langle\Psi\left[v\right]\right\vert$. In the free theory~(\ref{eq:Hsystem}), the factorisation~(\ref{eq:Psi:factorised}) gives rise to a similar one for the density matrix,
\bea
\label{eq:rho:factorization}
\hat{\rho}_v(\eta)=\prod_{\bm k\in \setR^{3+}}\prod_{s=\mathrm{R,I}}
\hat{\rho}_{\bm k}^{s}(\eta)\, .
\eea
As mentioned above, when non-linear interactions are introduced, this no longer holds.

The evolution of the system is controlled by the Schr\"odinger equation $\dd \left\vert\Psi\left[v\right]\right\rangle/\dd  t = - i \hat{H}_v \left\vert\Psi\left[v\right]\right\rangle$ or, equivalently, by the Liouville-von Neumann equation
\bea
\label{eq:Heisenberg:fullspace}
\frac{\dd \hat{\rho}_v}{\dd \eta}=
-i\left[\hat{H}_v,\hat{\rho}_v\right]\, .
\eea
If the state is factorisable, this can also be written in Fourier space as follows. Time differentiating \Eq{eq:rho:factorization}, one first has
\bea
\frac{\dd \hat{\rho}_v}{\dd  \eta} =\int_{\setR^{3+}}\dd ^3\bm k 
\left(\frac{{\mathrm d}\hat{\rho}_{\bm k}^{\mathrm R}}{{\mathrm d}\eta}
\hat{\rho}_{\bm k}^{\mathrm I}+\hat{\rho}_{\bm k}^{\mathrm R}
\frac{{\mathrm d}\hat{\rho}_{\bm k}^{\mathrm I}}{{\mathrm d}\eta}
\right)
\prod_{\bm k^\prime\neq\bm k}
\prod_{s=\mathrm{R},\mathrm{I}}
\hat{\rho}^{s}_{\bm k^\prime}\, .
\label{eq:ReciprocalLinbaldTerm1}
\eea
Then, using the fact that
$\hat{H}_v=\int_{\setR^{3+}} \dd ^3\bm
k\sum_{s=\mathrm{R,I}}\hat{\mathcal{H}}_{\bm k}^s$, the commutator in \Eq{eq:Heisenberg:fullspace} can be expressed as
\bea
\left[ \hat{H}_v,\hat{\rho}_v\right]&=
\int_{\setR^{3+}}\dd ^3\bm k\sum_{s=\mathrm{R,I}}\left[\hat{\mathcal{H}}_{\bm k}^{s},
\hat{\rho}_v\right]
%\nonumber
\\ &
=\int_{\setR^{3+}}\dd ^3\bm k
\left(\left[\hat{\mathcal{H}}_{\bm k}^{\mathrm{R}},\hat{\rho}_{\bm k}^{\mathrm{R}}
\right]\hat{\rho}_{\bm k}^{\mathrm{I}}
+\hat{\rho}_{\bm k}^{\mathrm{R}}
\left[\hat{\mathcal{H}}_{\bm k}^{\mathrm{I}},\hat{\rho}_{\bm k}^{\mathrm{I}}
\right]\right)
\prod_{\bm k^\prime\neq\bm k}\prod_{s=\mathrm{R,I}}\hat{\rho}_{\bm k^\prime}^{s}\, .
\label{eq:ReciprocalLinbaldTerm2}
\eea
Plugging \Eqs{eq:ReciprocalLinbaldTerm1} and~(\ref{eq:ReciprocalLinbaldTerm2}) into \Eq{eq:Heisenberg:fullspace}, one obtains
\bea
\label{eq:heisenbergfourier}
\frac{\dd \hat{\rho}_{\bm k}^s}{\dd \eta}=
-i\left[\hat{\mathcal{H}}_{\bm k}^s,\hat{\rho}_{\bm k}^s\right]\, .
\eea
This confirms that, in the absence of non-linear interactions, each Fourier subspace can be treated independently from the others.
\subsection{Including the interaction with an environment}
\label{sec:InteractionWithEnvironement}
The previous considerations assume that the cosmological perturbations
can be modelled as an isolated system. In practice however, there are
other degrees of freedom in the Universe that, on generic grounds,
interact with the perturbations. This is why cosmological
perturbations, here described by the set of variables
$v_{\bm k}(\eta)$ or, equivalently, $v(\eta,{\bm x})$, should rather
be modelled as an open system interacting with the ``environment''
comprising all other degrees of freedom associated with other fields,
cosmological perturbations outside our causal horizon, physics beyond
the UV or IR cutoffs of the theory, \etc~The total Hamiltonian can be
written as 
\bea \hat{H}=\hat{H}_{v}\otimes
\hat{\mathbb{I}}_{\mathrm{env}} + \hat{\mathbb{I}}_v \otimes
\hat{H}_{\mathrm{env}}+g\hat{H}_{\mathrm{int}}\, , 
\eea 
where $\hat{H}_v$ is the Hamiltonian~(\ref{eq:Hsystem}),
$\hat{H}_{\mathrm{env}}$ is the free evolution Hamiltonian for the
environment that we will not need to specify, $g$ is a dimensionless coupling
constant and $\hat{H}_{\mathrm{int}}$ is the interaction
Hamiltonian. Requiring that the system and the environment couple
through local interactions only, it can be expressed as 
\bea
\label{eq:localinter}
\hat{H}_{\mathrm{int}}\left(\eta\right)=\int\dd ^3\bm{x}\, 
\hat{A}\left(\eta,\bm{x}\right)\otimes\hat{R}
\left(\eta,\bm{x}\right)\, ,
\eea
where $\hat{A}$ belongs to the system sector and $\hat{R}$ belongs to the environment sector. 

In principle, $\hat{A}$ may involve the field operator $\hat{v}$ and its conjugated momentum $\hat{p}$. We however expect the interaction Hamiltonian to be dominated by terms depending on $\hat{v}$ only for several reasons. First, since one observes temperature fluctuations that are proportional to $\hat{v}$, the relevant pointer basis for decoherence must be given by field configurations~\cite{Kiefer:2006je}. Moreover, since the conjugated momentum $\hat{p}$ is proportional to the decaying mode, we expect its contribution to be subdominant. In addition, this is what is found in concrete examples. For instance, in \Refs{Burgess:2006jn, Nelson:2016kjm}, it is shown that cubic terms in the action for cosmological perturbations can induce decoherence of long wavelength fluctuations if the short wavelengths modes are collected as an environment. Some of these terms involve $\dot{\zeta}$, where $\zeta$ is the curvature perturbation, and can therefore be neglected as being proportional to the decaying mode. The remaining terms contain only spatial derivatives of $\zeta$, such as $\zeta(\partial_i\zeta)^2$, which implies that  $\hat{A}$ is proportional to the long wavelength part of $\hat{v}$ and $\hat{R}$ is proportional to the square of its short wavelength part (neglecting the spatial derivative of the long wavelength part). Here, we even consider the possibility of having higher-order terms in the action, leading to
\bea
\label{eq:A:vn}
\hat{A}=\hat{v}^n
\eea
where $n$ is a free index. This form is also obtained in the example detailed in \App{sec:massivescalarfield} where a massive test scalar field plays the role of the environment. Notice that if $\hat{A}$ is a more generic function of $\hat{v}$, the contributions from each term of its Taylor expansion can be computed from our result and summed up in the final result. In addition, in cases where the above generic arguments do not apply and $\hat{A}$ involves $\hat{p}$ explicitly, our method can still be employed as will be shown explicitly, see the discussion at the end of \Sec{subsec:linear:PowerSpectrum:Alternative}.
\subsubsection{Lindblad equation}
Let us notice that the interaction term~(\ref{eq:localinter}) is, strictly speaking, not of the form usually required to derive a Lindblad equation. However, in \App{sec:DerivingLindblad}, we show that the usual treatment can be generalised to an interaction term of the form~(\ref{eq:localinter}). In that Appendix, it is shown that, even if the full system starts off being described by a density matrix $\hat{\rho}$ for which there are no initial correlations between the system and the environment, $\hat{\rho}\left(\eta_\uin\right)=\hat{\rho}_v \left(\eta_\uin\right)\otimes \hat{\rho}_{\mathrm{env}}\left(\eta_\uin\right)$, then, at latter times, the system and the environment become entangled. Since we are only interested in tracking the evolution of the cosmological perturbations, let us introduce the reduced density matrix
\bea
\label{eq:rhov:reduced}
\hat{\rho}_v =  \underset{\mathrm{environment}}{\mathrm{Trace}}
\left(\left\vert v,\mathrm{env}\, \right\rangle\left\langle v, 
\mathrm{env}\, \right\vert\right)\, ,
\eea
where the environment degrees of freedom have been traced out. Under the assumption that the autocorrelation time of $\hat{R}$ in the environment, $\eta_{\uc}$, is much shorter than the time scale over which the system evolves, in \App{sec:DerivingLindblad} the reduced density matrix~(\ref{eq:rhov:reduced}) is shown to follow the non-unitary evolution equation 
\bea
\frac{\dd \hat{\rho}_v}{\dd  \eta} = -i\left[ \hat{H}_v,\hat{\rho}_v\right] 
-\frac{\gamma}{2}\int\dd ^3\bm x \, \dd ^3 {\bm y} 
\, C_{R}\left({\bm x},{\bm y}\right)
\left[\hat{A}(\bm x),\left[\hat{A}(\bm y),
\hat{\rho}_v\right]\right]\, ,
\label{eq:lindbladgeneral}
\eea
where $C_R$ is the same-time correlation function of $\hat{R}$ in the environment, $C_{R}({\bm x},{\bm y}) = \langle \hat{R}(\eta,{\bm x})\hat{R}(\eta,{\bm y}) \rangle $, and  the coefficient $\gamma $ is related to the coupling constant $g$ and to the autocorrelation (conformal) time $\eta_{\uc}$ of $\hat{R}$ in the environment, according to 
\bea
\label{eq:gamma:g:etac}
\gamma=2g^2\eta_{\uc}\, .
\eea
This parameter is, in general, time-dependent, and in what follows we assume that it is given by a power law in the scale factor
\bea
\label{eq:gamma}
\gamma =\gamma _*\left(\frac{a}{a_*}\right)^p,
\eea
where $p$ is a free index and a star refers to a reference time. For
convenience, we take it to be the time when the pivot scale
$k_* = 0.05\, \mathrm{Mpc}^{-1}$ crosses the Hubble radius during
inflation. The time dependence of $\gamma$ comes from the fact that
$g$ and $\eta_\uc$ are found to depend on time when it comes to
concrete models as will be exemplified below.

The Lindblad operator~(\ref{eq:lindbladgeneral}) is also a generator
of all quantum dynamical semigroups~\cite{Lindblad:1975ef}, \ie of the
transformations $F_\eta(\rho)$ of the density matrix indexed by the
time parameter $\eta$ that satisfy the Markovian property
$F_\eta[F_{\eta'}(\rho)]=F_{\eta+\eta'}(\rho)$. It therefore allows
one to investigate the dynamics of observable cosmological
fluctuations as an open quantum system on very generic grounds.

\subsubsection{Correlation function of the environment}
If the environment is in a statistically homogeneous configuration,
$C_R$ depends on $\bm{x}-\bm{y}$ only, and if statistical isotropy is
satisfied too, it simply depends on $\vert \bm{x}-\bm{y} \vert
$.
Assuming also that a single physical length scale $\lE$ is involved,
it is a function of $a\vert \bm{x}-\bm{y} \vert/\lE$, and in what
follows, for simplicity, we assume it to be a top-hat
function
\bea
\label{eq:tophatcorrelation}
C_R\left(\bm x , \bm y\right) = \bar{C}_R\, \Theta\left(\frac{a
    \left\vert \bm x - \bm y \right\vert}{\lE}\right)\, , 
\eea 
where $\Theta(x)$ is $1$ if $x<1$ and $0$ otherwise and $\bar{C}_R$ is a constant. Notice that the
appearance of the scale factor $a$ in the argument of the top-hat
function is due to the fact that $\bm x$ and $\bm y$ are comoving
coordinates while the correlation length $\lE$ is a physical scale.

With a given model for the environment, the correlation function can
in principle be calculated more precisely. In the following, the form
of $C_R(\bm{x},\bm{y})$ will be left unspecified as much as possible. In any case,
one expects the above approach to be a good approximation, since only
when physical scales are of the order of the correlation length of the
environment can our modelling be slightly inaccurate.
\subsubsection{A heavy scalar field as the environment}
In \App{sec:massivescalarfield}, the case where the environment consists of a scalar field $\psi$ with mass $M\gg H$ is investigated. In practice, the coupling between $\hat{v}$ and $\hat{\psi}$ is assumed to be of the form
\bea
\label{eq:Hint:massive:maintext}
\hat{H}_\mathrm{ int}=\lambda \mu^{4-n-m} \int \dd ^3 {\bm x} \sqrt{-g}\hat{\phi}^n\left(\eta,{\bm x}\right)
\hat{\psi}^m\left(\eta,{\bm x}\right)
\, ,
\eea
where $\mu$ is a fixed mass scale parameter, $\sqrt{-g}$ is the square root of minus the determinant of the metric and $\hat{\phi}=\hat{v}/a$. The correlation function of $\hat{\psi}^m$ can be calculated using renormalisation techniques for heavy scalar fields on de-Sitter space-times~\cite{Bunch:1977sq, Bunch:1978yq, Birrell:1982ix} and one finds
\bea
\bar{C}_R &= \left\lbrace \left(2m-1\right)!!-\sigma\left(m\right)\left[\left(m-1\right)!!\right]^2\right\rbrace
\left(\frac{37}{504\pi^2}\frac{H^6}{M^4}\right)^m\, ,\\
a\eta_\uc & = \lE = 2\sqrt{2}\sqrt{\frac{\left(2m-1\right)!!-\sigma\left(m\right)\left[\left(m-1\right)!!\right]^2}{m^2\left(2m-3\right)!!}} \frac{1}{M}\, .
\label{eq:MassivePsi:Cbar:tc:lE:maintext}
\eea
In this expression, $\sigma(m)$ is $1$ or $0$ depending on whether $m$ is even or odd, and ``$!!$'' denotes the double factorial. Defining the Lindblad operator as $\hat{A}=\hat{v}^n$, in agreement with \Eq{eq:A:vn}, the ansatz~(\ref{eq:gamma}) is realised with
\bea
\label{eq:gammastar:MassiveEnv}
\gamma_*&=
4\sqrt{2}\sqrt{\frac{\left(2m-1\right)!!-\sigma\left(m\right)\left[\left(m-1\right)!!\right]^2}{m^2\left(2m-3\right)!!}}
\frac{\lambda^2}{M} \mu^{8-2n-2m}a_*^{7-2n}
\, ,\\
p&=7-2n-6m\epsilon_{1*}\, .  
\eea 
The scaling of $\gamma$ with $a$, \ie $\gamma\propto a^p$ with $p=7-2n-6m\epsilon_{1*}$, can be understood as follows. In the
interaction Hamiltonian~(\ref{eq:Hint:massive:maintext}),
$\sqrt{-g}=a^4$ and the field redefinition $\phi=v/a$ contributes
$a^{-n}$, so the coupling constant $g$ introduced in
\App{sec:DerivingLindblad} (not to be confused with the determinant of
the metric) is time dependent and effectively scales as
$g\propto \lambda a^{4-n}$. Since the correlation cosmic time of the
environment $t_\uc$ is constant, $\eta_\mathrm{{c}}=t_\mathrm{{c}}/a$
scales as the inverse of the scale factor.\footnote{Notice that in
  \App{sec:DerivingLindblad}, the Lindblad equation is established in
  a non-cosmological setting in terms of the usual laboratory time
  $t$. In a cosmological context, this time corresponds to cosmic time
  $t$ (hence the notation). However, the cosmological Lindblad
  equation can also be written in terms of an arbitrary time label
  $\sigma$. This equation is given by
  \Eq{eq:Lindblad:finalformgeneralized} where $t$ is replaced by
  $\sigma$ and $t_\uc$ by $\sigma_\uc$, the correlation time measured
  in units of $\sigma$. As a consequence, \Eq{eq:gamma:g} reads
  $\gamma =2g^2\sigma_\uc$. Here, the cosmological Lindblad equation
  is established in conformal time which means $\sigma =\eta$.} We
conclude, using \Eq{eq:gamma:g:etac}, that
$\gamma\propto g^2\eta_\uc\propto a^{7-2n}$. Finally, since
$H\propto a^{-\epsilon_1}$ at first order in slow roll, $\bar{C}_R$ is
not strictly constant but scales as $a^{-6m\epsilon_1}$. This slow
time dependence can be absorbed in the definition of $\gamma$, by
shifting $p\rightarrow p-6m\epsilon_{1*}$, and in replacing $H$ with
$H_*$ in \Eq{eq:MassivePsi:Cbar:tc:lE:maintext}. For linear
interactions, $n=1$, $p\simeq 5$, for quadratic interactions, $n=2$,
$p\simeq 3$, and in \Sec{sec:powerspectrum} we will show why these
behaviours are in fact very remarkable.

The typical time scale over which the system evolves is of order the
Hubble time, so the assumption that it is much longer than the
environment autocorrelation time, which is necessary in order to
derive the Lindblad equation, amounts to $M\gg H$. In
\App{sec:massivescalarfield}, it is shown that this condition also
guarantees that $\psi$ is a test field (\ie does not substantially
contribute to the energy budget of the Universe). Another requirement
for the validity of the Lindblad approach is that the interaction
between the system and the environment does not affect much the
behaviour of the environment and only perturbatively affects the
system. In \App{sec:massivescalarfield} this is shown to be the case
if $\lambda\ll H^{6-3m} M^{2m-2}/ (\mu^{4-n-m} \phi^n)$. One concludes
that, if the additional field $\psi$ is sufficiently heavy, and if the
coupling constant $\lambda$ is sufficiently small, the influence of $\psi$
on the dynamics of the Mukhanov-Sasaki variable associated with $\phi$
can be studied with the Lindblad equation~(\ref{eq:lindbladgeneral}),
with the parameters given in
\Eqs{eq:gamma}-(\ref{eq:gammastar:MassiveEnv}).
\subsubsection{Evolving quantum mean values}
In order to extract observable predictions from the quantum state described by the density matrix $\hat{\rho}_v$, quantum expectation values
\bea
\left \langle \hat{O}\right\rangle =\mathrm{Tr}\left(\hat{\rho}_v\hat{O}\right)
\label{eq:meanO}
\eea
have to be calculated, where $\hat{O}$ is an arbitrary operator acting in the Hilbert space of the system. When the Lindblad equation~(\ref{eq:lindbladgeneral}) cannot be fully solved, it may also be convenient to restrict the analysis to (a subset of) such expectation values, as will be shown below. Differentiating \Eq{eq:meanO} with respect to time and plugging \Eq{eq:lindbladgeneral} in leads to
\bea
\label{eq:Lindblad:mean}
\frac{\dd \left\langle \hat{O}\right\rangle}
{\dd  \eta}&=
\left\langle\frac{\partial \hat{O}}{\partial \eta}\right\rangle
-i \left\langle \left[\hat{O},\hat{H}_{v}\right] \right\rangle
%\nonumber \\ &
-\frac{\gamma}{2}\int\dd ^3\bm x \, \dd ^3 {\bm y} 
\, C_{R}\left({\bm x},{\bm y}\right)
\left\langle\left[\left[\hat{O},\hat{A}(\bm x)\right],
\hat{A}({\bm y})\right]\right\rangle
\, .
\eea
In this expression, $\partial \hat{O}/\partial \eta$ accounts for a possible explicit time dependence of the operator $\hat{O}$.  The term describing  the interaction between the system and the environment can be written in Fourier space, and one obtains
\bea
\label{eq:Lindblad:mean:Fourrier}
\frac{\dd \left\langle \hat{O}\right\rangle}
{\dd  \eta}&=
\left\langle\frac{\partial \hat{O}}{\partial \eta}\right\rangle
-i \left\langle \left[\hat{O},\hat{H}_{v}\right] \right\rangle
%\nonumber \\ &
-\frac{\gamma}{2}(2\pi)^{3/2}\int_{\setR^3}\dd ^3\bm k  
\, \tilde{C}_{R}\left({\bm k}\right)
\left\langle\left[\left[\hat{O},\hat{A}_{\bm k}\right],
\hat{A}_{-{\bm k}}\right]\right\rangle\, .
\eea
To derive this expression, we have assumed that the environment is placed in a statistically homogeneous configuration such that, as stated above, $C_R(\bm{x},\bm{y})=C_R(\bm{x}-\bm{y})$, and the functions $C_R(\bm{x}-\bm{y})$ and $\hat{A}(\bm{x})$ have been Fourier expanded in a similar way as in \Eq{eq:v:Fourier}. Using the top-hat ansatz~(\ref{eq:tophatcorrelation}), one obtains
\bea
\tilde{C}_R\left({\bm k}\right) =
\sqrt{\frac{2}{\pi}}\frac{\bar{C}_R}{k^3}
\left[\sin\left(\frac{k\lE }{a}\right)
-\frac{k\lE }{a}\cos\left(\frac{k\lE }{a}\right)\right]\, ,
\label{eq:Ck:fromstep}
\eea
where $k$ stands for the modulus of $\bm{k}$. The fact that $\tilde{C}_R({\bm k}) $ depends only on $k$ is related to the statistical isotropy assumption behind \Eq{eq:tophatcorrelation}, namely the fact that $C_R(\bm{x}-\bm{y})$ depends only on $\vert \bm{x}-\bm{y} \vert$. The Fourier transform~(\ref{eq:Ck:fromstep}) can itself be approximated by a top-hat function of $k\lE/a$, 
\bea
\tilde{C}_R({\bm k}) \simeq \sqrt{\frac{2}{\pi}}\frac{ \bar{C}_R\lE^3}{3a^3} \Theta\left(\frac{k\lE}{a}\right)\, ,
\label{eq:Ck:appr}
\eea
where the amplitude at the origin has been matched.
\section{Power spectrum}
\label{sec:powerspectrum}
In the previous section we have shown how interactions with the environment can be modelled through the addition of a non-unitary term in the evolution equation of the density matrix for the system, which becomes of the Lindblad type~(\ref{eq:lindbladgeneral}). As explained in \Sec{sec:intro}, this new term leads to the dynamical suppression of the off-diagonal elements of the density matrix when written in the basis of the eigenstates of the operator through which the system couples to the environment [$\hat{A}$ in the notations of \Eq{eq:lindbladgeneral}, which here we take to be some power of the Mukhanov-Sasaki variable $\hat{v}$, see \Eq{eq:A:vn}]. This will be explicitly shown in \Sec{sec:decoherencetime} and is at the basis of the ``decoherence'' mechanism. However, \Eq{eq:lindbladgeneral} also has a unitary term which comes from the free Hamiltonian of the system. If that term couples the evolution of the diagonal elements of the density matrix to the non-diagonal ones, the Lindblad term also induces modifications of the diagonal elements of the density matrix, hence of the expected probabilities of observing given values of $\hat{v}$, that is to say of the observable predictions for measurements of the system. In particular, the power spectrum of cosmological curvature perturbations is altered by the presence of the Lindblad term, and in this section we calculate the corrected power spectrum. We then determine how observations constrain the size of this correction, and thus place bounds on the strength of interactions with the environment.
\subsection{Linear interaction}
\label{sec:PowerSpectrum:Linear}
Let us first consider the case where the system couples linearly to the environment, $\hat{A}(\bm{x})=\hat{v}(\bm{x})$, \ie $n=1$ in \Eq{eq:A:vn}. We will show that the Lindblad equation can be solved exactly in that case, \ie all elements of the density matrix will be given explicitly. In particular, we will show how the power spectrum can be extracted and how the correction it receives from the Lindblad term can be studied.
\subsubsection{Lindblad equation in Fourier space}
\label{subsec:eomlinear}
Let us first show that if $n=1$, the Lindblad
equation~(\ref{eq:lindbladgeneral}) decouples into a set of independent Lindblad
equations in each Fourier subspace. Since we have shown that this is
already the case in the free theory, see \Eq{eq:heisenbergfourier}, it is enough to consider the interaction term only and to show that it has the same property. From \Eq{eq:lindbladgeneral}, the interaction term is given by
\bea
 \int & \dd ^3\bm x \, \dd ^3 {\bm y} \, C_R({\bm x}-{\bm y})
\left[\hat{v}(\bm x),\left[\hat{v}(\bm y),\hat{\rho}_v\right]\right] =
(2\pi)^{3/2}\int _{\setR^3}
\dd  ^3 {\bm p}\, \tilde{C}_R({\bm p})
\left[\hat{v}_{\bm p}^{\dagger},\left[\hat{v}_{\bm p},\hat{\rho}_v\right]\right]
\\ &
\kern-1em =\frac{(2\pi)^{3/2}}{2} \int_{\setR^3}\,
\dd ^3\bm p \, \tilde{C}_R(\bm p)\biggl(
\left[\hat{v}_{\bm p}^{\mathrm{R}},
\left[\hat{v}_{\bm p}^{\mathrm{R}},\hat{\rho}_v\right]\right]
+
\left[\hat{v}_{\bm p}^{\mathrm{I}},\left[\hat{v}_{\bm p}^{\mathrm{I}},
\hat{\rho}_v\right]\right]
% \\ &
-i
\left[\hat{v}_{\bm p}^{\mathrm{I}},\left[\hat{v}_{\bm p}^{\mathrm{R}},
\hat{\rho}_v\right]\right]
+i
\left[\hat{v}_{\bm p}^{\mathrm{R}},\left[\hat{v}_{\bm p}^{\mathrm{I}},
\hat{\rho}_v\right]\right]\biggr)\, ,
\label{eq:inter:lin:FourierTransformed}
\eea
where in the first equality we have Fourier expanded $C_R$ and $\hat{v}$ and in the second equality we have used the decomposition $\hat{v}_{\bm p}= \left(\hat{v}_{\bm p}^{\mathrm{R}}+i\hat{v}_{\bm p}^{\mathrm{I}}\right)/\sqrt{2}$ introduced in \Sec{sec:quantising}. In \Eq{eq:inter:lin:FourierTransformed}, the two last terms vanish. Indeed, one can split the integral over $\setR^3$ into two
pieces, namely over $\setR^{3+}$ and $\setR^{3-}$, and, in the second piece,
perform the change of variable ${\bm p}\rightarrow -{\bm p}$.  Using
the symmetry relation
$\hat{v}_{-\bm p}^{\mathrm R}=\hat{v}_{\bm p}^{\mathrm R}$ and
$\hat{v}_{-\bm p}^{\mathrm I}=-\hat{v}_{\bm p}^{\mathrm I}$ together with the fact that
the correlation function of the environment only depends on the
modulus of the wavevector (hence is independent of its sign), one obtains that the two pieces of the integral cancel out each other. If the state is factorisable as in \Eq{eq:rho:factorization}, one then finds that the interaction term can be similarly factorised,
\bea
\int & \dd ^3{\bm x} \, \dd ^3{\bm y}\, C_R\left({\bm x}-{\bm y}\right)
\left[\hat{v}(\bm x),\left[\hat{v}(\bm y),\hat{\rho}_v\right]\right]
= \\ & \qquad\quad
(2\pi)^{3/2}
\int_{\setR^{3+}}\,\dd ^3\bm p\, \tilde{C}_R(\bm p)
\biggl(\left[\hat{v}_{\bm p}^{\mathrm{R}},\left[\hat{v}_{\bm p}^{\mathrm{R}},
\hat{\rho}_{\bm p}^{\mathrm{R}}\right]\right]
\hat{\rho}_{\bm p}^{\mathrm{I}}
+ \hat{\rho}_{\bm p}^{\mathrm{R}}
\left[\hat{v}_{\bm p}^{\mathrm{I}},\left[\hat{v}_{\bm p}^{\mathrm{I}},
\hat{\rho}_{\bm p}^{\mathrm{I}}\right]\right]\biggr)
\prod_{\bm p^\prime\neq\bm p}\prod_{s'=\mathrm{R,I}}
\hat{\rho}_{\bm p^\prime}^{s'}\, .
\label{eq:ReciprocalLinbaldTerm3}
\eea
The fact that the interaction term is linear in $\hat{v}(\eta,{\bm x})$
thus preserves the property that each Fourier mode evolves separately, and combining \Eqs{eq:ReciprocalLinbaldTerm1},
(\ref{eq:ReciprocalLinbaldTerm2}), and~(\ref{eq:ReciprocalLinbaldTerm3}), one obtains
\bea
  \frac{\dd \hat{\rho}_{\bm k}^s}{\dd  \eta}=
-i\left[ \hat{\mathcal{H}}_{\bm k}^{s},\hat{\rho}_{\bm k}^{s}\right]
-\frac{\gamma}{2}(2\pi)^{3/2}
\tilde{C}_R({\bm k})\left[\hat{v}_{\bm k}^{s},\left[\hat{v}_{\bm k}^{s},
\hat{\rho}_{\bm k}^{s}\right]\right]\, .
\label{eq:lindbladlinear}
\eea

Let us also notice that a particular comoving scale appears in the interaction term. Indeed, in order for \Eq{eq:lindbladlinear} to have the correct dimension, $\gamma \tilde{C}_R(\bm k)$ must be homogeneous to the square of a comoving wavenumber. In what follows, we denote this scale
$\kgamma$, and using the form~(\ref{eq:Ck:appr}), it can be written as
\bea
\label{eq:kbreak:def}
\kgamma \equiv \sqrt{\frac{8\pi}{3}\bar{C}_R\lE^3\frac{\gamma _*}
{a_*^3}}\, ,
\eea
where $8\pi/3$ has been included for future convenience. In terms of the microphysical parameters of the model described in \App{sec:massivescalarfield} where the environment is made of a heavy test scalar field, making use of \Eqs{eq:MassivePsi:Cbar:tc:lE:maintext} and~(\ref{eq:gammastar:MassiveEnv}), one has
\bea
\frac{\kgamma}{k_*} = 32\sqrt{\frac{\pi}{3}}\left(\frac{37}{504 \pi^2}\right)^{\frac{m}{2}}
 \frac{\left\lbrace \left(2m-1\right)!!-\sigma\left(m\right)\left[\left(m-1\right)!!\right]^2 \right\rbrace^{\frac{3}{2}}}{m^2{\left(2m-3\right)!!}} g
 \left(\frac{H_*}{M}\right)^{3m-1}
\left( \frac{M}{\mu}\right)^{m-3}\, ,
\label{eq:kbreak:n_eq_1:microParam}
\eea
where the relation $k_* = a_* H_*$ has been used and where one can check that the right-hand side is indeed dimensionless.  
\subsubsection{Solution to the Lindblad equation}
\label{subsec:LindbladSol:Linear}
We now show how \Eq{eq:lindbladlinear} can be solved exactly, leading to the solution of the full Lindblad equation~(\ref{eq:lindbladgeneral}). Let us introduce the eigenvectors $\vert v_{\bm k}^s\rangle$ of the operator $\hat{v}_{\bm k}^s$, \ie the states such that $\hat{v}_{\bm k}^s \vert v_{\bm k}^s\rangle = v_{\bm k}^s \vert v_{\bm k}^s\rangle$. By projecting \Eq{eq:lindbladlinear} onto $\langle v_{\bm k}^{s,(1)} \vert$ on the left and $\vert  v_{\bm k}^{s,(2)} \rangle$ on the right, one has
\bea
\label{eq:diffdensitymatrix}
\frac{\dd \left\langle v^{s,(1)}_{\bm k}\right\vert \hat{\rho}_{\bm k}^s 
\left\vert v^{s,(2)}_{\bm k}\right\rangle}{\dd \eta}= &
\Bigg{\lbrace}
 \frac{i}{2}\left[\frac{\partial^2}{\partial {v^{s,(1)}_{\bm k}}^2}
-\frac{\partial^2}{\partial {v^{s,(2)}_{\bm k}}^2}\right]
-i\frac{\omega^2\left(k\right)}{2}\left[{v^{s,(1)}_{\bm k}}^2
-{v^{s,(2)}_{\bm k}}^2\right]
 \\  &
- \frac{\gamma}{2}\left(2\pi\right)^{3/2}\tilde{C}_R({\bm k})
\left[{v^{s,(1)}_{\bm k}}-{v^{s,(2)}_{\bm k}}\right]^2
\Bigg\rbrace
\left\langle v^{s,(1)}_{\bm k}\right\vert \hat{\rho}^s_{\bm k} 
\left\vert v^{s,(2)}_{\bm k}\right\rangle\, ,
\eea
where \Eq{eq:Hsystem} for the free Hamiltonian has been used with the representation $\hat{p}_{\bm k}^s = -i \partial/(\partial v_{\bm{k}}^{s})$ of the momentum operator in position basis. If the generic element $\langle v^{s,(1)}_{\bm k}\vert \hat{\rho}_{\bm k}^s  \vert v^{s,(2)}_{\bm k}\rangle$ of the density matrix $\hat{\rho}_{\bm k}^s$ is seen as a function of $v^{s,(1)}_{\bm{k}}$, $v^{s,(2)}_{\bm{k}}$ and $\eta$, the above equation can be interpreted as a linear, second-order, partial differential equation. In \App{sec:dmlinearinteraction}, it is shown that expressed in terms of the variables $v^{s,(1)}_{\bm{k}}-v^{s,(2)}_{\bm{k}}$ and $v^{s,(1)}_{\bm{k}}+v^{s,(2)}_{\bm{k}}$, \Eq{eq:diffdensitymatrix} leads to a first-order partial differential equation when the coordinate $v^{s,(1)}_{\bm{k}}+v^{s,(2)}_{\bm{k}}$ is Fourier transformed, and can be solved with the method of characteristics. One obtains the solution given in \Eq{eq:finalrho}, namely
\bea
\label{eq:finalrhomaintext}
& \left\langle  v_{\bmk}^{s,(1)}  \right\vert \hat{\rho}_{\bm k}^s\left\vert v_{\bmk}^{s,(2)}\right\rangle  = 
\frac{\left(2\pi\right)^{-1/2}}
{\sqrt{\left\vert v_{\bm k}\right\vert^2+\mathcal{J}_{\bm k}}}
\exp\left\{-\frac{{v^{s,(2)}_{\bm k}}^2+
{v^{s,(1)}_{\bm k}}^2
+i{\left\vert v_{\bm k}\right\vert^2}^\prime
\left[{v^{s,(2)}_{\bm k}}^2-{v^{s,(1)}_{\bm k}}^2\right]}
{4\left(\left\vert v_{\bm k}\right\vert^2
+\mathcal{J}_{\bm k}\right)}\right\}
 \\ & \qquad\qquad \times
\exp\Biggl\{-\frac{\left[v^{s,(2)}_{\bm k}-v^{s,(1)}_{\bm k}\right]^2}{2\left(\left\vert v_{\bm k}\right\vert^2
+\mathcal{J}_{\bm k}\right)}
\left(\mathcal{I}_{\bm k}\mathcal{J}_{\bm k}
-\mathcal{K}_{\bm k}^2+\left\vert v_{\bm k}^\prime \right\vert^2
\mathcal{J}_{\bm k}+ \left\vert v_{\bm k}\right\vert^2\mathcal{I}_{\bm k}
-{\left\vert v_{\bm k}\right\vert^2}^\prime \mathcal{K}_{\bm k}\right)
 \\ &  \qquad\qquad
-\frac{i\mathcal{K}_{\bm k}}{2\left(\left\vert v_{\bm k}\right\vert^2
+\mathcal{J}_{\bm k}\right)}
\left[{v^{s,(2)}_{\bm k}}^2-{v^{s,(1)}_{\bm k}}^2\right]\Biggr\},
\eea
where the quantities ${\cal I}_{\bm k}$, ${\cal J}_{\bm k}$ and
${\cal K}_{\bm k}$ are defined by 
\begin{align}
\label{eq:defI}
\mathcal{I}_{\bm k}\left(\eta\right) &\equiv
4\left(2\pi\right)^{3/2}\int_{-\infty}^\eta \dd \eta'\gamma\left(\eta'\right) 
\tilde{C}_R\left(\bm k,\eta'\right) \mathrm{Im}^2\left[v_{\bm k}
\left(\eta'\right){v_{\bm k}^*}^\prime\left(\eta\right)\right]\, ,\\
\label{eq:defJ}
\mathcal{J}_{\bm k}\left(\eta\right) &\equiv 4\left(2\pi\right)^{3/2}
\int_{-\infty}^\eta \dd \eta'\gamma\left(\eta'\right) \tilde{C}_R\left(\bm k,
\eta'\right)
 \mathrm{Im}^2\left[v_{\bm k}\left(\eta'\right){v_{\bm k}^*}
\left(\eta\right)\right] \, ,\\
\label{eq:defK}
\mathcal{K}_{\bm k}\left(\eta\right) &\equiv 4\left(2\pi\right)^{3/2}
\int_{-\infty}^\eta \dd \eta'\gamma\left(\eta'\right) \tilde{C}_R
\left(\bm k,\eta'\right)
\mathrm{Im}\left[v_{\bm k}\left(\eta'\right){v_{\bm k}^*}^\prime
\left(\eta\right)\right]
 \mathrm{Im}\left[v_{\bm k}\left(\eta'\right){v_{\bm k}^*}
\left(\eta\right)\right]\, ,
\end{align}
and $v_{\bm k}(\eta)$ is the solution of the Mukhanov-Sasaki mode equation $v''_{\bm k}+\omega^2(k)v_{\bm{k}}=0$ with initial conditions set in the Bunch-Davies vacuum. With \Eq{eq:rho:factorization}, this provides an exact solution to the full Lindblad equation~(\ref{eq:lindbladgeneral}). Because of the linearity of the interaction term, the state is still Gaussian, and one can check that it is properly normalised, $\mathrm{Tr}(\rho_v)=1$. If one sets $\gamma=0$, \ie if one switches off the interaction with the environment, one can also check that 
\bea
\label{eq:purestate}
\left\langle v_{\bmk}^{s,(1)} \right\vert \hat{\rho}_{\bm k}^s \left\vert v_{\bmk}^{s,(2)}\right\rangle_{\gamma=0} = \Psi_{\bm{k}}^s\left({v^{s,(1)}_{\bm k}}\right){\Psi_{\bm{k}}^{s*}}\left({v^{s,(2)}_{\bm k}}\right)\, ,
\eea
with the wavefunction $\Psi_{\bm{k}}^s(v)\propto \ee^{i \frac{v_{\bm{k}}'}{2v_{\bm{k}}}v^2}$. The standard two-mode squeezed state, which is a pure state, is therefore recovered in that limit.

Notice that the ability to set initial conditions for the perturbations in the Bunch-Davies vacuum, one of the most attractive features of inflation, is preserved by the Lindblad equation, thanks to the presence of the environment correlation function. Indeed, as long as the mode has not crossed out the environment correlation length, it is unaffected by the presence of the environment, see \Eq{eq:Ck:appr}.
\subsubsection{Two-point correlation function}
\label{subsec:twopointlinear}
In the solution~(\ref{eq:finalrhomaintext}), one can check that the diagonal elements of the density matrix, \ie the coefficients obtained by setting $v_{\bm{k}}^{s,(1)}=v_{\bm{k}}^{s,(2)}$, are affected by the presence of the environment since they involve the quantity $\mathcal{J}_{\bm{k}}$ defined in \Eq{eq:defJ}, which depends on $\gamma$. This confirms that the observable predictions one can draw from the state~(\ref{eq:finalrhomaintext}) are modified by the interaction with the environment. Since this state is still Gaussian, this modification is entirely captured by the change in the two-point correlation function, \ie the power spectrum, that we now calculate. 

The quantum mean value of $\left(\hat{v}_{\bm{k}}^{s}\right)^2$ can be expressed as
\bea
P_{vv}\left(k\right) \equiv 
\left\langle \left\vert\hat{v}_{\bm{k}}\right\vert^2 \right\rangle=
\left\langle \left(\hat{v}_{\bm{k}}^s\right)^2 \right\rangle =
\mathrm{Tr}\left[\left(\hat{v}_{\bm{k}}^s\right)^2 \hat{\rho}_v \right] =
\int_{-\infty}^{\infty} \dd  v_{\bmk}^{s} \left\langle v_{\bmk}^{s} \left\vert
    \hat{\rho}_{\bmk}^{s}\right\vert v_{\bmk}^{s}\right\rangle \left(v_{\bmk}^{s}\right)^2\, .
\eea
Making use of \Eq{eq:finalrhomaintext}, this integral is Gaussian and can be performed easily, leading to
\bea
\label{eq:solPvv}
P_{vv}\left(k\right) =\left\vert v_{\bm k}\right\vert^2+\mathcal{J}_{\bm k} \, .  
\eea
In the absence of interaction with the environment,
$\mathcal{J}_{\bm k}=0$ and one recovers the standard result. The
power spectrum of curvature perturbations can be directly obtained
from the relation $\zeta_{\bm k} = v_{\bm k}/(a\sqrt{2\epsilon_1}\Mp)$,
and this leads to
\bea
\label{eq:modifiedPzeta}
\calP_{\zeta}=\frac{k^3}{2\pi^2}
\frac{P_{vv}}{2a^2\epsilon_1\Mp^2}
=\left.\calP_\zeta\right\vert_{\mathrm{standard}}
\left(1+\Delta {\cal P}_{\bm k}\right),
\eea
with
\bea
\label{eq:defDPlinear}
\Delta {\cal P}_{\bm k}\equiv \frac{{\cal J}_{\bm k}}{\left\vert v_{\bm k}\right\vert^2}\, .
\eea
\subsubsection{Alternative derivation of the power spectrum}
\label{subsec:linear:PowerSpectrum:Alternative}
Before proceeding to the ``concrete'' calculation of the modified
power spectrum, \ie of the quantity ${\cal J}_{\bm k}/\vert v_{\bm k}\vert^2$, let us
notice that the above result can also be obtained without solving for the Lindblad equation~(\ref{eq:lindbladgeneral}) entirely, but by restricting the analysis to two-point correlators. This technique will be of particular convenience in the case of quadratic interaction with the environment since there, no explicit solution to the Lindblad equation can be found, see \Sec{sec:PowerSpectrum:Quadratic}. 

Let us first consider the case of one-point correlators. Making use of \Eq{eq:Lindblad:mean:Fourrier} with $\hat{O}=\hat{v}_{\bm k}$ and $\hat{O}=\hat{p}_{\bm k}$, one has
\bea
\label{eq:linearlinear}
\frac{\dd  \left\langle \hat{v}_{\bm k}\right\rangle}{\dd  \eta } =
\left\langle \hat{p}_{\bm k} \right\rangle \, , \qquad 
\frac{\dd  \left\langle \hat{p}_{\bm k} \right\rangle}{\dd  \eta } =
-\omega^2(k) \left\langle \hat{v}_{\bm k} \right\rangle\, .
\eea
This is nothing but the Ehrenfest theorem since no correction due to
the interaction with the environment appears in these
equations. Combined together, they lead to $\langle \hat{v}_{\bm k}\rangle'' + \omega^2(k)\langle \hat{v}_{\bm k}\rangle=0$, \ie the classical Mukhanov-Sasaki equation for $\langle \hat{v}_{\bm k}\rangle$. Since the state is initially symmetric, $\langle \hat{v}_{\bm k}\rangle=\langle \hat{p}_{\bm k}\rangle=0$, it remains so at all times.

If one then considers two-point correlators of the form
$\langle \hat{O} \rangle = \langle \hat{O}_{{\bm k}_1} \hat{O}_{{\bm k}_2}\rangle$ with $\hat{O}_{{\bm k}_i}=\hat{v}_{\bm{k}_i}$ or $\hat{p}_{\bm{k}_i}$, one obtains 
\bea
\label{eq:vvlinear:vplinear:pvlinear:pplinear}
\frac{\dd  }{\dd  \eta} \left\langle \hat{v}_{{\bm k}_1}
\hat{v}_{{\bm k}_2}\right\rangle &=\left \langle 
\hat{v}_{{\bm k}_1}
\hat{p}_{{\bm k}_2}
+\hat{p}_{{\bm k}_1}
\hat{v}_{{\bm k}_2}
\right\rangle, \\
%\label{eq:vplinear}
\frac{\dd  }{\dd  \eta} \left\langle \hat{v}_{{\bm k}_1}
\hat{p}_{{\bm k}_2}\right\rangle &=\left \langle 
\hat{p}_{{\bm k}_1}
\hat{p}_{{\bm k}_2}\right \rangle
-\omega^2(k_2)\left\langle
\hat{v}_{{\bm k}_1}
\hat{v}_{{\bm k}_2}
\right\rangle,\\
%\label{eq:pvlinear}
\frac{\dd  }{\dd  \eta} \left\langle \hat{p}_{{\bm k}_1}
\hat{v}_{{\bm k}_2}\right\rangle &=\left \langle 
\hat{p}_{{\bm k}_1}
\hat{p}_{{\bm k}_2}\right \rangle
-\omega^2(k_1)\left\langle
\hat{v}_{{\bm k}_1}
\hat{v}_{{\bm k}_2}
\right\rangle,\\
%\label{eq:pplinear}
\frac{\dd  }{\dd  \eta} \left\langle \hat{p}_{{\bm k}_1}
\hat{p}_{{\bm k}_2}\right\rangle &=
-\omega^2(k_2)\left\langle\hat{p}_{{\bm k}_1}
\hat{v}_{{\bm k}_2}\right \rangle
-\omega^2(k_1)\left\langle
\hat{v}_{{\bm k}_1}
\hat{p}_{{\bm k}_2}
\right\rangle
+\gamma (2\pi)^{3/2}
\tilde{C}_{R}({\bm k}_1)\delta ({\bm k}_1+{\bm k}_2).
\eea
One can see that the environment induces a new term proportional to $\gamma$ only in the evolution equation for $\left\langle \hat{p}_{\bm{k}_1}\hat{p}_{{\bm k}_2}\right\rangle$, but since the previous equations are all coupled together, it affects the evolution of all two-point correlators. The appearance of a Dirac function $\delta({\bm k}_1+{\bm k}_2)$ in the modification to the last equation is important since it means that the interaction with the environment preserves statistical homogeneity, \ie solutions of the form
\bea
\label{eq:homoP}
\left\langle \hat{O}_{{\bm k}_1}\hat{O}'_{{\bm k}_2}\right \rangle
=P_{OO'}(\bm{k}_1)\, \delta ({\bm k}_1+{\bm k}_2).
\eea
can be found.\footnote{Indeed, if one calculates the two-point correlation function in real space and
uses \Eq{eq:homoP}, due to the presence of the Dirac function,
one obtains 
\bea
\left\langle \hat{O}({\bm x}_1)\hat{O}'({\bm x}_2)
\right \rangle=\frac{1}{(2\pi)^3}
\int_{\setR^3} \dd ^3{\bm k}_1 P_{OO'}(\bm k_1)\, \mathrm{e}^{i{\bm k}_1\cdot ({\bm x}_1-{\bm x}_2)},
\eea
which is a function of ${\bm x}_1-{\bm x}_2$ only and is invariant by translating both ${\bm x}_1$ and ${\bm x}_2$ with a constant displacement vector $\bm{u}$. Conversely, one can show that if $\langle \hat{O}({\bm x}_1)\hat{O}'({\bm x}_2) \rangle = \langle \hat{O}({\bm x}_1+\bm{u})\hat{O}'({\bm x}_2+\bm{u}) \rangle$ for all $\bm{u}$, then $\langle \hat{O}_{{\bm k}_1}\hat{O}'_{{\bm k}_2}\rangle=\langle \hat{O}_{{\bm k}_1}\hat{O}'_{{\bm k}_2}\rangle \ee^{i\bm{u}(\bm{k}_1+\bm{k}_2)}$ for all $\bm{u}$, hence \Eq{eq:homoP} must be true.} If the environment correlator also preserves statistical isotropy, $\tilde{C}_{R}({\bm k})=\tilde{C}_{R}(k)$, as is the case of the ansatz~(\ref{eq:Ck:appr}) adopted in this work, the above system admits statistically homogeneous and isotropic solutions where $P_{OO'}$ in \Eq{eq:homoP} is a function of the modulus of the wavenumber only. Plugging
\Eq{eq:homoP} into \Eqs{eq:vvlinear:vplinear:pvlinear:pplinear}, one then obtains 
\bea
\frac{\dd  P_{vv}(k)}{\dd  \eta}&=P_{vp}(k)+P_{pv}(k)\, , \\
\frac{\dd  P_{vp}(k)}{\dd  \eta} &= \frac{\dd  P_{pv}(k)}{\dd  \eta}
=P_{pp}(k)-w^2(k)P_{vv}(k)\, , \\
\frac{\dd  P_{pp}(k)}{\dd  \eta}&= -\omega^2(k)\left[P_{pv}(k)+P_{vp}(k)\right]
+\gamma (2\pi)^{3/2}\tilde{C}_{R}\left(k\right)\, .
\eea
These equations can be combined into a single third-order equation for $P_{vv}$ only, which reads
\bea
\label{eq:thirdvlinear}
P_{vv}'''+4\omega^2P_{vv}'+4\omega'\omega P_{vv}
=S_1\, ,
\eea
with $S_1$ being defined as
\bea
\label{eq:source:linear:def}
S_1\left(\bm{k},\eta\right)\equiv 2 (2\pi)^{3/2}\gamma
\tilde{C}_{R}\left(k\right)\, .
\eea

The consistency check is then to verify that the solution obtained in \Eq{eq:solPvv} is indeed a solution of \Eq{eq:thirdvlinear}. Using the explicit form of ${\cal J}_{\bm k}$ given by \Eq{eq:defJ}, we see that differentiating \Eq{eq:solPvv} requires to differentiate a function of the form $\int _{-\infty}^\eta \dd  \eta' g(\eta',\eta)$, which gives $g(\eta,\eta)+\int _{-\infty}^\eta \dd  \eta' \partial [g(\eta',\eta)]/\partial \eta $.
This leads to
\bea
P_{vv}' =v_{\bm k}'v_{\bm k}^*+v_{\bm k}v_{\bm k}^*{}'
+4\left(2\pi\right)^{3/2}
\int_{-\infty}^\eta \dd \eta'\gamma\left(\eta'\right) \tilde{C}_R\left(k,
\eta'\right)
 \frac{\partial}{\partial \eta} 
\mathrm{Im}^2\left[v_{\bm k}\left(\eta'\right){v_{\bm k}^*}
\left(\eta\right)\right] \, ,
\eea
the term corresponding to $g(\eta,\eta)$ being absent since
proportional to
$\mathrm{Im}^2\left[v_{\bm k}\left(\eta\right){v_{\bm
      k}^*(\eta)}\right]=0$.
Similar considerations lead to the third derivative of $P_{vv}$, which can be expressed
as
\bea
P_{vv}''' = &v_{\bm k}'''v_{\bm k}^*+3v_{\bm k}''v_{\bm k}^*{}'
+3v_{\bm k}'v_{\bm k}^*{}''+v_{\bm k}v_{\bm k}^*{}'''
+2\gamma (2\pi)^{3/2}\tilde{C}_R\left(k\right)
 \\ &
+
4\left(2\pi\right)^{3/2}
\int_{-\infty}^\eta \dd \eta'\gamma\left(\eta'\right) \tilde{C}_R\left(k,
\eta'\right)
 \frac{\partial ^3}{\partial \eta^3} 
\mathrm{Im}^2\left[v_{\bm k}\left(\eta'\right){v_{\bm k}^*}
\left(\eta\right)\right] \, .
\eea
Using the mode equation for $v_{\bm k}$, $v_{\bm k}''+\omega^2v_{\bm k}=0$, it is
then easy to show that $P_{vv}$ given in \Eq{eq:solPvv} is indeed a solution of
\Eq{eq:thirdvlinear}.

As explained in \Sec{sec:InteractionWithEnvironement}, it is natural to consider that $\hat{A}$ depends on $\hat{v}$ only. It was also mentioned that our technique was applicable even if $\hat{A}$ involves $\hat{p}$. In the linear case discussed here, this implies $\hat{A}=\hat{p}$. One can then show that the evolution of the system is still controlled by an equation of the form~(\ref{eq:thirdvlinear}), the only difference being the source term that is now given by $S_1(\bm{k},\eta)=(2\pi)^{3/2}[(\gamma \tilde{C}_R)''+2\omega^2(k)\gamma\tilde{C}_R]$.
\subsubsection{Power spectrum in the slow-roll approximation}
\label{sec:linear:PS:SR}
Our final move is to use the slow-roll approximation to calculate $\Delta \calP_{\bm{k}}=\mathcal{J}_{\bm{k}}/\vert v_{\bm{k}}\vert^2$, where ${\cal J}_{\bm k}$ is defined by \Eq{eq:defJ} and $v_{\bm{k}}$ is the solution to the Mukhanov-Sasaki equation that is normalised to the Bunch-Davies vacuum in the sub-Hubble limit. This is done in details in \App{subsec:dmsr} and here we simply quote the results. Defining $\nu\equiv 3/2+\epsilon_{1*}+\epsilon_{2*}/2$, where $\epsilon_{1*}$ and $\epsilon_{2*}$ are the first and second slow-roll parameters calculated at Hubble exit time of the pivot scale $k_*$, at first order in slow roll the solution to the mode equation is given by a Bessel function of index $\nu$, see \Eq{eq:modeFunction:SR}. Inserting this mode function into \Eq{eq:defJ}, and making use of \Eq{eq:gamma} with $a\propto\eta^{-1-\epsilon_{1*}}$, one obtains \Eq{eq:calculJ}, where the integrals can be performed explicitly in terms of generalised hypergeometric functions, see \Eqs{eq:I1} and~(\ref{eq:I2}). 

The result can then be expanded in two limits. The first one consists in using the requirement that the environment autocorrelation time $t_\uc$ is much shorter than the typical time scale over which the system evolves, $H^{-1}$. If the environment correlation time and length are similar, $t_\uc\sim \lE$, as is the case for the example discussed in \App{sec:massivescalarfield} where the degrees of freedom contained in a heavy scalar field play the role of the environment, this amounts to $H\lE \ll 1$. The second limit consists in evaluating the power spectrum at the end of inflation, where all modes of astrophysical interest today are far outside the Hubble radius, \ie are such that $-k\eta\ll 1$. 

In these limits, the dominant contribution to the power spectrum depends on the value of $p$ and three cases need to be distinguished. If $p>3+(2+2\nu)/(1+\epsilon_{1*})$,
referred to as case one if what follows, then
\bea
\label{eq:dP1linear}
\Delta {\cal P}_{\bm k}\bigl \vert_1
\simeq&\frac{\pi 2^{-1-2\nu}}{\nu^2\Gamma^2(\nu)}
\left(\frac{\kgamma}{k_*}\right)^2
\left(\frac{k}{k_*}\right)^{2\nu}
\left(\frac{\eta}{\eta_*}\right)^{2+2\nu-(p-3)(1+\epsilon_1)}
\biggl[\frac{2}{2-\left(p-3\right)\left(1+\epsilon_{1*}\right)}
 \\ &
-\frac{1}{2\left(1+\nu\right)-\left(p-3\right)\left(1+\epsilon_{1*}\right)}
-\frac{1}{2\left(1-\nu\right)-\left(p-3\right)
\left(1+\epsilon_{1*}\right)}\biggr].
\eea
The second case is when
$3+1/(1+\epsilon_{1*})<p<3+(2+2\nu)/(1+\epsilon_{1*})$, for which the
quantity $\Delta {\cal P}_{\bm k}$ reads
\bea
\label{eq:dP2linear}
\Delta {\cal P}_{\bm k}\bigl \vert_2\simeq&
\frac{ \sqrt{\pi}}{4 }
\left(\frac{\kgamma}{k_*}\right)^2
\left(\frac{k}{k_*}\right)^{\left(p-3\right)
\left(1+\epsilon_{1*}\right)-2}
\frac{\Gamma\left[\frac{\left(p-3\right)
\left(1+\epsilon_{1*}\right)-1}{2}\right]
\Gamma\left[1+\nu-\frac{\left(p-3\right)
\left(1+\epsilon_{1*}\right)}{2}\right]
}{\Gamma\left[\frac{\left(p-3\right)\left(1+\epsilon_{1*}\right)}
{2}\right]\Gamma\left[\frac{\left(p-3\right)\left(1
+\epsilon_{1*}\right)}{2}+\nu\right]}\, .
\eea
Finally remains the third case where $p<3+1/(1+\epsilon_{1*})$ and one
obtains
\bea
\label{eq:dP3linear}
\Delta {\cal P}_{\bm k}\bigl \vert_3\simeq&
\left(\frac{\kgamma}{k_*}\right)^2
\left(\frac{k}{k_*}\right)^{\left(p-3\right)\left(1+\epsilon_{1*}\right)\left(1-\epsilon_{1*}\right)+\epsilon_{1*}-2}
\frac{\left(1+\epsilon_{1*}\right)^{1-(p-3)(1+\epsilon_{1*})}}{2-2\left(p-3\right)
\left(1+\epsilon_{1*}\right)}\\ & 
\left(H_*\lE\right)^{\left[\left(p-3\right)\left(1+\epsilon_{1*}\right)-1\right]\left(1-\epsilon_{1*}\right)}
\, .
\eea

For complete consistency, these expressions must also be expanded in slow roll since we have used this
approximation before. At first order, the form of the result for the three cases
($i=1,2,3$) is given by
\bea
\label{eq:DeltaPzeta:slowroll}
\Delta {\cal P}_{\bm k}\bigl \vert_i\simeq&
{\cal A}_i(k)\left[1+{\cal B}_i\epsilon_{1*}
+{\cal C}_i\epsilon_{2*}+\left({\cal D}_i\epsilon_{1*}+{\cal E}_i\epsilon_{2*}\right)
 \ln \left(\frac{k}{k_*}\right)
\right]\, ,
\eea
where ${\cal A}_i(k)$, ${\cal B}_i$, ${\cal C}_i$ and ${\cal D}_i$ can be calculated from \Eqs{eq:dP1linear}, (\ref{eq:dP2linear}) and~(\ref{eq:dP3linear}) and are given by 
\begin{align}
\label{eq:lin:amplitude1}
{\cal A}_1(k) &=\left(\frac{\kgamma}{k_*}\right)^2
\left(\frac{k}{k_*}\right)^{3}
\left(\frac{\eta}{\eta_*}\right)^{2+2\nu-(p-3)(1+\epsilon_{1*})}
\frac{2}{(p-2)(p-5)(p-8)},
\\
{\cal B}_1 &=2\gamma_\mathrm{E}+\ln 4
-7+\frac{1}{2-p}+\frac{3}{8-p}+\frac{2}{5-p},
\\
{\cal C}_1&=\gamma_\mathrm{E}+\ln 2 -2+\frac{6}{(p-2)(p-8)},
\quad {\cal D}_1 =2,\quad {\cal E}_1 =1,
\\
\label{eq:amplitude2}
{\cal A}_2(k) &=\left(\frac{\kgamma}{k_*}\right)^2
\left(\frac{k}{k_*}\right)^{p-5}
\frac{(6-p)\pi }{2^{6-p}(p-2)\sin(\pi p/2)
\Gamma(p-3)},
\\
{\cal B}_2 &=-2\frac{(p-1)(p-3)}{(p-4)(p-2)}
-\frac12 (p-5)\psi\left(4-\frac{p}{2}\right)
-\psi\left(-2+\frac{p}{2}\right)
\nonumber \\ &
-\frac12 (p-3) \psi\left(-\frac32+\frac{p}{2}\right),
\\
{\cal C}_2 &=
\frac12 \psi\left(4-\frac{p}{2}\right)
-\frac12 \psi\left(\frac{p}{2}\right), 
\quad
{\cal D}_2 = p-3,\quad {\cal E}_2=0, 
\\
\label{eq:aplitude3}
{\cal A}_3(k) &= \left(\frac{\kgamma}{k_*}\right)^2
\left(\frac{k}{k_*}\right)^{p-5}
\frac{\left(H_*\lE \right)^{p-4}}
{2(4-p)},
\\
{\cal B}_3 &=3-p+\frac{1}{4-p}+\ln \left(H_*\lE \right),
\quad
{\cal C}_3 =0,\quad {\cal D}_3 =1,\quad {\cal E}_3=0\, ,
\label{eq:lin:C3:D3:E3}
\end{align}
where $\gamma_{\mathrm E}\simeq 0.577$ is the Euler-Masheroni constant and
$\psi(z)$ is the digamma function. 

Let us note that in the first and second cases, the result is independent of $\lE$, since the main contribution to the integral~(\ref{eq:defJ}) comes from the neighbourhood of its upper bound. It means that the details of the shape of the environment correlator $C_R$ are irrelevant in these cases, and the top-hat ansatz~(\ref{eq:tophatcorrelation}) we have employed does not lead to any loss of generality. In the third case however, the result is directly dependent on $\lE$. The slow-roll corrections given in \Eq{eq:lin:C3:D3:E3} may therefore lie beyond the accuracy level of the present calculation where the environment correlator is approximated with a top-hat function. In any case, the observational constraints derived below make use of the expression~(\ref{eq:aplitude3}) for the amplitude $\mathcal{A}_3$ only in this case, and are therefore robust. 

Another remark is that in the second and third cases, the power spectrum settles to a stationary value at late time since none of the expressions~(\ref{eq:dP2linear}) and~(\ref{eq:dP3linear}) depends on time. In the first case however, the power spectrum is not frozen on
large scales and continues to evolve as is revealed by the amplitude
${\cal A}_1$ in \Eq{eq:lin:amplitude1}, which can also be written as
\bea
\label{eq:amplitude1}
{\cal A}_1(k)=\left(\frac{\kgamma}{k_*}\right)^2
\left(\frac{k}{k_*}\right)^{3}
\frac{2}{(p-2)(p-5)(p-8)}
\exp\left\{\left(N-N_*\right) \left[p-3-\frac{2(1+\nu)}{1+\epsilon_{1*}}\right]
\right\} \, .
\eea
In this expression, $N-N_*$ is the number of
\efolds~elapsed since the pivot scale crosses the Hubble radius. The
exponential dependence on $N-N_*$ explains why we have not
expanded this term, and the time-dependent term of
\Eq{eq:lin:amplitude1}, in slow roll. Let us note that the power of
$\ee^{N-N_*}$ is positive since the condition
$p>3+(2+2\nu)/(1+\epsilon_{1*})$ is precisely what defines case number
one. The correction to the standard result is therefore enhanced by a
very large factor in this case.
\subsubsection{Observational constraints}
\label{sec:lin:obsConstraints}
\begin{figure}[t]
\begin{center}
\includegraphics[width=0.48\textwidth]{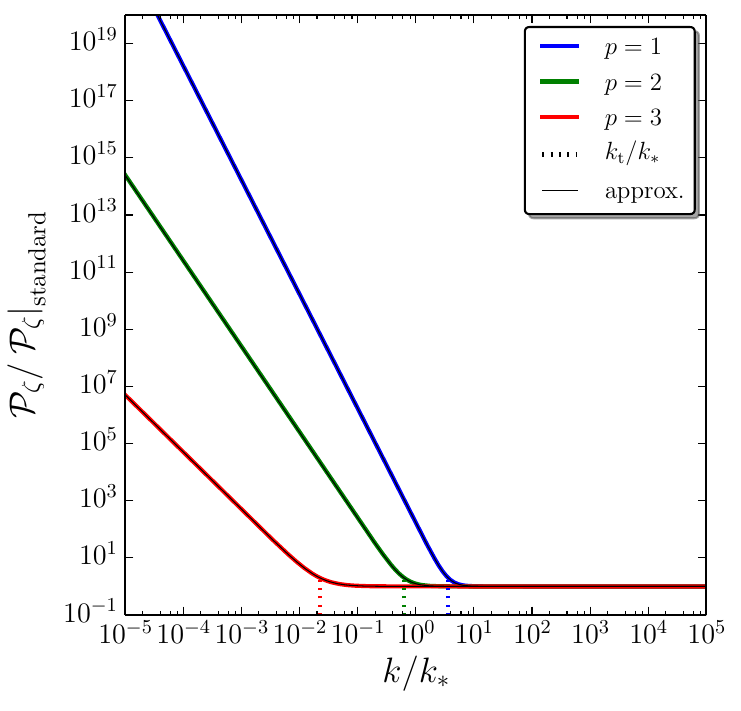}
\includegraphics[width=0.48\textwidth]{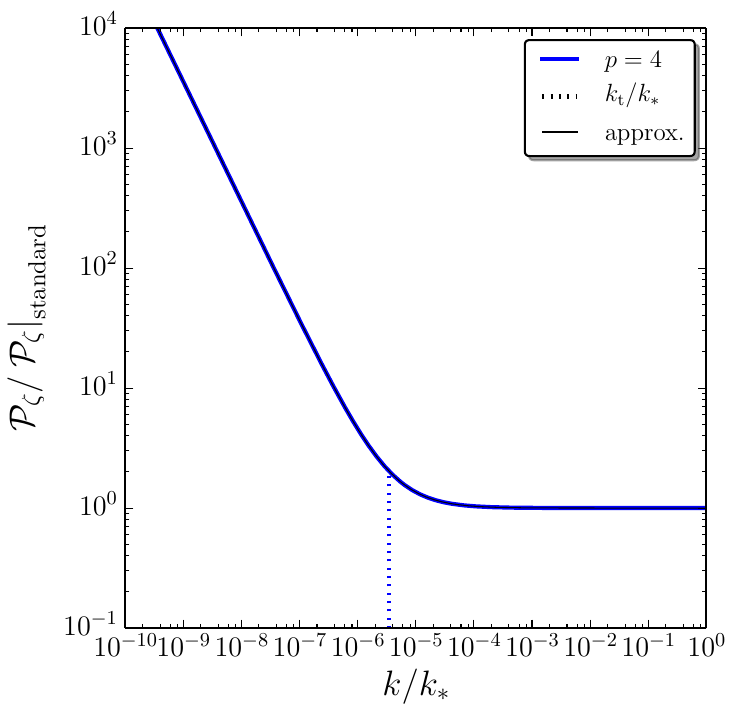}
\includegraphics[width=0.48\textwidth]{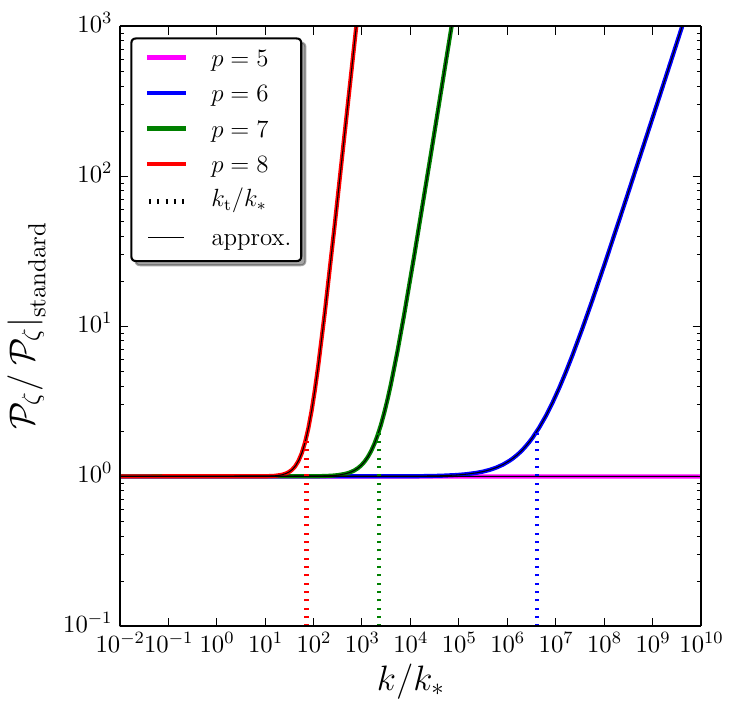}
\includegraphics[width=0.48\textwidth]{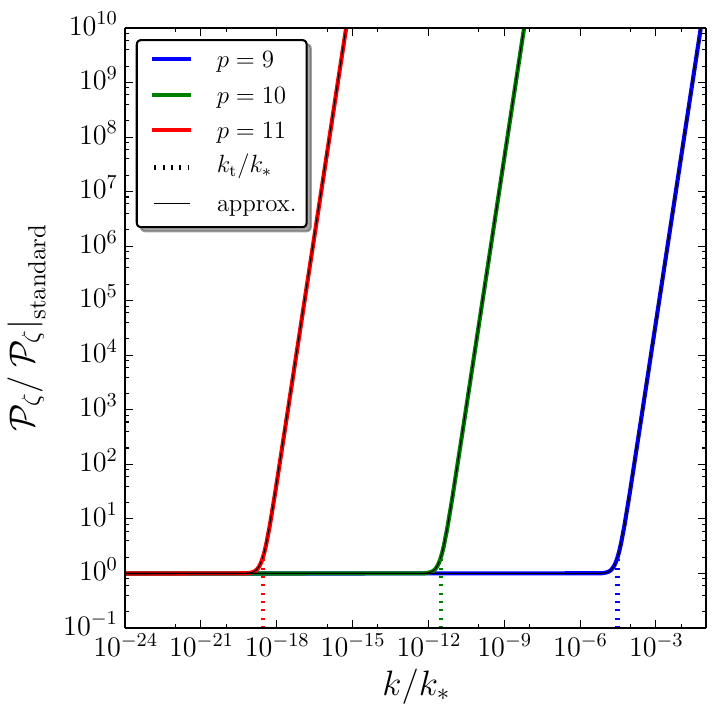}
\caption{Numerical integration of the power spectrum in the case of a
  linear interaction with the environment, for different values of $p$ as indicated on the
  plots. The coloured lines represent the exact result while the black
  lines correspond to our analytical approximation~(\ref{eq:lin:amplitude1})-(\ref{eq:lin:C3:D3:E3}) (because of the perfect overlap they are difficult to distinguish). The
  vertical dotted lines indicate the position of the transition scale
  $k_{\mathrm t}$ as given in \Eqs{eq:lin:kt}. The values chosen for the parameters are
  $\epsilon_{1*}=10^{-4}$, $\epsilon_{2*}=1-0.96-2\epsilon_{1*}$ (such that $\nS\simeq 0.96$ in the standard branch),
  $H_*\lE =10^{-3}$, $\kgamma/k_*=10^{-3}$, and
  $\Delta N_*=50$.}
\label{fig:Pzeta}
\end{center}
\end{figure}
In order to check the validity of the above calculations, we have
numerically integrated the power spectrum for different values of $p$,
and compared the result to our analytical approximations. The
comparison is displayed in \Fig{fig:Pzeta}, where one can check that the analytical approximations~(\ref{eq:lin:amplitude1})-(\ref{eq:lin:C3:D3:E3})
match very well the numerical result (because of the perfect overlap they can in fact not even be distinguished).

The structure of \Eq{eq:modifiedPzeta} implies that the power spectrum in the presence of decoherence is made of two branches, the
standard, almost scale-invariant one, and a new branch which  strongly deviates from scale invariance (except for the case
$p\simeq 5$ discussed below). The scale
$k_\mathrm{t}$ at which the transition between the two branches occurs
is such that ${\cal A}_i(k_\mathrm{t})\sim 1$, and \Eqs{eq:amplitude1}, (\ref{eq:amplitude2}) and~(\ref{eq:aplitude3}) give rise to
\bea
\label{eq:lin:kt}
\frac{k_\mathrm{t}}{k_*}\biggl\vert_1 & \simeq
\left(\frac{\kgamma}{k_*}\right)^{-\frac{2}{3}}
\exp\left\{-\frac{\Delta N_*}{3} 
\left[p-3-\frac{2(1+\nu)}{1+\epsilon_1}\right]
\right\},\\
\frac{k_\mathrm{t}}{k_*}\biggl\vert_2 & \simeq 
\left(\frac{\kgamma}{k_*}\right)^{-\frac{2}{p-5}}, \\
\frac{k_\mathrm{t}}{k_*}\biggl\vert_3 & \simeq 
\left(\frac{\kgamma}{k_*}\right)^{-\frac{2}{p-5}}
\left(H_*\lE \right)^{-\frac{p-4}{p-5}}\, ,
\eea
where $\Delta N_*$ corresponds to $N-N_*$ evaluated at the end of inflation, \ie $\Delta N_* = N_\uend-N_*$.

More precisely, in the first case, \ie when $p>8-3\epsilon_{1*}+\epsilon_{2*}$, \Eq{eq:amplitude1} shows that the correction to the standard power spectrum scales as $k^3$ regardless
of the value of $p$. This can be checked on the bottom-right panel of \Fig{fig:Pzeta} where, for $p=9$, $10$ and $11$,
the corrections grow for large values of $k$ (small scales) and have
the same slope. Since we observe an almost scale-invariant power spectrum, the part $k>k_\mathrm{t}$ of the power spectrum has to be outside the observable window, $k_\mathrm{t}\gg k_*$. Together with \Eq{eq:lin:kt}, it gives rise to
\bea
\label{eq:lin:kgammaUpperBound:1}
\left.\frac{k_\gamma}{k_*}\right\vert_1 \ll \ee^{-\frac{1}{2}\left(p-8+3\epsilon_{1*}-\epsilon_{2*}\right)\Delta N_*}\, .
\eea
Through \Eq{eq:kbreak:def}, this directly constrains the interaction strength with the environment, to very small values.

The second case corresponds to $4-\epsilon_{1*}<p<8-3\epsilon_{1*}+\epsilon_{2*}$, and \Eq{eq:amplitude2} implies that the correction scales as $k^{p-5}$, such that it modifies the power spectrum at $k>k_\mathrm{t}$ if $p>5$ and $k<k_\mathrm{t}$ if $p<5$ (the case $p=5$ is singular and is discussed separately below), with a $p$-dependent slope. This can be checked on the top-right and bottom-left panels
of \Fig{fig:Pzeta}. For instance, for $p=4$ (top-right panel),
${\cal A}_2(k)\propto k^{-1}$ and, indeed, the correction
grows on large scales. On the contrary, if we consider, say $p=6$ and
$p=7$ (bottom-right panel), one has ${\cal A}_2(k)\propto k^{1}$ and
${\cal A}_2(k)\propto k^{2}$ respectively, and, indeed,
the corrections now grow on small scales and with different slopes. In order for observed modes to lie outside the modified, non scale-invariant part of the power spectrum, one needs to have $k_{\mathrm{t}}\ll k_*$ for $p<5$ and $k_{\mathrm{t}}\gg k_*$ for $p>5$. In both cases, with \Eq{eq:lin:kt} it leads to
\bea
\left.\frac{k_\gamma}{k_*}\right\vert_2 \ll 1\, .
\label{eq:lin:kgammaUpperBound:2}
\eea

Finally, the
third case is defined by $p<4-\epsilon_{1*}$. The
scaling of ${\cal A}_3(k)$ is the same as for ${\cal A}_2(k)$ as can be checked
on the top-left panel in \Fig{fig:Pzeta}. For
instance, $p=2$ corresponds to ${\cal A}_3(k)\propto k^{-3}$ and $p=3$
to ${\cal A}_2(k)\propto k^{-2}$. Observational constraints on scale invariance impose $k_{\mathrm{t}}\ll k_*$, which, with \Eq{eq:lin:kt}, translates into
 \bea
\left.\frac{k_\gamma}{k_*}\right\vert_3 \ll \left(H_* \lE\right)^{\frac{4-p}{2}}\, .
\label{eq:lin:kgammaUpperBound:3}
\eea

Notice that the constraint $k_{\mathrm{t}} \ll k_*$ (when $p<5$) is conceptually different from the constraint $k_{\mathrm{t}} \gg k_*$ (when $p>5$). Indeed, in the former case, one removes the corrections to the power spectrum outside our observational window, while in the later case, one pushes the corrections to scales that are smaller than the ones probed in the CMB but that could still be of astrophysical interest.
\subsubsection{Case of a heavy scalar field as the environment}
\label{sec:lin:PowerSpectrumConstraints:MassiveFieldAsEnv}
As can be seen in \Fig{fig:Pzeta}, of particular interest is the case $p=5$, see the pink line in the bottom-left panel, where the power spectrum is almost scale invariant even in the modified branch. A crucial remark is that the microphysical example studied in \App{sec:massivescalarfield}, where the environment is made of the degrees of freedom contained in a heavy test scalar field, precisely gives $p\simeq 5$. More precisely, from \Eq{eq:gammastar:MassiveEnv}, one has $p=5-6m\epsilon_{1*}$ in that case. Let us study the observational predictions of this model in more details. Combining the standard expression of the power spectrum, namely
\bea
\label{eq:Pzeta:standard}
\calP_\zeta\vert_{\mathrm{standard}}
=\frac{H_*^2}{8\pi^2\epsilon_1\Mp^2}
\left[1-2(C+1)\epsilon_{1*}-C\epsilon_{2*}
-(2\epsilon_{1*}+\epsilon_{2*})\ln \left(\frac{k}{k_*}\right)\right],
\eea
where $C=\gamma_\mathrm{E}+\ln 2-2\simeq -0.7296$ is a numerical constant, with \Eqs{eq:modifiedPzeta} and~(\ref{eq:dP2linear}), one
has
\bea
\label{eq:lin:massivepsi:Pzeta:mod}
\calP_{\zeta}=&\frac{H_*^2\left(1+\frac{\pi}{6}\frac{\kgamma^2}{ k_*^2}\right)}{8\pi^2\epsilon_1\Mp^2}
\left\lbrace
1-\frac{2\left(C+1\right)+\frac{\pi}{9}\frac{\kgamma^2}{k_*^2}\left[2-3C+3\left(3C-1\right)m\right]}{1+\frac{\pi}{6}\frac{\kgamma^2}{k_*^2}}\epsilon_{1*}
\right. \\ & \left.
-\left[C+\frac{\frac{\pi}{6}\frac{ k_\gamma^2}{ k_*^2}}{3\left(1+\frac{\pi}{6}\frac{k_\gamma^2}{ k_*^2}\right)}\right]\epsilon_{2*}
-\left[2\epsilon_{1*}+\epsilon_{2*}+\frac{\frac{\pi}{6}\frac{\kgamma^2}{k_*^2}\left(6m-2\right)}{1 +\frac{\pi}{6}\frac{\kgamma^2}{k_*^2}} \epsilon_{1*}\right] \ln\left(\frac{k}{k_*}\right)
\right\rbrace\, .
\eea
In this expression, recall that $k_\gamma/k_*$ is given by \Eq{eq:kbreak:n_eq_1:microParam}. This formula differs in two ways from the standard one~(\ref{eq:Pzeta:standard}). First, the amplitude is no longer a function of the inflationary energy scale $H_*$ and the first slow-roll parameter $\epsilon_{1*}$ only, but now also depends on the ratio $\kgamma/k_*$. If one assumes that tensor perturbations are not affected by the presence of the environment 
%(which seems realistic at leading order in perturbation theory since the environment contains scalar degrees of freedom only) 
and that their power spectrum is still given by the standard formula $\calP_h \simeq 2H_*^2/(\pi^2\Mp^2)$, the tensor-to-scalar ratio $r\equiv \calP_h/\calP_\zeta$, which is the standard case in given by $\left. r\right\vert_{\mathrm{standard}}=16 \epsilon_{1*}$, now reads
\bea
r=\frac{\left. r\right\vert_{\mathrm{standard}}}{1+\frac{\pi}{6}\frac{\kgamma^2}{ k_*^2}}\, .
\eea
When $k_\gamma/ k_* \ll 1$, one recovers the standard result, otherwise $r$ is reduced compared to the free theory. Second, the spectral index $\nS\equiv 1+\dd\ln\calP_\zeta/\dd\ln k$ is also modified, and instead of the standard formula
$\left.\nS\right\vert_{\mathrm{standard}}=1-2\epsilon_{1*}-\epsilon_{2*}$, we now have
\bea
\label{eq:newns}
\nS=\left.\nS\right\vert_{\mathrm{standard}}
-\frac{\frac{\pi}{6}\frac{\kgamma^2}{k_*^2}}{1 +\frac{\pi}{6}\frac{\kgamma^2}{k_*^2}} \left(6m-2\right) \epsilon_{1*}\, .
\eea
When $k_\gamma/ k_* \ll 1$, one recovers the standard result $\nS\simeq\left.\nS\right\vert_{\mathrm{standard}}$ but if $k_\gamma/ k_* \gg 1$, one obtains $\nS\simeq\left.\nS\right\vert_{\mathrm{standard}} - (6m-2)\epsilon_{1*}$. The shift in the spectral index $\Delta \nS$ is negative (at least for $m>1/3$) and proportional to $\epsilon_{1*}$, so it is still compatible with quasi scale invariance but it may have consequences for particular models of inflation. 
\begin{figure}[t]
\begin{center}
\includegraphics[width=0.50\textwidth]{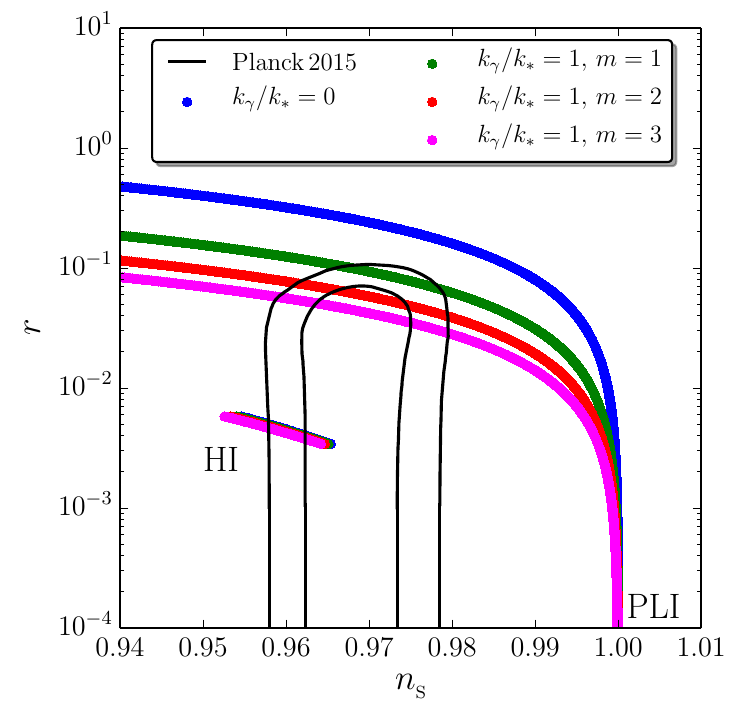}
\includegraphics[width=0.48\textwidth]{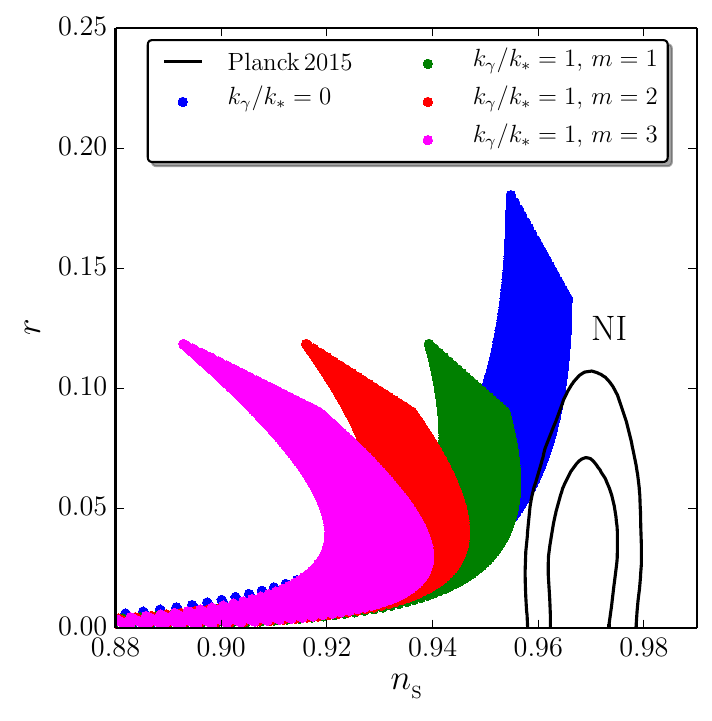}
\caption{Spectral index $\nS$ and tensor-to-scalar ratio $r$ for three single-field models of inflation, Higgs inflation (or the Starobinsky model) ``HI'' and power-law inflation ``PLI'' in the left panel, and natural inflation ``NI'' in the right panel. The black lines correspond to the one- and two-sigma contours obtained from the Planck 2015 $TT$ data combined with the high-$\ell$ $C_\ell^{TE}+C_\ell^{EE}$ likelihood and the low-$\ell$ temperature plus polarisation likelihood (PlanckTT,TE,EE+lowTEB in the notations of \Refa{Aghanim:2015xee}) together with the BICEP2-Keck/Planck likelihood described in \Refa{Ade:2015tva}. The blue disks correspond to the standard results in absence of decoherence, while the green, red and pink disks correspond to the modified power spectrum~(\ref{eq:lin:massivepsi:Pzeta:mod}) with $k_\gamma/k_*=1$, for $m=1$, $2$ and $3$ respectively.}
\label{fig:nsr}
\end{center}
\end{figure}

To illustrate this result, in \Fig{fig:nsr} we have displayed the
spectral index and the tensor-to-scalar ratio predicted by three
single-field models of inflation: Higgs
inflation~\cite{Bezrukov:2007ep} (or the Starobinsky
model~\cite{Starobinsky:1980te}, HI) where the potential is given by
$V(\phi)\propto [1-\exp(-\sqrt{2/3}\phi/\Mp)]^2$, power-law
inflation~\cite{Abbott:1984fp} (PLI) where
$V(\phi)\propto \exp(-\alpha\phi/\Mp)$, and natural
inflation~\cite{Freese:1990rb} (NI) where
$V(\phi)\propto 1+\cos(\phi/f)$.  The predictions are calculated with
the \texttt{ASPIC} library~\cite{Martin:2013tda, aspic}, where the
free parameters appearing in the potential are varied together with
the reheating temperature\footnote{In the case where $k_\gamma/k_*$
  does not vanish, the amplitude of the power spectrum is modified and
  the constraint on the inflationary energy scale $H_*$ arising from
  the measured amplitude~\cite{Ade:2015lrj} of the power spectrum,
  $\calP_\zeta\simeq 2\times 10^{-9}$, is modified too. This implies
  that the upper bound on the reheating temperature is decreased and
  that the uncertainty coming from the reheating expansion history is
  reduced. This effect is however minimal in \Fig{fig:nsr} where
  $k_\gamma/k_*=1$ (it would be larger for $k_\gamma/k_*\gg 1$, except
  for power-law inflation where the potential is conformally invariant
  and gives predictions that are independent of the reheating
  expansion history).} according to the priors of \Refa{Martin:2013nzq}, and one assumes the averaged
equation-of-state parameter during reheating to vanish.  The black
lines are the one- and two-sigma contours of the BICEP2-Keck/Planck
2015 likelihood~\cite{Aghanim:2015xee, Ade:2015tva}. The blue disks
correspond to the standard predictions~(\ref{eq:Pzeta:standard}),
while the green, red and pink disks correspond to the modified
predictions~(\ref{eq:lin:massivepsi:Pzeta:mod}) $k_\gamma/k_*=1$, for
$m=1$, $2$ and $3$ respectively. For the plateau model of Higgs
inflation, because the predicted tensor-to-scalar ratio in the free
theory is small, $\epsilon_{1*}$ is small and the shift in the
spectral index is also small, such that the model still provides a
good fit to the data in the presence of decoherence. For power-law
inflation, in the absence of decoherence the model is strongly
disfavoured, since it predicts a too large value for $\nS$ when $r$ is
sufficiently small. In the presence of decoherence however, the
decrease in the value of both $\nS$ and $r$ is such that the model
becomes viable for some values of its free parameter $\alpha$. For
natural inflation, the standard version of the model is disfavoured
since it predicts a too small value for the spectral index. By further
decreasing the spectral index, decoherence makes it worse and the
model is even more disfavoured.

One concludes that for the interaction model proposed in \App{sec:massivescalarfield}, the observational constraint on the interaction strength, here parametrised by $k_\gamma/k_*$, depends on the specific model of inflation. For models predicting the right value of the spectral index and very low values for the tensor-to-scalar ratio, such as Higgs inflation, decoherence does not change much the predictions and as a result there is no strong constraint on $k_\gamma/k_*$. For models predicting too large values for the spectral index, such as power-law inflation, the model becomes viable only in the presence of decoherence and observations place a lower bound on the interaction strength. On the contrary, for models predicting too small values for the spectral index, such as natural inflation, or models predicting values for the spectral index that are in agreement with the data and values for the tensor-to-scalar ratio that are not far from the current upper bound, decoherence can only make the model worse, and observations impose an upper bound on the interaction strength.
\subsection{Quadratic interaction}
\label{sec:PowerSpectrum:Quadratic}
Let us now consider the case where the system couples quadratically to the environment, $\hat{A}(\bm{x})=\hat{v}^2(\bm{x})$, \ie $n=2$ in \Eq{eq:A:vn}. Contrary to the case of linear interactions, because of mode coupling, the Lindblad equation~(\ref{eq:lindbladgeneral}) does not decouple into a set of independent Lindblad equations for each Fourier mode, and cannot be solved entirely. However, the power spectrum can still be calculated by making use of the technique presented in \Sec{subsec:linear:PowerSpectrum:Alternative}. 
\subsubsection{Two-point correlation functions}
\label{subsec:eomquadratic}
Let us start by deriving the equation of motion for the two-point correlation functions. Using the fact that the Fourier transform of a squared function is the convolution product of its Fourier transform, namely
\bea
\hat{v}^2(\eta,{\bm x})=\frac{1}{(2\pi)^3}
\int_{\setR^3} {\mathrm d}^3{\bm k}'\, {\mathrm d}^3{\bm p}\, 
\hat{v}_{\bm k'}\hat{v}_{\bm p- \bm k'}
\, \mathrm{e}^{i{\bm p}\cdot {\bm x}},
\eea
one obtains that \Eq{eq:Lindblad:mean:Fourrier} leads to
\bea
\label{eq:Linbald:meanfourierquadratic}
\frac{\dd \left\langle \hat{\rho}_{v}\hat{O}\right\rangle}
{\dd  \eta}= &
-i\left\langle \left[\hat{O},\hat{H}_{v}\right] \right\rangle
+\left\langle \frac{\partial \hat{O}}
{\partial \eta}\hat{\rho}_v\right\rangle
 \\
&-\frac{\gamma}{2(2\pi)^{3/2}}\int_{\setR^3}\dd ^3\bm k\,
\dd ^3{\bm k}'\, \dd ^3{\bm q}  
\, \tilde{C}_{R}(\bm k)
\left\langle \left[\left[\hat{O},\hat{v}_{{\bm k}'}
\hat{v}_{{\bm k}-{\bm k}'}
\right],
\hat{v}_{{\bm q}}\hat{v}_{-{\bm k}-{\bm q}}\right]\hat{\rho}_v\right\rangle
\, .
\eea
As explained in \Sec{subsec:linear:PowerSpectrum:Alternative}, one can derive the equations
of motion for one-point correlators and, as before, find that, in agreement with Ehrenfest theorem, the
standard equations are unmodified, see
\Eqs{eq:linearlinear}. For two-point correlators, as in
\Sec{subsec:linear:PowerSpectrum:Alternative}, only the equation for
$\langle \hat{p}_{{\bm k}_1}\hat{p}_{{\bm k}_2}\rangle$ is changed and one finds
\bea
\label{eq:vvquadratic:vpquadratic:pvquadratic:ppquadratic}
\frac{\dd  }{\dd  \eta} \left\langle \hat{v}_{{\bm k}_1}
\hat{v}_{{\bm k}_2}\right\rangle &=\left \langle 
\hat{v}_{{\bm k}_1}
\hat{p}_{{\bm k}_2}
+\hat{p}_{{\bm k}_1}
\hat{v}_{{\bm k}_2}
\right\rangle, \\
%\label{eq:vpquadratic}
\frac{\dd  }{\dd  \eta} \left\langle \hat{v}_{{\bm k}_1}
\hat{p}_{{\bm k}_2}\right\rangle &=\left \langle 
\hat{p}_{{\bm k}_1}
\hat{p}_{{\bm k}_2}\right \rangle
-\omega^2(k_2)\left\langle
\hat{v}_{{\bm k}_1}
\hat{v}_{{\bm k}_2}
\right\rangle,\\
%\label{eq:pvquadratic}
\frac{\dd  }{\dd  \eta} \left\langle \hat{p}_{{\bm k}_1}
\hat{v}_{{\bm k}_2}\right\rangle &=\left \langle 
\hat{p}_{{\bm k}_1}
\hat{p}_{{\bm k}_2}\right \rangle
-\omega^2(k_1)\left\langle
\hat{v}_{{\bm k}_1}
\hat{v}_{{\bm k}_2}
\right\rangle,\\
%\label{eq:ppquadratic}
\frac{\dd  }{\dd  \eta} \left\langle \hat{p}_{{\bm k}_1}
\hat{p}_{{\bm k}_2}\right\rangle &=
-\omega^2(k_2)\left\langle\hat{p}_{{\bm k}_1}
\hat{v}_{{\bm k}_2}\right \rangle
-\omega^2(k_1)\left\langle
\hat{v}_{{\bm k}_1}
\hat{p}_{{\bm k}_2}
\right\rangle
\\ &
+\frac{4\gamma}{(2\pi)^{3/2}}
\int \dd ^3 {\bm k}\, \tilde{C}_{R}(\bm k)
\left \langle \hat{v}_{{\bm k}+{\bm k}_1}
\hat{v}_{-{\bm k}+{\bm k}_2}\right\rangle.
\eea
An important remark is that the term involving $\gamma $ is not
explicitly proportional to $\delta ({\bm k}_1+{\bm k}_2)$ as was
the case for linear interactions, where the presence of the Dirac function guaranteed that the Lindblad term preserved statistical homogeneity. One may therefore be concerned that the above system generates violations to statistical homogeneity. However, if the system is solved through a perturbative expansion in $\gamma$, during the first iteration the Lindblad term contains the correlator $\langle \hat{v}_{{\bm k}+{\bm k}_1}
\hat{v}_{-{\bm k}+{\bm k}_2}\rangle$ evaluated in the free theory, which is proportional to $\delta(\bm{k}_1+\bm{k}_2)$. This guarantees that the solution that is obtained at the first iteration is statistically homogeneous. Since it sources the equation at the second iteration, the solution is again statistically homogeneous, so on and so forth. The result is therefore statistically homogeneous,\footnote{This property can also be seen in physical space, where upon using \Eq{eq:Lindblad:mean}, one has
\bea
\label{eq:Lindbladphyiscal}
\frac{\dd }
{\dd  \eta}
\left\langle \hat{p}({\bm x}_0)\hat{p}({\bm x}_0+{\bm d})\right\rangle
= 
-i\left\langle\left[\hat{p}({\bm x}_0)\hat{p}({\bm x}_0+{\bm d}),\hat{H}_{v}\right]
\hat{\rho}_{v}\right\rangle
+4\gamma C_{R}(\bm d)
\left\langle \hat{v}({\bm x}_0)\hat{v}({\bm x}_0+{\bm d})\right\rangle\, .
\eea
Since the correlators in the free theory are statistically homogeneous, \ie independent of $\bm{x}_0$, a solution to the system based on a perturbative expansion in $\gamma$  can only give statistically homogeneous correlators.
}
and \Eq{eq:vvquadratic:vpquadratic:pvquadratic:ppquadratic} admits a solution of the form given by
\Eq{eq:homoP}. If $\tilde{C}_R$ is also isotropic, $\tilde{C}_R(\bm{k})=\tilde{C}_R(k)$, the system for isotropic solutions then reduces to
\bea
\frac{\dd  P_{vv}(k)}{\dd  \eta}&=P_{vp}(k)+P_{pv}(k), \\
\frac{\dd  P_{vp}(k)}{\dd  \eta} &= \frac{\dd  P_{pv}(k)}{\dd  \eta}
=P_{pp}(k)-w^2(k)P_{vv}(k), \\
\frac{\dd  P_{pp}(k)}{\dd  \eta}&= -\omega^2(k)\left[P_{pv}(k)+P_{vp}(k)\right]
+\frac{4\gamma}{(2\pi)^{3/2}}
\int_{\setR^3} \dd ^3 {\bm k}' \, \tilde{C}_{R}\left( 
k'\right)P_{vv}\left(\left\vert \bm{k}'+\bm{k}\right\vert\right).
\eea
As was done in the linear interaction case, one can combine
the above equations in order to get a third order equation for
$P_{vv}$ only,
\bea
\label{eq:thirdvquadratic}
P_{vv}'''+4\omega^2P_{vv}'+4\omega'\omega P_{vv}
-\frac{8\gamma}{(2\pi)^{3/2}}
\int_{\setR^3} \dd  ^3 {\bm k}'\, \tilde{C}_{R}\left({ k}'\right)
P_{vv}\left(\left\vert \bm{k}'+\bm{k}\right\vert\right)=0\, .
\eea
Compared with the corresponding equation~(\ref{eq:thirdvlinear}) for linear interactions, where the power spectrum is sourced by the Fourier transform of the environment correlation function, in the present case it is sourced by the convolution product of the Fourier transform of the environment correlation function with the power spectrum itself. This makes clear that, as announced, quadratic interactions yield mode coupling, since the power spectrum at a given mode is sourced by its value at all other modes.

As was mentioned for the linear case, our technique can still be employed if $\hat{A}$ involves $\hat{p}$. In the quadratic case discussed here, this either implies $\hat{A}=\hat{p}^2$ or  $\hat{A}=\hat{v}\hat{p}$. In that situtation, $P_{vv}$ obeys the same equation as \Eq{eq:thirdvquadratic}, with a different source function that we do not give here since its concrete form is not especially illuminating but that can be readily obtained.
\subsubsection{Solving the equation for the power spectrum}
\label{subsec:psquadratic}
The third-order differential equation for the power spectrum $P_{vv}$,
\Eq{eq:thirdvquadratic}, is of the form
\bea
\label{eq:thirdPvv:source}
P_{vv}'''+4\omega^2 P_{vv}'+4\omega\omega' P_{vv} = S_2\, ,
\eea
where the source $S_2$ is a function of time that involves the power spectrum
$P_{vv}$ itself evaluated at all scales, namely
\bea
\label{eq:source:quadratic:def}
S_2\left(\bm{k},\eta\right)=\frac{8\gamma}{(2\pi)^{3/2}}
\int_{\setR^3} \dd  ^3 {\bm k}'\, \tilde{C}_{R}\left({ k}'\right)
P_{vv}\left(\left\vert \bm{k}'+\bm{k}\right\vert\right)\, .
\eea
This is therefore an integro-differential equation, that couples all modes together, and that is very difficult to solve in full generality. 
However, at leading order in $\gamma$, one can use the free theory to calculate $S_2$, which becomes a fixed function of time. Then, \Eq{eq:thirdPvv:source} is turned into an ordinary differential equation that can
be solved as we now explain. If one wants to go to higher orders in $\gamma$, one can repeat the procedure and embed it in a recursive expansion in $\gamma$, but that would be inconsistent with the standard derivation of the Lindblad equation, see \App{sec:DerivingLindblad}, which is valid at linear order in $\gamma$ only.

Inspired by the fact that \Eq{eq:defJ} provides a solution to \Eq{eq:thirdvlinear}, let us introduce the function ${\cal S}_{\bm k}$ defined by 
\bea
\label{eq:Pvv:generic:ParticularSolution}
  \mathcal{S}_{\bm k}\left(\eta\right) = -\frac{2}{\left(v^*_{\bm k} v_{\bm k}' 
- v'^*_{\bm k}
  v_{\bm k}\right)^2} \int_{\eta_0}^\eta S_2\left(\eta'\right)
  \mathrm{Im}^2\left[v_{\bm k}\left(\eta'\right)v^*_{\bm k}
  \left(\eta\right)\right]
  \dd\eta'\, ,
\eea
where $v_{\bm k}(\eta)$ is a mode function, that is to say a solution of the Mukhanov-Sasaki equation $v''_{\bm k}+\omega^2 v_{\bm k}=0$. Since the Wronskian of $v_{\bm k}$ is conserved through the Mukhanov-Sasaki equation, the factor in front the integral in \Eq{eq:Pvv:generic:ParticularSolution} is a constant.  Using similar techniques as the ones used below \Eq{eq:thirdvlinear}, it is straightforward to check that $\mathcal{S}_{\bm k}$ is a particular solution of the equation of motion~(\ref{eq:thirdPvv:source}) for $P_{vv}$. 

In fact, one can show that this particular solution is independent of the mode function $v_{\bm k}(\eta)$ one has chosen. Indeed, $\mathcal{S}_{\bm{k}}(\eta)$ is \emph{the} solution of \Eq{eq:thirdPvv:source} that, as can be shown from \Eq{eq:Pvv:generic:ParticularSolution}, satisfies $\mathcal{S}_{\bm{k}}(\eta_0)=\mathcal{S}'_{\bm{k}}(\eta_0)=\mathcal{S}''_{\bm{k}}(\eta_0)=0$. It is therefore unique. In practice, to evaluate it, one can use the Bunch-Davies normalised mode function, for which $v_{\bm{k}}^* v_{\bm{k}} - v_{\bm{k}}'^{*}v_{\bm{k}} = i$, but the result is independent of that choice, and $\mathcal{S}_{\bm{k}}(\eta)$ therefore carries a single integration constant, namely $\eta_0$. The full solution can be obtained by adding the solution of the homogeneous equation (\ie without the source term) $P_{vv}'''+4\omega^2 P_{vv}'+4\omega\omega' P_{vv}=0$. As
already mentioned for the linear case, one can check that
$v_{\bm k}(\eta) v^*_{\bm k}(\eta)$ satisfies this equation if
$v_{\bm k}$ is a solution [not necessarily the same as the one used to write down \Eq{eq:Pvv:generic:ParticularSolution}] of the mode equation. The
complete solution can then be expressed as
\bea
\label{eq:Pvv:sol:gen}
P_{vv}\left(k\right) = \vert v_{\bm k}\vert^2 + \mathcal{S}_{\bm k}.
\eea
If one sets the mode function appearing in the first term of the right-hand side of \Eq{eq:Pvv:sol:gen} to the Bunch-Davies normalised one, $P_{vv}$ matches the Bunch-Davies result in the infinite past if $\eta_0=-\infty$, and this leads to
\bea
\label{eq:exactsolquadratic} 
P_{vv} = v_{\bm k}\left(\eta\right)
v^*_{\bm k}\left(\eta\right) +2\int_{-\infty}^\eta S_2\left(\eta'\right)
\mathrm{Im}^2\left[v_{\bm k}\left(\eta'\right)
v^*_{\bm k}\left(\eta\right)\right]
\dd\eta'\, .
\eea
In this expression, let us stress that $v_{\bm k}$ is now Bunch-Davies normalised. Notice that since this does not rely on the concrete form of $S_2$, \Eq{eq:exactsolquadratic} is a solution of \Eq{eq:thirdPvv:source} for any source function. In particular, if the source term is given by $S_1$, see \Eq{eq:source:linear:def}, then \Eq{eq:Pvv:generic:ParticularSolution} for $\mathcal{S}_{\bm k}$ reduces to \Eq{eq:defJ} that defines ${\cal J}_{\bm k}$, and \Eq{eq:Pvv:sol:gen} matches \Eq{eq:solPvv}.
\subsubsection{Calculation of the source term}
\label{sec:quad:source}
The next step is to calculate the source term, that is to say the convolution product between the power spectrum in the free theory and the Fourier transform of the environment correlator. This is done in details in \App{subsec:source} and here, we simply quote the result. By performing the angular integration, one can first show that
\bea 
\int_{\setR^3} \dd ^3 {\bm k}' \, \tilde{C}_{R}\left( 
k'\right)P_{vv}\left(\left\vert \bm{k}'+\bm{k} \right\vert\right) &= \frac{\pi}{k}\int_0^\infty \dd  p\, p \, P_{vv}(p)
\int_{\left(k-p\right)^2}^{\left(k+p\right)^2} \dd  z
\, \tilde{C}_R\left(\sqrt{z}\right)\, ,
\label{eq:integral:Cz:maintext}
\eea
see \Eq{eq:integral:Cz}. The integral over $p$ contains a UV part ($p>aH$, sub-Hubble scales) and an IR part ($p<aH$, super-Hubble scales). 

The UV part is finite, but is removed by adiabatic subtraction~\cite{Bunch:1980vc, Markkanen:2017rvi}. This amounts to setting the upper bound of the integral over $p$ in \Eq{eq:integral:Cz:maintext} to the comoving Hubble scale $-1/\eta$.

The IR part is divergent, and can be regularised by imposing an IR cutoff, that corresponds to the comoving Hubble scale at the onset of inflation and that we call $-1/\eta_{\mathrm{IR}}$. In the integral over $z$, $\tilde{C}_R$ is such that it vanishes when its argument is larger than the comoving correlation wavenumber of the environment, $a/\lE$, and goes to a constant value in the opposite limit. Since the integral over $p$ is now restricted to super-Hubble modes, $p<aH$, and given that we assumed $\lE  \ll H^{-1}$ when deriving the Lindblad equation (at least if $\lE\sim t_\uc$, see \App{sec:DerivingLindblad}), one has $p\ll a/\lE$. Two cases can then be distinguished. If $k\ll a/\lE$, both $\vert k-p \vert$ and $k+p$ are much smaller than $a/\lE$, and $\tilde{C}_R(\sqrt{z})$ is constant over its integration range. If  $k\gg a/\lE$, both $\vert k-p \vert$ and $k+p$ are much larger than $a/\lE$, and $\tilde{C}_R(\sqrt{z})$ vanishes over its full integration range. This means that in \Eq{eq:integral:Cz:maintext}, one can simply replace $\tilde{C}_R(\sqrt{z})$ by $\tilde{C}_R(k)$, and one obtains
\bea 
\int_{\setR^3} \dd ^3 {\bm k}' \, \tilde{C}_{R}\left( 
k'\right)P_{vv}\left(\left\vert \bm{k}'+\bm{k} \right\vert\right) &\simeq
4\pi \tilde{C}_R\left(k\right)
\int_{-1/\eta_{\mathrm{IR}}}^{-1/\eta} \dd p p^2 P_{vv}\left(p\right)\, .
\label{eq:integral:Cz:maintext:appr}
\eea
In \App{subsec:source}, it is shown how this formula arises as a leading order expansion in $H \lE$ starting from the ansatz~(\ref{eq:tophatcorrelation}), here we simply gave a heuristic argument. Finally, on super-Hubble scales and neglecting slow-roll corrections, the power spectrum in the free theory is given by $P_{vv}(p) = (2p)^{-1} (-p\eta)^{-2}$. Plugging this formula into \Eq{eq:integral:Cz:maintext:appr}, one is led to
\bea 
\int \dd ^3 {\bm k}' \, \tilde{C}_{R}\left( 
k'\right)P_{vv}\left(\left\vert \bm{k}'+\bm{k} \right\vert \right) &\simeq
\frac{2\pi}{\eta^2} \tilde{C}_R\left(k\right)
\ln\left(\frac{\eta_{\mathrm{IR}}}{\eta}\right)\, .
\label{eq:integral:Cz:maintext:final}
\eea

Notice that this expression was obtained neglecting all slow-roll
corrections. Indeed, since the integral of the power spectrum over all
modes appears in the source function, a slow-roll expansion around the
pivot scale $k_*$ cannot be used consistently to describe the entire
set of modes, and one would have to specify the details of the
inflationary potential over the entire inflating
domain~\cite{Hardwick:2017fjo} in order to calculate the source term
beyond the de-Sitter limit. For this reason, we will ignore all
following slow-roll corrections since their inclusion would be
inconsistent. A more refined calculation would have to be carried out on
a model-by-model basis. From \Eqs{eq:thirdvquadratic}
and~(\ref{eq:integral:Cz:maintext:final}), the source function is then given by
\bea
\label{eq:quad:source:renormalised}
S_2\left(\eta\right)=\frac{8\gamma}{\sqrt{2\pi}\eta^2}
\tilde{C}_R\left(k\right)
\ln\left(\frac{\eta_{\mathrm{IR}}}{\eta}\right)\, .
\eea
\subsubsection{Power spectrum and observational constraints}
\label{sec:quad:PSconstraints}
\begin{figure}[t]
\begin{center}
\includegraphics[width=0.495\textwidth]{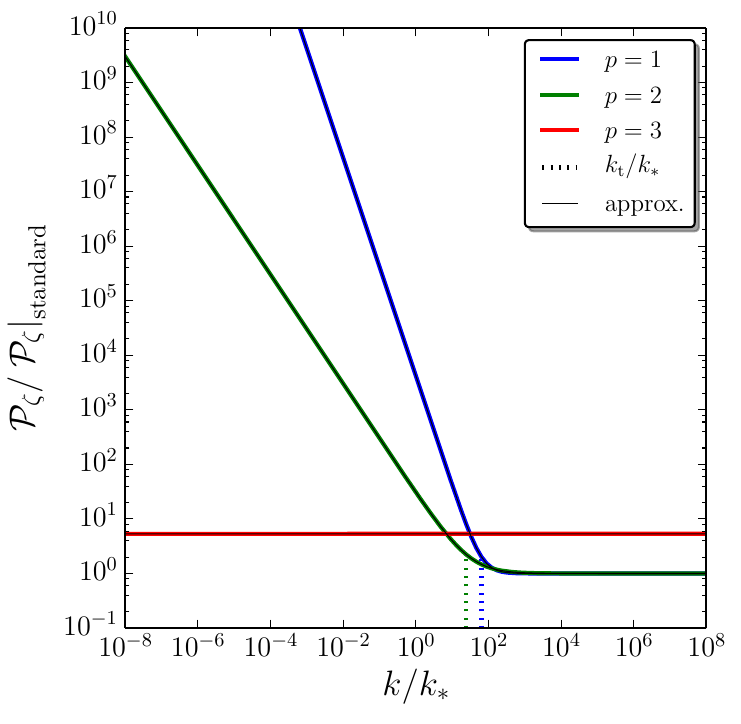}
\includegraphics[width=0.48\textwidth]{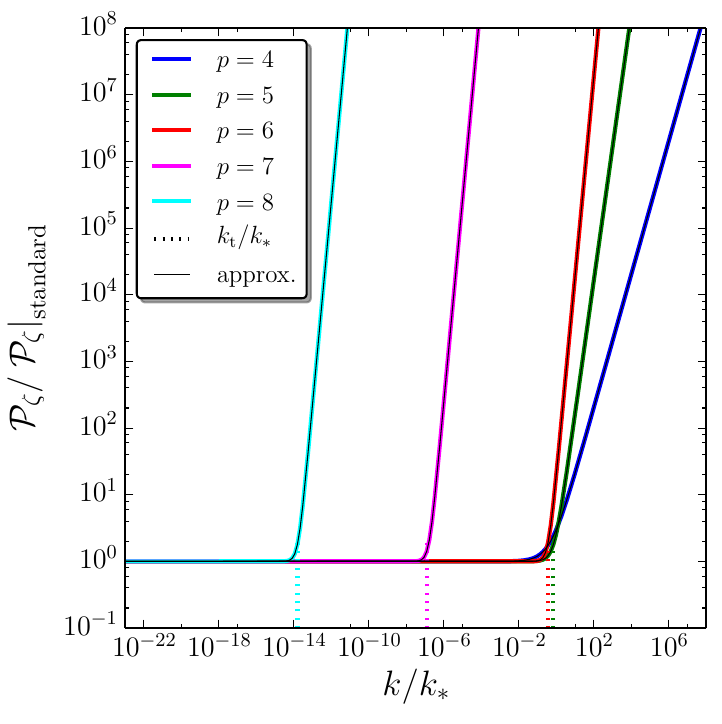}
\caption{Power spectrum at the end of inflation, numerically integrated in the case of a
  quadratic interaction with the environment, for different values of $p$ as indicated on the
  plots. The coloured lines represent the exact result while the black
  lines correspond to our analytical approximations~(\ref{eq:dP1quadratic})-(\ref{eq:dP:singular:quadratic}) (because of the perfect overlap they are difficult to distinguish). The
  vertical dotted lines indicate the position of the transition scale
  $k_{\mathrm t}$ defined by $\Delta \calP_{\bm{k}_\mathrm{t}} =1$. The values chosen for the parameters are $H_*\lE =10^{-3}$, $\sigma_\gamma=10^{-3}$, $\Delta N_*=50$, and $N_{_{\mathrm T}} = N_\uend - N_{\mathrm{IR}}= 10^4$.}
\label{fig:Pzetaquadratic}
\end{center}
\end{figure}

The final step consists in inserting
this source with \Eq{eq:Ck:appr} into \Eq{eq:exactsolquadratic}, which is done in
\App{subsec:solvingthirdquadratic} and results in
\Eqs{eq:exactPvvquadratic}-(\ref{eq:defI3I4}). As for the linear case, the integrals can be performed in terms of generalised hypergeometric functions, see \Eqs{eq:primitiveI3} and~(\ref{eq:primitiveI4}). One can then
simplify the expression of the power spectrum when evaluated on super-Hubble scales and using
the fact that the environment has a correlation length much smaller than
the Hubble radius, see \Eq{eq:finalPvvquadratic}. The modified power spectrum can still be defined by \Eq{eq:modifiedPzeta},
where the equivalent of \Eq{eq:defDPlinear} for the quadratic
interaction reads
$\Delta {\cal P}_{\bm k}\equiv {\cal S}_{\bm k}/\vert v_{\bm
  k}\vert^2$.
One finds that three regimes have to be distinguished. The first regime corresponds to when $p>6$, and gives
\bea
\label{eq:dP1quadratic}
\left. \Delta {\cal P}_{\bm k}\right \vert _1 &\simeq
\frac{8\sigma_\gamma}{27\pi}
\left(\frac{k}{k_*}\right)^{3}
\ee^{(p-6)\left(N-N_*\right)}
\Biggl[\frac{1}{p^2}
-\frac{2}{(p-3)^2}+\frac{1}{(p-6)^2}+\frac{18\left(N-N_{\mathrm{IR}}\right)}{p(p-3)(p-6)}
\Biggr].
\eea
The second case is when $2<p<6$, for which one has
\bea
\label{eq:dP2quadratic}
\left. \Delta {\cal P}_{\bm k}\right \vert_2 \simeq &
\frac{2^{p-1}(p-4)}{3p\Gamma(p-1)\sin(\pi p/2)} 
\sigma_\gamma
\left(\frac{k}{k_*}\right)^{p-3}\biggl[
N_*-N_{\mathrm{IR}}+\frac{1}{p-4}-\frac{2(p-1)}{p(p-2)}
 \\ &
-\frac{\pi}{2}\cot \left(\frac{\pi p}{2}\right)
+\ln 2 -\psi(p-2)+
\ln \left(\frac{k}{k_*}\right)
\biggr].
\eea
Finally, the third regime is when $p<2$. In that case, the
modification to the power spectrum reads
\bea
\label{eq:dP3quadratic}
\left. \Delta {\cal P}_{\bm k}\right \vert _3 \simeq &
\frac43 \frac{\left(H_*\lE \right)^{p-2}}
{\pi(2-p)}
\sigma_\gamma
\left(\frac{k}{k_*}\right)^{p-3}
\biggl[\frac{1}{2-p}+N_*-N_{\mathrm{IR}}
+\ln \left(H_*\lE \right)
+ \ln \left(\frac{k}{k_*}\right)\biggr].
\eea
In these expressions, $N-N_{\mathrm{IR}} \equiv \ln(\eta_{\mathrm{IR}}/\eta)$ denotes the number of \efolds~elapsed since the onset of inflation, and we have introduced the dimensionless coefficient
\bea
\label{eq:sigmagamma:def}
\sigma_\gamma\equiv \bar{C}_R\frac{\lE ^3}{a_*^3}\gamma_*
\eea
that characterises the strength of the interaction with the environment. One notices that the cases $p=2$ and $p=6$ are singular and must be treated separately, giving rise to
\bea
\label{eq:dP:singular:quadratic}
\left. \Delta {\cal P}_{\bm k}\right \vert _{p=2} \simeq &
\frac{\sigma_\gamma}{18 \pi}\left(\frac{k}{k_*}\right)^{-1}\Bigg\lbrace
12 - \pi^2 + 12 C \left(2 + C \right)
 - 12 \ln^2\left(H_*\lE\right) 
 \\ & 
+24\left[C+1-\ln\left(H_*\lE\right)\right] \left[N_*-N_{\mathrm{IR}} + \ln\left(\frac{k}{k_*}\right)\right]
 \Bigg\rbrace  \, ,  \\     
 \left. \Delta {\cal P}_{\bm k}\right \vert _{p=6} \simeq &  
 \frac{\sigma_\gamma}{162\pi}\left(\frac{k}{k_*}\right)^{3}
 \Bigg\lbrace
 2 \pi^2 -21  -  12C \left(1 + 2 C \right)
 -12\left(3+4C\right) \left(N-N_{\mathrm{IR}}\right)
     \\ &
      + 12 \left(1+4C\right) \left(N-N_*\right)
 + 24  \left(N-N_*\right) \left[2 \left(N-N_{\mathrm{IR}}\right)-\left(N-N_*\right)\right]
 \\ &
 -  12 \ln\left(\frac{k}{k_*}\right)\left[1 + 4 \left(C +  N_*-N_{\mathrm{IR}}\right) + 2 \ln\left(\frac{k}{k_*}\right)\right]
 \Bigg\rbrace
 \, .
\eea
These analytical formulas are superimposed with a numerical integration of the power spectrum in \Fig{fig:Pzetaquadratic}, where one can check that the agreement is excellent (because of the perfect overlap, they are even hard to distinguish). 

As for the linear case, the power spectrum is made of two branches, one which matches the standard formula, and one where scale invariance is strongly broken (with the exception of the case $p=3$ that is discussed below separately). The transition between these two branches occurs at the scale $k_\mathrm{t}$ such that $\Delta \calP_{\bm{k}_\mathrm{t}} =1$, and expressions similar to \Eq{eq:lin:kt} can be derived from \Eqs{eq:dP1quadratic}-(\ref{eq:dP:singular:quadratic}) (we do not write them down here for display convenience but they are straightforward). A case-by-case analysis reveals several other similarities with the linear interaction results.

In the first case indeed, \ie for large values of $p$, the power spectrum is not frozen on large scales and continues to increase, leading to a very large enhancement of the correction to the standard power spectrum at late time. The modified branch of the power spectrum scales as $k^3$, similarly to what was seen in \Eq{eq:amplitude1}. The requirement that observed modes are scale invariant, $k_\mathrm{t} \gg k_*$, leads to the constraint
\bea
\left.\sigma_\gamma\right\vert_1 \ll \frac{\ee^{\left(6-p\right)
\Delta N_*}}{N_{{}_\mathrm{T}}}\, ,
\eea
where $N_{{}_\mathrm{T}}$ corresponds to $N-N_{\mathrm{IR}}$ evaluated at the end of inflation, \ie $N_{{}_\mathrm{T}} = N_\uend - N_{\mathrm{IR}}$ and stands for the total duration of inflation. Let us notice that the $1/N_{{}_\mathrm{T}}$ term originates from the last term in \Eq{eq:dP1quadratic}, which dominates when $N_{{}_\mathrm{T}}$ is very large.

In the second case, \ie for intermediate values of $p$, the power spectrum is frozen on large scales and independent of the shape and correlation length of the environment correlator. This is again similar to the linear case. The power spectrum is modified on small scales, $k>k_{\mathrm{t}}$, if $p>3$, and on large scales, $k<k_{\mathrm{t}}$, if $p<3$. The requirement that the modified, non-scale invariant branch of the power spectrum is unobserved leads to
\bea
\left.\sigma_\gamma\right\vert_2 \ll 
\frac{1}{N_{{}_\mathrm{T}}-\Delta N_*}
\, .
\eea
The exception evading this constraint is the case $p=3$, where the power spectrum is almost scale invariant even in the modified branch. Strikingly, $p\simeq 3$ again corresponds to the scaling expected in the model proposed in \App{sec:massivescalarfield}, where the environment is made of a heavy test scalar field. In this case, $\sigma_\gamma$ is related to the microphysical parameters of the model according to
\bea
\sigma_\gamma = 128\left(\frac{37}{504 \pi^2}\right)^m
\frac{\left\lbrace \left(2m-1\right)!!-\sigma\left(m\right)\left[\left(m-1\right)!!\right]^2 \right\rbrace^3}{\left[m^2\left(2m-3\right)!!\right]^2} \lambda^2 \left(\frac{\mu}{M}\right)^{4-2m}\left(\frac{H_*}{ M}\right)^{6m}\, ,
\eea
where we have combined \Eqs{eq:MassivePsi:Cbar:tc:lE:maintext},~(\ref{eq:gammastar:MassiveEnv}) and~(\ref{eq:sigmagamma:def}). Similarly to what was done in \Sec{sec:lin:PowerSpectrumConstraints:MassiveFieldAsEnv}, constraints on $\gamma$ could be placed from the current observational bounds on $\nS$ and $r$ within specific models of inflation, but this would require to include slow-roll corrections to the calculation of the source term, as explained above.

In the third case finally, \ie for small values of $p$, the power spectrum freezes on large scales. The amplitude of the correction depends explicitly on the environment correlation length $\lE$, and it scales with the wavenumber in the same manner as for intermediate values of $p$. This is again similar to the linear case. The power spectrum is modified on large scales $k<k_{\mathrm{t}}$, and requiring that the non scale-invariant part is unobserved leads to the constraint
\bea
\left.\sigma_\gamma\right\vert_3 \ll \frac{ \left(H\lE\right)^{2-p}}{N_{_{\mathrm T}}-\Delta N_*
+\ln \left(H_*\lE \right)}\, . 
\eea
\section{Decoherence}
\label{sec:decoherencetime}
In this section, we show how the addition of a non-unitary term in the evolution equation~(\ref{eq:lindbladgeneral}) of the density matrix of the system, that models the interaction with environmental degrees of freedoms, leads to the dynamical suppression of its off-diagonal elements in the basis of the eigenstates of the interaction operator (here the Mukhanov-Sasaki variable $\hat{v}$). Since this decoherence mechanism is thought to play a role in the quantum-to-classical transition of primordial cosmological perturbations, as explained in \Sec{sec:intro}, we calculate the required strength of interaction that leads to decoherence at the end of inflation. We then compare this lower bound with the upper bound derived in the previous section from the requirement that quasi scale invariance is preserved. We thus identify the models for which successful decoherence occurs without spoiling standard observational predictions.
\subsection{Linear interaction}
\label{subsec:decotimelinear}
In the case of linear interactions with the environment, in \Sec{subsec:LindbladSol:Linear} (see also \App{sec:dmlinearinteraction}) we have shown how \Eq{eq:lindbladgeneral} can be solved exactly and the density matrix was given in \Eq{eq:finalrhomaintext}. Since it is written in the basis of the eigenstates of $\hat{v}_{\bm{k}}$, through which the system couples to the environment, the suppression of its off-diagonal elements directly allows us to study decoherence. 
\subsubsection{Decoherence criterion}
Let us consider an off-diagonal element of the density matrix $\hat{\rho}_{\bm k}^s$, away from the diagonal by a distance $\Delta v_{\bm{k}}$, that we write $\langle v_{\bmk}^s +  \Delta v_{\bmk}/2  \vert \hat{\rho}_{\bm k}^s \vert v_{\bmk}^s -\Delta v_{\bmk}/2\rangle$. From \Eq{eq:finalrhomaintext}, its amplitude can be expressed as
\bea
\left\vert \left\langle v_{\bmk}^s + \frac{ \Delta v_{\bmk}}{2}  \right \vert \hat{\rho}_{\bm k}^s \left\vert v_{\bmk}^s -\frac{\Delta v_{\bmk}}{2}\right\rangle \right\vert =
\left\vert \left\langle v_{\bmk}^s  \right \vert \hat{\rho}_{\bm k}^s \left\vert v_{\bmk}^s \right\rangle \right\vert
\exp\left[-\frac{\delta_{\bm k}+\frac{1}{4}}{2} \frac{\Delta v_{\bmk}^2}{P_{vv}(k)}\right]\, ,
\label{eq:defdeltak}
\eea
where \Eq{eq:solPvv} has been used for the power spectrum $P_{vv}(k)$, and where we have introduced the parameter
\bea
\label{eq:defdeltak:integrals}
\delta_{\bm k}(\eta) &\equiv \mathcal{I}_{\bm k}\mathcal{J}_{\bm k}
-\mathcal{K}_{\bm k}^2+\left\vert v_{\bm k}^\prime 
\right\vert^2\mathcal{J}_{\bm k}+ \left\vert v_{\bm k}\right
\vert^2\mathcal{I}_{\bm k}-{\left\vert v_{\bm k}\right\vert^2}^\prime 
\mathcal{K}_{\bm k}\, .
\eea
In \Eq{eq:defdeltak}, the factor $1/4$ has been separated out from the definition of $\delta_{\bm{k}}$ since it is present even in the absence of interactions with the environment (contrary to $\delta_{\bm{k}}$), such that $\delta_{\bm{k}}$ characterises the \emph{additional} decrease of the off-diagonal elements produced by the environment. Successful decoherence is characterised by the condition $\delta_{\bm{k}} \gg 1$, which can be justified by either of the three following reasons.

First, the typical distance between two realisations $v_{\bm{k}}^{s,(1)}$  and $v_{\bm{k}}^{s,(2)}$ of the Mukhanov-Sasaki variable $v_{\bm{k}}$ is by definition given by the square root of its expected second moment, $\sqrt{P_{vv}(k)}$. If one replaces $\Delta v_{\bm{k}}$ by $\sqrt{P_{vv}(k)}$ in \Eq{eq:defdeltak}, one can see that a large suppression of the off-diagonal element corresponds to a large value of $\delta_{\bm{k}}$.

Second, $\delta_{\bm{k}}$ appears in \Eq{eq:defdeltak} added to $1/4$, which corresponds to the standard decrease of the off-diagonal term in the absence of an environment. It seems therefore natural to compare the environment-induced suppression of the off-diagonal element with the standard one, and to define ``decoherence'' by the requirement that the former is larger than the later, $\delta_{\bm{k}} \gg 1$.

Third, decoherence is sometimes characterised by the so-called purity of the state, defined as $\mathrm{Tr}\left(\hat{\rho}_{\bm k}^s{}^2\right)$ (for other measures of coherence see \eg \Refa{2014PhRvL.113n0401B}). Making use of \Eq{eq:finalrhomaintext}, this can be expressed as a Gaussian integral and one obtains
\bea
\label{eq:delta:purity}
\mathrm{Tr}\left(\hat{\rho}_{\bm k}^s{}^2\right) =
 \int_{-\infty}^\infty\dd  v_{\bmk}^{s,(1)}
 \int_{-\infty}^\infty\dd  v_{\bmk}^{s,(2)}
\left\langle v_{\bmk}^{s,(1)}  \right\vert \hat{\rho}_{\bm k}^s\left\vert v_{\bmk}^{s,(2)}\right\rangle=
\frac{1}{\sqrt{1+4\delta_{\bm k}}}\, .  
\eea
When $\delta_{\bm k}\ll 1$, $\mathrm{Tr}\left(\hat{\rho}_{\bm k}^s{}^2\right) \simeq 1$ and the
state remains pure, while when $\delta_{\bm k}\gg 1$,
$\mathrm{Tr}\left(\hat{\rho}_{\bm k}^s{}^2\right)\ll 1$ and the state
becomes highly mixed. This is another reason why we define decoherence with the criterion $\delta_{\bm{k}}\gg 1$.
\subsubsection{Calculation of the decoherence parameter}
\label{sec:lin:delta:calc}
\begin{figure}[t]
\begin{center}
\includegraphics[width=0.49\textwidth]{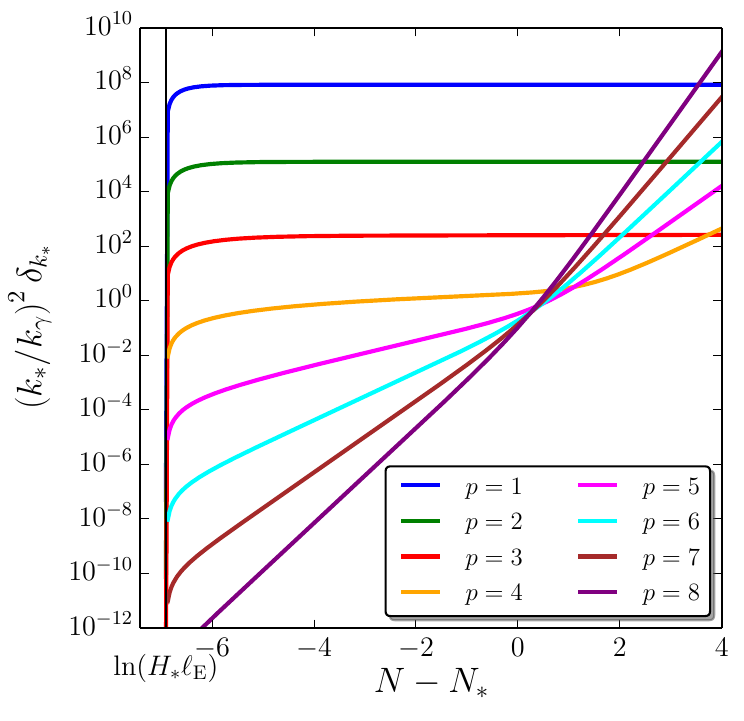}
\includegraphics[width=0.49\textwidth]{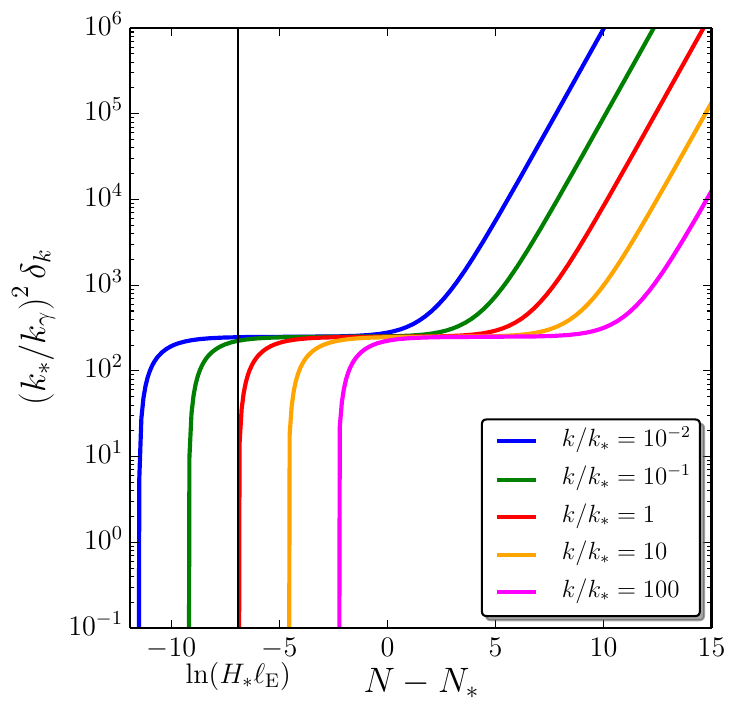}
\caption{Decoherence parameter $\delta_{\bm{k}}$ [rescaled by $(k_\gamma/k_*)^2$], computed with \Eq{eq:exactdelta}, as a function of time (labeled with the number of \efolds~since Hubble exit of the pivot scale), for a few values of $p$ and a
  fixed value of $k=k_*$ (left panel) and for a few values of $k$ and a fixed value of $p=3$ (right panel). The values chosen for the other parameters are $\epsilon_{1*}=10^{-4}$,
  $\epsilon_{2*}=1-0.96-2\epsilon_{1*}$ and
  $H_*\lE =10^{-3}$. }
\label{fig:deltatime}
\end{center}
\end{figure}
At linear order in $\gamma$, which is the order at which the Lindblad equation has been established in \App{sec:DerivingLindblad}, \Eq{eq:defdeltak:integrals} gives rise to
\bea
\delta_{\bm k}(\eta) 
&\simeq \left\vert v_{\bm k}^\prime \right\vert^2\mathcal{J}_{\bm k}
+ \left\vert v_{\bm k}\right\vert^2\mathcal{I}_{\bm k}
-{\left\vert v_{\bm k}\right\vert^2}^\prime
\mathcal{K}_{\bm k}, 
\label{eq:defdeltak:integrals:appr}
\eea
since one recalls that $\mathcal{I}_{\bm{k}}$, $\mathcal{J}_{\bm{k}}$ and $\mathcal{K}_{\bm{k}}$ all carry a factor $\gamma$, see \Eqs{eq:defI}-(\ref{eq:defK}). As shown in \App{subsec:dmsr}, the integrals $\mathcal{I}_{\bm{k}}$, $\mathcal{J}_{\bm{k}}$ and $\mathcal{K}_{\bm{k}}$ can be related to the quantities $I_1$ and $I_2$ defined in \Eq{eq:I1:I2}, see \Eqs{eq:calculI}-(\ref{eq:calculK}). Plugging \Eqs{eq:calculI}-(\ref{eq:calculK}) into \Eq{eq:defdeltak:integrals:appr}, many cancellations occur (the reason behind all these cancellations will be made explicit in \Sec{sec:linear:DecoherenceParam:alternative}) and the following expression is obtained
\bea
\label{eq:exactdelta}
\delta_{\bm k} \left(\eta\right)= &
\frac{\pi}{8\sin^2\left(\pi\nu\right)}
\left(\frac{\kgamma}{k_*}\right)^2
\left(\frac{k}{k_*}\right)^{\left(p-3\right)\left(1+\epsilon_{1*}\right)-2}
\left[I_1\left(\nu\right)+I_1\left(-\nu\right)
-2\cos\left(\pi\nu\right)I_2\left(\nu\right)\right]\, .
\eea 
The corresponding time evolution of
$\delta_{\bm k}$ is displayed on the left panel of
\Fig{fig:deltatime} for different values of $p$ and a fixed value of
$k$ ($k=k_*$), and on the right panel for a fixed value
of $p$ ($p=3$) and different values of $k$. On the left panel, one can see that $\delta_{\bm k}$ takes off very rapidly as soon as the mode under consideration crosses out the correlation length of the environment, and either settles to a stationary value afterwards (if $p \leq 2$) or continues to grow (if $p> 2$). The case $p=3$ for different values of $k$ is displayed on the right panel, where one can see that after crossing out the environment correlation length, $\delta_{\bm k}$ remains stationary for some transient period of time and starts to increase again after a few \efolds. Since $\delta_{\bm k}$ always increases and since it takes off later for smaller scales (larger values of $k$), it is larger on larger scales (smaller values of $k$).

\begin{figure}[t]
\begin{center}
\includegraphics[width=0.49\textwidth]{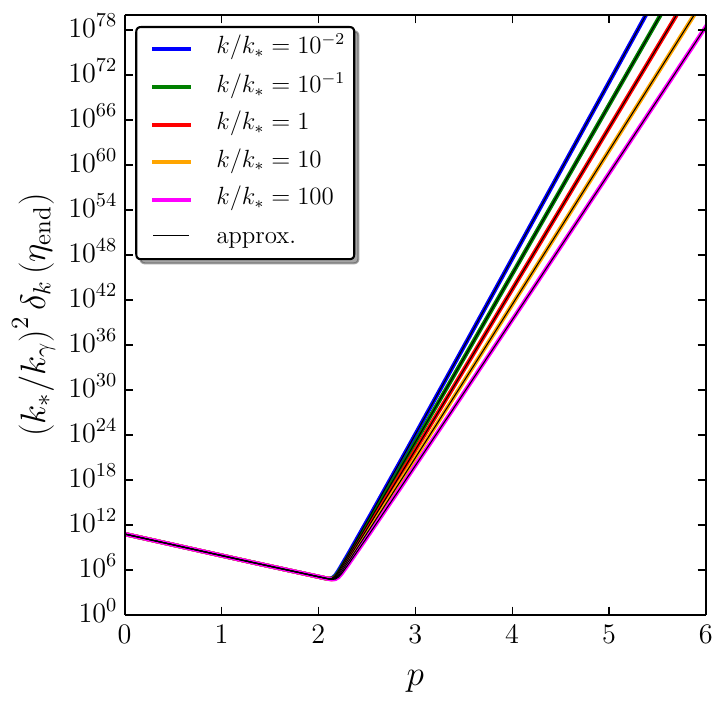}
\includegraphics[width=0.49\textwidth]{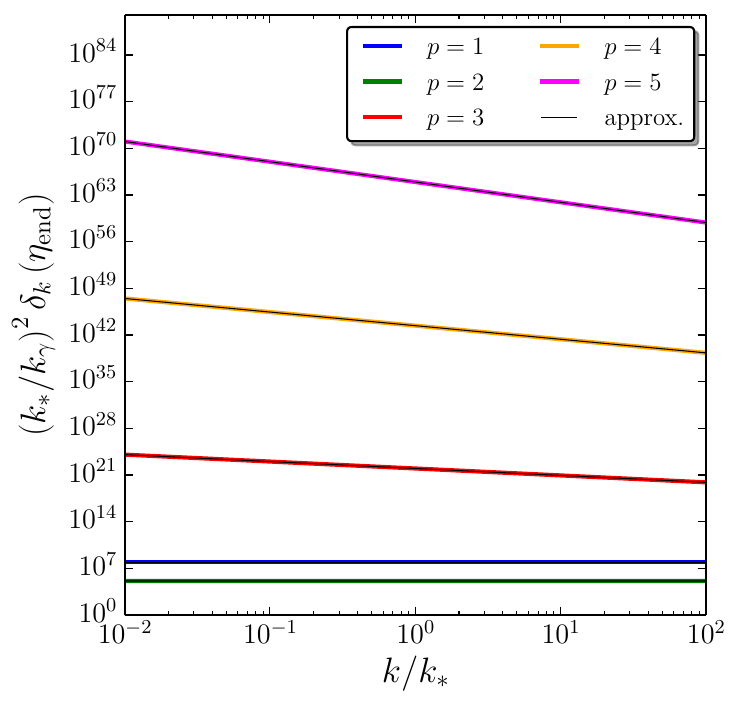}
\caption{Decoherence parameter $\delta_{\bm{k}}$ [rescaled by $(k_\gamma/k_*)^2$] at the end of inflation, as a function of $p$ and for a few values of $k$ (left panel) and as a function of $k$ for a few values of $p$ (right panel). The values chosen for the parameters are $\epsilon_{1*}=10^{-4}$, $\epsilon_{2*}=1-0.96-2\epsilon_{1*}$ and $H_*\lE =10^{-3}$. The coloured lines correspond to the exact formula~(\ref{eq:exactdelta}) while the black lines stand for the analytical approximation~(\ref{eq:deltapprox}) (they are hard to distinguish because of the perfect matching).}
\label{fig:deltak}
\end{center}
\end{figure}
This behaviour can be analytically understood by computing $\delta_{\bm{k}}$ at the end of inflation, when the modes of astrophysical interest today are well outside the Hubble radius and \Eq{eq:exactdelta} can be expanded in the limit $-k\eta \ll 1$. Further assuming that $\lE  \ll H_*^{-1}$ (which was necessary to derive the Lindblad equation in \App{sec:DerivingLindblad}, at least if $\lE\sim t_\uc$), one obtains
\bea
\label{eq:deltapprox}
\delta _{\bm k}(N)\simeq &
\frac14 \left(\frac{\kgamma}{k_*}\right)^2
\Biggl\{\frac{\left(H_*\lE \right)^{\left(p-3\right)\left(1+\epsilon_{1*}\right)-1}}
{1-\left(p-3\right)\left(1+\epsilon_{1*}\right)} \left(\frac{k}{k_*}\right)^{\left(p-3\right)\left(1+\epsilon_{1*}\right)-2}
\\ &
  -\frac{\Gamma^2\left(\nu\right) \ee^{\left(p-3-2\frac{1-\nu}{1+\epsilon_{1*}}\right)(N-N_*)}}
{2^{1-2\nu}\pi\left[2\left(1-\nu\right)-\left(p-3\right)\left(1+\epsilon_{1*}\right)\right]}
\left(\frac{k}{k_*}\right)^{-2\nu}
\Biggr\}\, . 
\eea
In this equation, one can see that  if the coefficient in the argument of the exponential is positive, namely $p>3+2(1-\nu)/(1+\epsilon_1)$ (or, if one neglects slow-roll corrections, $p\gtrsim 2$), then $\delta _{\bm k}$ grows on large scales. If, on the contrary, $p\lesssim 2$, then the exponential becomes very quickly negligible and one is left with the first term, which is constant. This is agreement with the above discussion about \Fig{fig:deltatime}.

In \Fig{fig:deltak}, we have represented $\delta _{\bm k}$ calculated at the end of inflation using the exact result~(\ref{eq:exactdelta}) and the analytical approximation~(\ref{eq:deltapprox}). In the left panel, $\delta _{\bm k}(\eta_\uend)$ is plotted as a function of $p$ and for a few values of $k$, and in the right panel it is plotted as a function of $k$ and for a few values of $p$. The coloured lines correspond to the exact result~(\ref{eq:exactdelta}) while the black lines stand for the analytical approximation~(\ref{eq:deltapprox}). Evidently, they match very well (and are in fact hard to distinguish).

Since decoherence at observable scales is characterised by the condition $\delta_{\bm{k}_*} \gg 1$, \Eq{eq:deltapprox} allows us to calculate the minimum interaction strength that is required for decoherence to complete before the end of inflation. One obtains that decoherence occurs when
\bea
\label{eq:lin:condkgamma:decoherence}
\frac{k_\gamma}{k_*} \gg
\begin{cases}
(H_*\lE)^{\frac{1-\left(p-3\right)\left(1+\epsilon_{1*}\right)}{2}} \quad&\mathrm{if}\quad p<3+\frac{2-2\nu}{1+\epsilon_{1*}}\, ,\\
\ee^{\left(\frac{1-\nu}{1+\epsilon_{1*}}-\frac{p-3}{2}\right)\Delta N_*}\quad&\mathrm{if}\quad p>3+\frac{2-2\nu}{1+\epsilon_{1*}}\, .
\end{cases}
\eea
\subsubsection{Combining with observational constraints}
\begin{figure}[t]
\begin{center}
\includegraphics[width=0.95\textwidth]{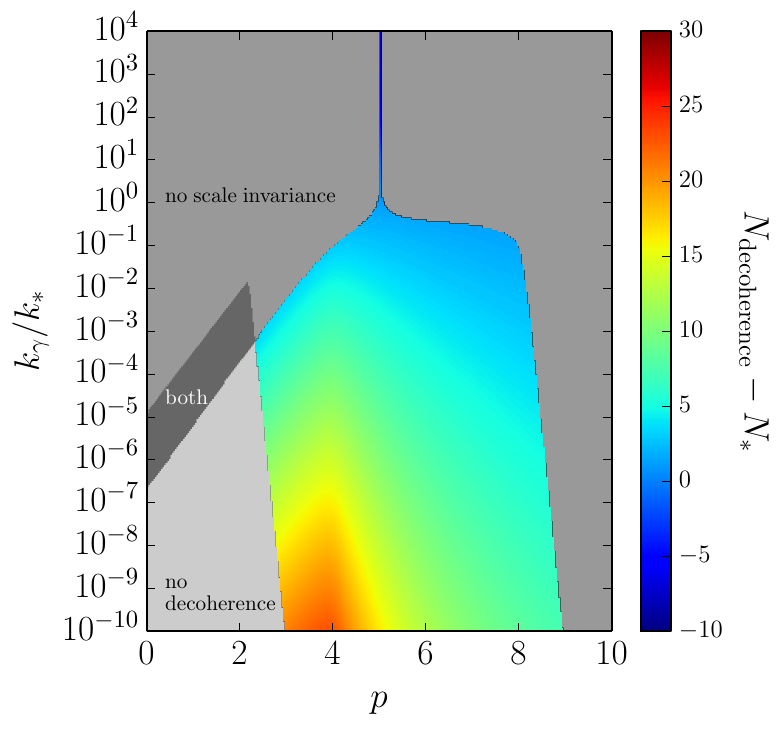}
\caption{Regions in parameter space $(p,\kgamma/k_*)$ where
  decoherence and quasi scale invariance can or cannot be realised. The light grey region corresponds to values of $p$ and $\kgamma/k_*$ where the interaction strength with the environment, parametrised by $\kgamma/k_*$, see \Eq{eq:kbreak:def}, is too small to lead to decoherence. The medium grey region is where it is too large to preserve quasi scale invariance, and the dark grey region is where both problems occur (no decoherence and scale invariance breaking). The coloured region corresponds to parameters where perturbations decohere and scale invariance is preserved. The colour code, indicated by the vertical bar, quantifies how many \efolds~since Hubble crossing it takes before complete decoherence is reached. Here, decoherence is characterised by the condition $\delta_{\bm{k}_*}>10$, and quasi scale invariance by the condition $\vert \nS - \bar{n}_{{}_\mathrm{S}} \vert < 5\sigma_{\nS}$, where $\bar{n}_{{}_\mathrm{S}}\simeq 0.96$ and $\sigma_{\nS} \simeq 0.006$ are the mean value and standard deviation of the Planck spectral index measurement~\cite{Ade:2015lrj} respectively. The spectral index and the decoherence parameter are computed numerically without approximations.}
\label{fig:deltamap}
\end{center}
\end{figure}
We have shown that decoherence occurs in the presence of linear interactions with an environment if the interaction strength is sufficiently large and satisfies \Eq{eq:lin:condkgamma:decoherence}. However, in \Sec{sec:lin:obsConstraints}, it was explained that if the interaction strength is too large, quasi scale invariance is lost, which is observationally excluded. The upper bounds~(\ref{eq:lin:kgammaUpperBound:1})-(\ref{eq:lin:kgammaUpperBound:3}) on the interaction strength were then obtained. When the lower bound provided by \Eq{eq:lin:condkgamma:decoherence} is smaller than the upper bound from \Eqs{eq:lin:kgammaUpperBound:1}-(\ref{eq:lin:kgammaUpperBound:3}), there is a range of values for the interaction strength such that decoherence is obtained without spoiling scale invariance. This range is given by
\bea
\label{eq:Constraint:linear:finalAppr}
\ee^{-\frac{\Delta N_*}{2}\left(p-2+\epsilon_{1*}+\epsilon_{2*}\right)}
\ll \frac{\kgamma}{k_*} \ll 
\begin{cases}
\left(H_*\lE\right)^{\frac{4-p}{2}}\  &\mathrm{if}\quad 2-\epsilon_{1*}-\epsilon_{2*}<p<4-\epsilon_{1*}\, ,\\
1\  &\mathrm{if}\quad 4-\epsilon_{1*}<p<8-3\epsilon_{1*}+\epsilon_{2*}\, ,\\
\ee^{-\frac{\Delta N_*}{2}\left(p-8+3\epsilon_{1*}-\epsilon_{2*}\right)} \  &\mathrm{if}\quad 8-3\epsilon_{1*}+\epsilon_{2*} <p\, .
\end{cases}
\eea
In particular, one can see that when $p<3+(2-2\nu)/(1+\epsilon_{1*})\simeq 2$, decoherence and quasi scale invariance cannot be achieved simultaneously, since \Eq{eq:lin:kgammaUpperBound:3} and the first of \Eq{eq:lin:condkgamma:decoherence} directly contradict each other. This is because, from \Eqs{eq:dP3linear} and~(\ref{eq:deltapprox}), one can see that $\Delta\calP_{\bm{k}} = 2\delta_{\bm{k}}$ in that case.

This allows one to understand \Fig{fig:deltamap} where the situation is summarised. This figure represents the regions in parameter space $(p,\kgamma/k_*)$ where quasi scale invariance and decoherence can or cannot be realised.  The light grey region corresponds to situations where the interaction strength with the environment is too small to yield decoherence. The medium grey region is where it is too large to preserve quasi scale invariance, and the dark grey region is where both problems occur (no decoherence and scale invariance breaking). For $p \lesssim 2$, one can see that decoherence cannot be realised without spoiling the quasi scale invariance of the power spectrum, in agreement with the above discussion. For $p \gtrsim 2$, there are intermediate values of $\kgamma/k_*$ for which decoherence is obtained while preserving quasi scale invariance. This region is coloured in \Fig{fig:deltamap}, and the colour code quantifies how many \efolds~since Hubble crossing are needed in order to complete decoherence. One can check that the larger the interaction strength, the fewer \efolds~it takes to reach decoherence. In general, decoherence happens after Hubble crossing (light blue to red regions), but it can also occur soon after crossing out the environment correlation length and before crossing out the Hubble radius (dark blue regions). 

The most striking feature of the plot is probably the thin vertical line centred at $p=5$.  As shown in \Sec{sec:lin:PowerSpectrumConstraints:MassiveFieldAsEnv}, the origin of this line is the fact that for $p=5$, the correction to the power spectrum caused by the Lindblad term is itself scale invariant. As a consequence, even a very large value of the coupling constant can lead to quasi scale invariance. Let us also recall that $p\simeq 5$ is precisely the case that corresponds to the model described in \App{sec:massivescalarfield} where inflationary perturbations couple to heavy scalar degrees of freedom.  \Fig{fig:deltamap} highlights again the remarkable property of this type of environment, which allows cosmological perturbations to decohere during inflation without spoiling their scale invariance.
\subsubsection{Can we neglect the free Hamiltonian?}
It is sometimes argued that decoherence can be estimated without taking the free Hamiltonian into account, since no decoherence occurs in the free theory. In principle, this is true only if decoherence is so rapid that the free evolution of the system can be neglected while it occurs. Here we re-calculate the decoherence parameter in the absence of free evolution in order to determine when this is indeed the case. 

Neglecting the free Hamiltonian terms in \Eq{eq:diffdensitymatrix}, one has
\bea
\frac{\dd \left\langle v^{s,(1)}_{\bm k}\right\vert \hat{\rho}_{\bm k}^s 
\left\vert v^{s,(2)}_{\bm k}\right\rangle}{\dd \eta}&=
- \frac{\gamma}{2}\left(2\pi\right)^{3/2}\tilde{C}_R({\bm k})
\left[{v^{s,(1)}_{\bm k}}-{v^{s,(2)}_{\bm k}}\right]^2
\left\langle v^{s,(1)}_{\bm k}\right\vert \hat{\rho}^s_{\bm k} 
\left\vert v^{s,(2)}_{\bm k}\right\rangle\, .
\label{eq:rho12approx}
\eea
Using the Bunch-Davies vacuum as the initial state, this can be readily integrated and gives rise to \Eq{eq:defdeltak} if one replaces $\delta_{\bm{k}}$ with $\delta_{\bm{k}}^{\mathrm{no}\,\mathrm{free}}$, defined as
\bea
\delta_{\bm k}^{\mathrm{no}\,\mathrm{free}}(\eta)= \left(2\pi\right)^{3/2}P_{vv}(k)
\int_{-\infty}^\eta \gamma\left(\eta'\right) 
\tilde{C}_R\left(\bm k,\eta'\right) \dd \eta'\, .
\label{eq:deltak:nofree:sol}
\eea
In this expression, $P_{vv} = \vert v_{\bm k} \vert^2 + \mathcal{J}_{\bm k}$, see \Eq{eq:solPvv}. Since $\mathcal{J}_{\bm k}$ is proportional to $\gamma$, it only adds terms of second order in $\gamma$ in \Eq{eq:deltak:nofree:sol}, so if one wants to calculate $\delta_{\bm k}^{\mathrm{no}\,\mathrm{free}}$ at leading order in $\gamma$, it is enough to keep only the standard contribution to the power spectrum. Expanding $\vert v_{\bm k} \vert^2$ given by \Eq{eq:modeFunction:SR} in the super-Hubble limit, and making use of \Eqs{eq:Ck:appr} and~(\ref{eq:gamma:SR}), this gives rise to
\bea
\label{eq:deltak:superH:nofree}
\delta_{\bm k}^{\mathrm{no}\,\mathrm{free}}(N )\simeq & \frac14 \left(\frac{\kgamma}{k_*}\right)^2
\Biggl\{\frac{\Gamma^2(\nu)\left(H\lE 
\right)^{\left(p-3\right)\left(1+\epsilon_{1*}\right)-1}}
{2^{1-2\nu}\pi[1-(p-3)(1+\epsilon_1)]}
\left(\frac{k}{k_*}\right)^{\left(p-3\right)\left(1+\epsilon_{1*}\right)-1-2\nu}
\ee^{\frac{2\nu-1}{1+\epsilon_{1*}}\left(N-N_*\right)}
\\ &
-\frac{\Gamma^2(\nu) \ee^{\left(p-3-2\frac{1-\nu}{1+\epsilon_{1*}}\right)(N-N_*)}}{2^{1-2\nu}
\pi[1-(p-3)(1+\epsilon_{1*})]}
\left(\frac{k}{k_*}\right)^{-2\nu}
\Biggr\}\, .
\eea
This expression needs to be compared with \Eq{eq:deltapprox}. Like \Eq{eq:deltapprox}, it is made of two terms and which one dominates at late time depends on the value of $p$.  By comparing the exponentials of each term, one finds that if $p<3+1/(1+\epsilon_{1*})\simeq 4$, the first term dominates. In \Eq{eq:deltapprox}, the dominant term depends on whether $p$ is smaller or larger than $3+2(1-\nu)/(1+\epsilon_{1*})\simeq 2$ but since neither term matches the first term of \Eq{eq:deltak:superH:nofree}, the two expressions strongly disagree in that case. If $p>3+1/(1+\epsilon_{1*})\simeq 4$, however, the second term in \Eq{eq:deltak:superH:nofree} dominates, which matches the second term of \Eq{eq:deltapprox}, up to a factor $\left[\left(p-3\right)\left(1+\epsilon_{1*}\right)-1\right]/ \left[\left(p-3\right)\left(1+\epsilon_{1*}\right) -2\left(1-\nu\right)\right]$ that is of order one, and the two expressions agree in that case. Interestingly, $p=3+1/(1+\epsilon_{1*})$ corresponds to the limiting value between cases 2 and 3 in the classification introduced in \Sec{sec:linear:PS:SR}, where it was shown that $p>3+1/(1+\epsilon_{1*})$ is the condition for the power spectrum to be independent of the environment correlation shape and length. We conclude that only in this case can the free Hamiltonian be neglected when computing the decoherence parameter. Further insight into this result is provided at the end of \Sec{sec:linear:DecoherenceParam:alternative}.
\subsubsection{Alternative derivation of the decoherence parameter}
\label{sec:linear:DecoherenceParam:alternative}
Before moving on to study decoherence in the presence of quadratic interactions, let us show how the above result can be obtained without solving for the Lindblad equation~(\ref{eq:lindbladgeneral}) entirely. Indeed, in the case of quadratic interactions, a full solution to \Eq{eq:lindbladgeneral} is not available and we will need an alternative technique. 

The starting point is to use \Eq{eq:delta:purity}, $\mathrm{Tr} (\hat{\rho}_{\bm k}^s{}^2)=(1+4\delta_{\bm{k}})^{-1/2}$, as a definition of the decoherence parameter $\delta_{\bm{k}}$. Let us recall that when the state is pure, $\hat{\rho}_{\bm k}^s{}^2=\hat{\rho}_{\bm k}^s$ such that $\mathrm{Tr} (\hat{\rho}_{\bm k}^s{}^2)=1$ and $\delta_{\bm{k}}=0$, while in the presence of decoherence, $\delta_{\bm{k}}>0$ and $\mathrm{Tr} (\hat{\rho}_{\bm k}^s{}^2)<1$. Using the linearity and the cyclicity of the trace operator, one has 
\bea
\frac{\dd}{\dd\eta}\mathrm{Tr} \left(\hat{\rho}_{\bm k}^s{}^2\right) & = 
2 \mathrm{Tr} \left( \hat{\rho}_{\bm k}^s \frac{\dd \hat{\rho}_{\bm k}^s}{\dd\eta} \right)
\\ & =
- 2 i  \mathrm{Tr} \left( \hat{\rho}_{\bm k}^s \left[ \hat{\mathcal{H}}_{\bm k}^{s},\hat{\rho}_{\bm k}^{s}\right] \right)
-\gamma (2\pi)^{3/2} \tilde{C}_R({\bm k}) \mathrm{Tr} \left( \hat{\rho}_{\bm k}^s \left[\hat{v}_{\bm k}^{s},\left[\hat{v}_{\bm k}^{s},
\hat{\rho}_{\bm k}^{s}\right]\right] \right)
\, ,
\label{eq:Trroh2:1}
\eea
where in the second equality, the Lindblad equation~(\ref{eq:lindbladlinear}) written in Fourier subspaces has been used. In this expression, using the cyclicity of the trace operator, one finds that the first term vanishes and the second one can be written as $\mathrm{Tr} ( \hat{\rho}_{\bm k}^s [\hat{v}_{\bm k}^{s},[\hat{v}_{\bm k}^{s},\hat{\rho}_{\bm k}^{s}]] ) = 2 \mathrm{Tr} ( \hat{\rho}_{\bm k}^s [ \hat{\rho}_{\bm k}^s,\hat{v}_{\bm k}^{s}]  \hat{v}_{\bm k}^s)$, so that \Eq{eq:Trroh2:1} becomes
\bea
& \frac{\dd}{\dd\eta}\mathrm{Tr} \left(\hat{\rho}_{\bm k}^s{}^2\right)  = 
-2\gamma (2\pi)^{3/2} \tilde{C}_R({\bm k}) \mathrm{Tr} \left( \hat{\rho}_{\bm k}^s \left[ \hat{\rho}_{\bm k}^s,\hat{v}_{\bm k}^{s}\right]  \hat{v}_{\bm k}^s\right)
\\ & \qquad =
-2\gamma (2\pi)^{3/2} \tilde{C}_R({\bm k}) \int\dd v_{\bm k}^{s,(1)} \int\dd v_{\bm k}^{s,(2)}
v_{\bm k}^{s,(1)} \left[v_{\bm k}^{s,(1)}-v_{\bm k}^{s,(2)}\right] 
\left\vert \left\langle v_{\bm k}^{s,(1)} \right\vert  \hat{\rho}_{\bm k}^s \left\vert v_{\bm k}^{s,(2)} \right\rangle\right\vert^2
\, ,
\label{eq:Trroh2:2}
\eea
where in the second equality, $\mathrm{Tr} ( \hat{\rho}_{\bm k}^s [ \hat{\rho}_{\bm k}^s,\hat{v}_{\bm k}^{s}]  \hat{v}_{\bm k}^s)$ has been written explicitly in terms of the elements of the density matrix $\hat{\rho}_{\bm k}^s$. 

The next step is to notice that at linear order in $\gamma$, it is enough to evaluate the right-hand side of the above equation in the free theory, where the density matrix is given by \Eq{eq:purestate}. In this case, the integrals appearing in \Eq{eq:Trroh2:2} are Gaussian and can be performed explicitly, and one obtains $\mathrm{Tr} ( \hat{\rho}_{\bm k}^s [ \hat{\rho}_{\bm k}^s,\hat{v}_{\bm k}^{s}]  \hat{v}_{\bm k}^s) = P_{vv}(k)$. The relation~(\ref{eq:Trroh2:2}) can then be readily integrated, and expanding \Eq{eq:delta:purity} at leading order in $\gamma$, $\mathrm{Tr} (\hat{\rho}_{\bm k}^s{}^2)\simeq 1-2\delta_{\bm{k}}$, one obtains
\bea
\label{eq:lin:delta:intPvv}
\delta_{\bm{k}}\left(\eta\right) &= \left(2\pi\right)^{3/2}\int_{-\infty}^\eta \gamma\left(\eta'\right)\tilde{C}_R\left(k,\eta'\right) P_{vv}\left(k,\eta'\right)\dd\eta'\\
&= \frac{1}{2}\int_{-\infty}^{\eta} S_1\left(\bm{k},\eta'\right) P_{vv}\left(\bm{k},\eta'\right) \dd \eta'
\, ,
\eea
where in the second line we have recast the result in terms of the source function $S_1$ defined in \Eq{eq:source:linear:def}. Several remarks about this expression are in order.

First, let us notice that \Eq{eq:lin:delta:intPvv} is valid beyond the linear expansion in $\gamma$. Indeed, if one uses the exact solution~(\ref{eq:finalrhomaintext}) to the Lindblad equation to calculate the integrals appearing in the right-hand side of \Eq{eq:Trroh2:2}, one obtains $\mathrm{Tr} ( \hat{\rho}_{\bm k}^s [ \hat{\rho}_{\bm k}^s,\hat{v}_{\bm k}^{s}]  \hat{\rho}_{\bm k}^s) = P_{vv}(k)(1+4\delta_{\bm{k}})^{-3/2} = P_{vv}(k) \mathrm{Tr}^3(\hat{\rho}_{\bm k}^s{}^2)$. This allows one to write \Eq{eq:Trroh2:2} as a linear differential equation for $\mathrm{Tr}(\hat{\rho}_{\bm k}^s{}^2)$, the solution of which gives rise to \Eq{eq:lin:delta:intPvv} when combined with \Eq{eq:delta:purity}. The formula~(\ref{eq:lin:delta:intPvv}) is therefore exact.

Second, at linear order in $\gamma$, the right-hand side of \Eq{eq:lin:delta:intPvv} can be evaluated in the free theory where $P_{vv} = \vert v_{\bm{k}} \vert^2$, where the mode function $ v_{\bm{k}}$ is given by \Eq{eq:modeFunction:SR} in terms of Bessel functions. Making use of \Eqs{eq:Ck:appr} and~(\ref{eq:gamma:SR}), one obtains \Eq{eq:exactdelta}. This elucidates the numerous cancellations that appeared when going from \Eq{eq:defdeltak:integrals:appr} to \Eq{eq:exactdelta}, and which result from the equivalence between \Eq{eq:defdeltak:integrals:appr}  and \Eq{eq:lin:delta:intPvv}. In fact, without expanding in $\gamma$, the right-hand side of \Eq{eq:lin:delta:intPvv} can be evaluated with $P_{vv}=\vert v_{\bm k}\vert^2+\mathcal{J}_{\bm k}$, see \Eq{eq:solPvv}, and this gives rise to \Eq{eq:defdeltak:integrals}. This explains why, since $P_{vv}$ is linear in $\gamma$, $\delta_{\bm{k}}$ is quadratic in $\gamma$.

Third, it is interesting to compare \Eq{eq:lin:delta:intPvv} with \Eq{eq:deltak:nofree:sol}, which was obtained by neglecting the influence of the free Hamiltonian. One can see that the only difference between these two expressions is that in \Eq{eq:deltak:nofree:sol}, the power spectrum is taken out of the integral and evaluated at the time $\eta$. This explains why only the contribution from the upper bound of the integral of \Eq{eq:deltak:nofree:sol} is correctly computed, up to a prefactor of order one.
\subsection{Quadratic interaction}
\label{subsec:decotimequadratic}
Let us now study decoherence as produced by quadratic interactions with an environment. As already explained, because of mode coupling, the Lindblad equation~(\ref{eq:lindbladgeneral}) does not decouple into a set of independent Lindblad equations for each Fourier mode, and cannot be solved entirely. This means that the calculation of \Sec{sec:lin:delta:calc} cannot be reproduced here, and that the alternative technique presented in \Sec{sec:linear:DecoherenceParam:alternative} must instead be employed. Since the state is not factorisable into Fourier subspaces in the presence of quadratic interactions, \ie \Eq{eq:rho:factorization} does not apply, the effective density matrix on the space $\left\lbrace s \atop {\bm{k}} \right\rbrace$ has first to be defined by tracing over all other degrees of freedom,
\bea
\label{eq:quad:rhoks:def}
\hat{\rho}_{\bm{k}}^s \equiv \mathrm{Tr}_{
\left\lbrace
 s' \atop {\bm{k}'\neq \bm{k}}
\right\rbrace ,
\left\lbrace
 \bar{s} \atop {\bm{k}}
\right\rbrace
}
\left(\hat{\rho}_v\right)\, ,
\eea
where $\bar{s}=\mathrm{I}$ if $s=\mathrm{R}$ and $\bar{s}=\mathrm{R}$ if $s=\mathrm{I}$. The decoherence parameter $\delta_{\bm{k}}$ is then defined according to \Eq{eq:delta:purity} through $\mathrm{Tr}_{ \left\lbrace s \atop {\bm{k}} \right\rbrace} (\hat{\rho}_{\bm{k}}^s{}^2)$.
\subsubsection{Decoherence criterion}
\label{sec:DecoherenceCriterion}
Making use of the linearity and ciclycity of the trace operator, one has
\bea
\frac{\dd}{\dd \eta}\mathrm{Tr}_{ \left\lbrace s \atop {\bm{k}} \right\rbrace} \left(\hat{\rho}_{\bm{k}}^s{}^2\right) & =
2 \mathrm{Tr}_{ \left\lbrace s \atop {\bm{k}} \right\rbrace} \left[
\mathrm{Tr}_{
\left\lbrace
 s' \atop {\bm{k}'\neq \bm{k}}
\right\rbrace ,
\left\lbrace
 \bar{s} \atop {\bm{k}}
\right\rbrace
}
\left(\hat{\rho}_v\right)
\frac{\dd}{\dd\eta}
\mathrm{Tr}_{
\left\lbrace
 s' \atop {\bm{k}'\neq \bm{k}}
\right\rbrace ,
\left\lbrace
 \bar{s} \atop {\bm{k}}
\right\rbrace
}
\left(\hat{\rho}_v\right)
\right]
\\ & = 
2 \mathrm{Tr}_{ \left\lbrace s \atop {\bm{k}} \right\rbrace} \left[
\hat{\rho}_{\bm{k}}^s
\mathrm{Tr}_{
\left\lbrace
 s' \atop {\bm{k}'\neq \bm{k}}
\right\rbrace ,
\left\lbrace
 \bar{s} \atop {\bm{k}}
\right\rbrace
}
\left(\frac{\dd\hat{\rho}_v}{\dd\eta}\right)
\right]\, .
\label{eq:quad:decoherence:dTrrho2deta:1}
\eea
In this expression, $\dd\hat{\rho}_v/\dd\eta$ needs to be replaced by the Lindblad equation~(\ref{eq:lindbladgeneral}) (with $\hat{A}=\hat{v}^2$), which contains two terms. The first one involves the free Hamiltonian~(\ref{eq:Hsystem}) and is, in practice, difficult to incorporate in the following calculation. However, the contribution from this term to the decoherence parameter only reflects the correlations that develop between the Fourier subspace $\left\lbrace s \atop {\bm{k}} \right\rbrace$ under consideration and the other Fourier subspaces, over which we have traced over, see \Eq{eq:quad:rhoks:def}. The reason for this partial trace is not that we do not ``observe'' the other Fourier degrees of freedom, but is because we want to define a decoherence parameter for each Fourier subspace, by analogy with the linear case. If the state were factorisable, the Hamiltonian contribution to \Eq{eq:quad:decoherence:dTrrho2deta:1} would vanish, which is why it is simply discarded in what follows. The second term coming from \Eq{eq:lindbladgeneral} is the Lindblad term. Fourier expanding $\hat{v}$ and $C_R$, it gives rise to
\bea
\mathrm{Tr}_{
\left\lbrace
 s' \atop {\bm{k}'\neq \bm{k}}
\right\rbrace ,
\left\lbrace
 \bar{s} \atop {\bm{k}}
\right\rbrace
}
\left(\frac{\dd\hat{\rho}_v}{\dd\eta}\right)= &
-\frac{\gamma}{2}\left(2\pi\right)^{-3/2}\int_{\setR^3} \dd\bm{k}_1 \dd\bm{k}_2 \dd\bm{k}_3 \tilde{C}_R\left(\bm{k}_1\right)
\\ & \quad
\mathrm{Tr}_{
\left\lbrace
 s' \atop {\bm{k}'\neq \bm{k}}
\right\rbrace ,
\left\lbrace
 \bar{s} \atop {\bm{k}}
\right\rbrace
}
\left(
\left[\hat{v}_{\bm{k}_2}\hat{v}_{-\bm{k}_1-\bm{k}_2}
,\left[\hat{v}_{\bm{k}_3}\hat{v}_{\bm{k}_1-\bm{k}_3},
\hat{\rho}_v
\right]\right]
\right)\, .
\label{eq:quad:decoherence:dTrrho2deta:3}
\eea
At leading order in $\gamma$, the second line of the above expression can be evaluated in the free theory, where the density matrix is factorisable and given by \Eqs{eq:rho:factorization} and~(\ref{eq:purestate}).

Let us consider the case where $\bm{k}\notin \lbrace \pm \bm{k}_2, \pm (\bm{k}_1+\bm{k}_2), \pm \bm{k}_3, \pm (\bm{k}_1-\bm{k}_3) \rbrace$. By removing $\hat{\rho}_{\bm{k}}^{s}$ from the trace in the second line of \Eq{eq:quad:decoherence:dTrrho2deta:1} (since it commutes with all $\hat{v}$ operators), one is left with a full (as opposed to partial) trace, that vanishes. This means that $\bm{k}$ must be equal, up to a sign (recall that $\bm{k}$ and $\bm{k}'$ live in $\setR^{3+}$ while $\bm{k}_1$, $\bm{k}_2$ and $\bm{k}_3$ live in $\setR^3$), to one of the wavenumbers that index the $\hat{v}$ operators in \Eq{eq:quad:decoherence:dTrrho2deta:3}. If it is equal to one such wavenumber only and the other three are different,  one can show that the trace vanishes again, such that $\bm{k}$ must be equal to 2, 3 and all 4 wavenumbers that index the $\hat{v}$ operators in \Eq{eq:quad:decoherence:dTrrho2deta:3}. Let us discuss these three possibilities separately.

We first examine the situation where $\bm{k}$ is equal to two of the wavenumbers that index the $\hat{v}$ operators in \Eq{eq:quad:decoherence:dTrrho2deta:3}. For instance, let us consider the case where $\bm{k}=\bm{k}_2=-\bm{k}_3$ and $\bm{k}_1\neq - 2\bm{k}$, for which one has
\bea
&\kern -5em
\mathrm{Tr}_{
\left\lbrace
 s' \atop {\bm{k}'\neq \bm{k}}
\right\rbrace ,
\left\lbrace
 \bar{s} \atop {\bm{k}}
\right\rbrace
}
\left(
\left[\hat{v}_{\bm{k}}\hat{v}_{-\bm{k}_1-\bm{k}}
,\left[\hat{v}_{-\bm{k}}\hat{v}_{\bm{k}_1+\bm{k}},
\hat{\rho}_v
\right]\right]
\right)=
\\ &
\mathrm{Tr}_{\left\lbrace
 \bar{s} \atop {\bm{k}}
\right\rbrace
}
\left( \hat{v}_{\bm{k}} \hat{v}_{-\bm{k}} \hat{\rho}_{\bm{k}}^s \hat{\rho}_{\bm{k}}^{\bar{s}}   \right)
\mathrm{Tr}_{\left\lbrace
 s,\bar{s} \atop {\bm{k}+\bm{k}_1}
\right\rbrace}
\left(\hat{v}_{-\bm{k}-\bm{k}_1}\hat{v}_{\bm{k}+\bm{k_1}} \hat{\rho}_{\bm{k}+\bm{k_1}}^s \rho_{\bm{k}+\bm{k_1}}^{\bar{s}} \right)
\\ - &
\mathrm{Tr}_{\left\lbrace
 \bar{s} \atop {\bm{k}}
\right\rbrace
}
\left( \hat{v}_{\bm{k}}  \hat{\rho}_{\bm{k}}^s \hat{\rho}_{\bm{k}}^{\bar{s}} \hat{v}_{-\bm{k}}  \right)
\mathrm{Tr}_{\left\lbrace
 s,\bar{s} \atop {\bm{k}+\bm{k}_1}
\right\rbrace}
\left(\hat{v}_{-\bm{k}-\bm{k}_1} \hat{\rho}_{\bm{k}+\bm{k_1}}^s \rho_{\bm{k}+\bm{k_1}}^{\bar{s}} \hat{v}_{\bm{k}+\bm{k_1}} \right)
\\ - &
\mathrm{Tr}_{\left\lbrace
 \bar{s} \atop {\bm{k}}
\right\rbrace
}
\left( \hat{v}_{-\bm{k}}  \hat{\rho}_{\bm{k}}^s \hat{\rho}_{\bm{k}}^{\bar{s}} \hat{v}_{\bm{k}}  \right)
\mathrm{Tr}_{\left\lbrace
 s,\bar{s} \atop {\bm{k}+\bm{k}_1}
\right\rbrace}
\left(\hat{v}_{\bm{k}+\bm{k}_1} \hat{\rho}_{\bm{k}+\bm{k_1}}^s \rho_{\bm{k}+\bm{k_1}}^{\bar{s}} \hat{v}_{-\bm{k}-\bm{k_1}} \right)
\\ + &
\mathrm{Tr}_{\left\lbrace
 \bar{s} \atop {\bm{k}}
\right\rbrace
}
\left(   \hat{\rho}_{\bm{k}}^s \hat{\rho}_{\bm{k}}^{\bar{s}} \hat{v}_{-\bm{k}} \hat{v}_{\bm{k}}  \right)
\mathrm{Tr}_{\left\lbrace
 s,\bar{s} \atop {\bm{k}+\bm{k}_1}
\right\rbrace}
\left( \hat{\rho}_{\bm{k}+\bm{k_1}}^s \rho_{\bm{k}+\bm{k_1}}^{\bar{s}} \hat{v}_{\bm{k}+\bm{k}_1} \hat{v}_{-\bm{k}-\bm{k_1}} \right)
\\ = &
\frac{1}{2} \left[ \hat{v}_{\bm{k}}^s, \left[  \hat{v}_{\bm{k}}^s, \hat{\rho}_{\bm{k}}^s\right]\right] P_{vv}\left({\bm{k}}+{\bm{k}}_1\right)\, ,
\label{eq:quad:decoherence:Tr:22}
\eea
where we have used the decomposition $\hat{v}_{\bm{k}} = (\hat{v}_{\bm{k}}^{\mathrm{R}}+i\hat{v}_{\bm{k}}^{\mathrm{I}})/\sqrt{2}$. The same result is obtained with $\bm{k}=\bm{k}_2=\bm{k}_3-\bm{k}_1$, $ \bm{k}=-\bm{k}_1-\bm{k_2}=-\bm{k}_3$ or $\bm{k} =- \bm{k}_1-\bm{k_2}=- \bm{k}_1+\bm{k_3} $, if the condition $\bm{k}_1\neq - 2\bm{k}$ is enforced. In the same manner, if $\bm{k}=-\bm{k}_2=\bm{k}_3$, $\bm{k}=-\bm{k}_2=\bm{k}_1-\bm{k}_3$, $\bm{k}=\bm{k}_1+\bm{k}_2=\bm{k}_3$ or $\bm{k}=\bm{k}_1+\bm{k}_2=\bm{k}_1-\bm{k}_3$, and if the condition $\bm{k}_1\neq  2\bm{k}$ is enforced, the same result as in \Eq{eq:quad:decoherence:Tr:22} is obtained, except that the power spectrum is evaluated at $\bm{k}-\bm{k}_1$ instead of $\bm{k}+\bm{k}_1$. One can check that all other configurations give a vanishing result.

Then, one can show that the case where $\bm{k}$ is equal to three of the wavenumbers that index the $\hat{v}$ operators in \Eq{eq:quad:decoherence:dTrrho2deta:3} gives contributions that are always proportional to the quantum mean value of a single mode function operator and therefore vanish, see the discussion below \Eq{eq:linearlinear}. Finally remains the situation where all wavenumbers are, up a sign, equal. For instance, let us consider the case where $\bm{k_1}=-2\bm{k}$, $\bm{k}_2 = \bm{k}$ and $\bm{k}_3=-\bm{k}$, for which one has
\bea
\mathrm{Tr}_{
\left\lbrace
 s' \atop {\bm{k}'\neq \bm{k}}
\right\rbrace ,
\left\lbrace
 \bar{s} \atop {\bm{k}}
\right\rbrace
}
\left(
\left[\hat{v}_{\bm{k}}\hat{v}_{\bm{k}}
,\left[\hat{v}_{-\bm{k}}\hat{v}_{-\bm{k}},
\hat{\rho}_v
\right]\right]
\right)&=
\mathrm{Tr}_{
\left\lbrace
 \bar{s} \atop {\bm{k}}
\right\rbrace
}
\left(
\left[\hat{v}_{\bm{k}}\hat{v}_{\bm{k}}
,\left[\hat{v}_{-\bm{k}}\hat{v}_{-\bm{k}},
\hat{\rho}_{\bm{k}}^s\hat{\rho}_{\bm{k}}^{\bar{s}}
\right]\right]
\right)
\\ & =
\frac{1}{4}\left[\hat{v}_{\bm{k}}^s{}^2,\left[\hat{v}_{\bm{k}}^s{}^2,\hat{\rho}_{\bm{k}}^s\right]\right]
+P_{vv}\left(\bm{k}\right)\left[\hat{v}_{\bm{k}}^s,\left[\hat{v}_{\bm{k}}^s,\hat{\rho}_{\bm{k}}^s\right]\right]\, .
\eea
The only other non-vanishing configuration of this type is when $\bm{k_1}=2\bm{k}$, $\bm{k}_2 =- \bm{k}$ and $\bm{k}_3=\bm{k}$, which gives the same result. These contributions are however suppressed by a volume factor with respect to the ones giving \Eq{eq:quad:decoherence:Tr:22}, since they do not vanish only for a single configuration of the wavenumbers $\bm{k}_1$, $\bm{k}_2$ and $\bm{k}_3$, while the configurations leading to \Eq{eq:quad:decoherence:Tr:22} leave one wavenumber free. This is why, in the factorisable state, one obtains that
\bea
 & \kern -2em
 \mathrm{Tr}_{
\left\lbrace
 s' \atop {\bm{k}'\neq \bm{k}}
\right\rbrace ,
\left\lbrace
 \bar{s} \atop {\bm{k}}
\right\rbrace
} 
\left(
\left[\hat{v}_{\bm{k}_2}\hat{v}_{-\bm{k}_1-\bm{k}_2}
,\left[\hat{v}_{\bm{k}_3}\hat{v}_{\bm{k}_1-\bm{k}_3},
\hat{\rho}_v
\right]\right]
\right)= \frac{1}{2} \left[ \hat{v}_{\bm{k}}^s, \left[  \hat{v}_{\bm{k}}^s, \hat{\rho}_{\bm{k}}^s\right]\right] 
\times \\
\Big\lbrace
 & 
 P_{vv}\left({\bm{k}}+{\bm{k}}_1\right)\left[
\delta\left({\bm{k}}_2-{\bm{k}}\right)+\delta\left({\bm{k}}_2+{\bm{k}}_1+{\bm{k}}\right)
\right]\left[
\delta\left({\bm{k}}_3+{\bm{k}}\right)+\delta\left({\bm{k}}_3-{\bm{k}}_1-{\bm{k}}\right)
\right]
\\ + & 
P_{vv}\left({\bm{k}}-{\bm{k}}_1\right)\left[
\delta\left({\bm{k}}_2+{\bm{k}}\right)+\delta\left({\bm{k}}_2-{\bm{k}}_1+{\bm{k}}\right)
\right]\left[
\delta\left({\bm{k}}_3-{\bm{k}}\right)+\delta\left({\bm{k}}_3-{\bm{k}}_1+{\bm{k}}\right)
\right]
\Big\rbrace
\, .
\eea
Plugging back this expression into \Eq{eq:quad:decoherence:dTrrho2deta:3}, the integrals over $\bm{k}_2$ and $\bm{k}_3$ can be performed, and one finds that the right-hand side of \Eq{eq:quad:decoherence:dTrrho2deta:3} is given by $-[ \hat{v}_{\bm{k}}^s, [  \hat{v}_{\bm{k}}^s, \hat{\rho}_{\bm{k}}^s]]  S_2/4$, where $S_2$ is the source function defined in \Eq{eq:source:quadratic:def} and computed in \Eq{eq:quad:source:renormalised}. Inserting the result into \Eq{eq:quad:decoherence:dTrrho2deta:1}, one obtains
\bea
\frac{\dd}{\dd \eta}\mathrm{Tr}_{ \left\lbrace s \atop {\bm{k}} \right\rbrace} \left(\hat{\rho}_{\bm{k}}^s{}^2\right) & = -S_2\left(\bm{k},\eta\right) \mathrm{Tr}_{ \left\lbrace s \atop {\bm{k}} \right\rbrace} \left(\hat{\rho}_{\bm{k}}^s  \left[ \hat{v}_{\bm{k}}^s, \left[  \hat{v}_{\bm{k}}^s, \hat{\rho}_{\bm{k}}^s\right]\right]\right)
=
- S_2\left(\bm{k},\eta\right) P_{vv}\left(\bm{k},\eta\right)
\, ,
\eea
where in the second equality we have used the formula $\mathrm{Tr}(\hat{\rho}_{\bm{k}}^s[ \hat{v}_{\bm{k}}^s, [  \hat{v}_{\bm{k}}^s, \hat{\rho}_{\bm{k}}^s]]) =  P_{vv}(k)$ derived around \Eq{eq:Trroh2:2}. At leading order in $\gamma$, $\mathrm{Tr}_{ \left\lbrace s \atop {\bm{k}} \right\rbrace} \left(\hat{\rho}_{\bm{k}}^s{}^2\right) \simeq 1-2\delta_{\bm{k}}$, and this gives rise to
\bea
\label{eq:quad:delta:intPvv}
\delta_{\bm{k}} = \frac{1}{2} \int_{-\infty}^\eta S_2\left(\bm{k},\eta'\right) P_{vv}\left(\bm{k},\eta'\right) \dd \eta'\, .
\eea
which is analogous to \Eq{eq:lin:delta:intPvv}. This suggests that the above formula is in fact generic, in the same way that \Eq{eq:exactsolquadratic} for the power spectrum applies for any source function. 
\subsubsection{Calculation of the decoherence parameter}
\begin{figure}[t]
\begin{center}
\includegraphics[width=0.49\textwidth]{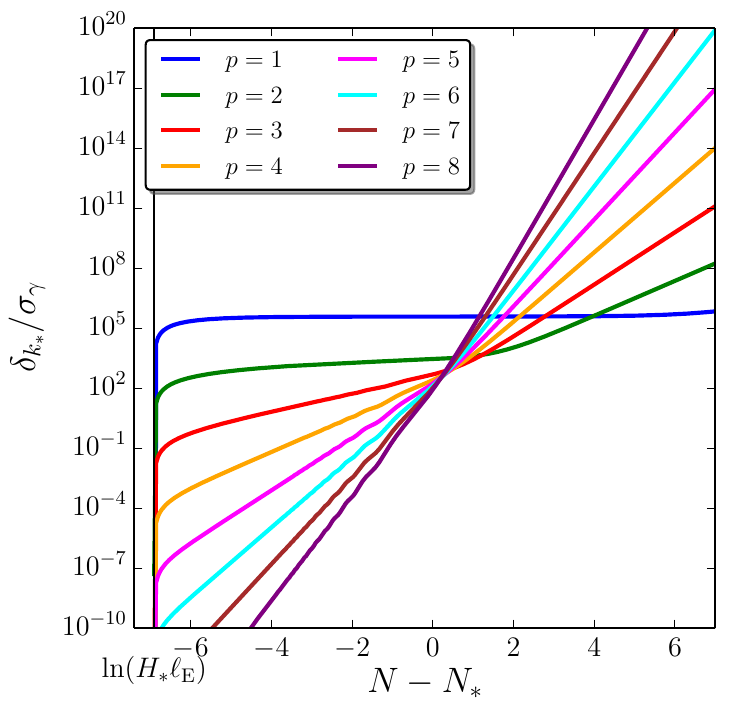}
\includegraphics[width=0.49\textwidth]{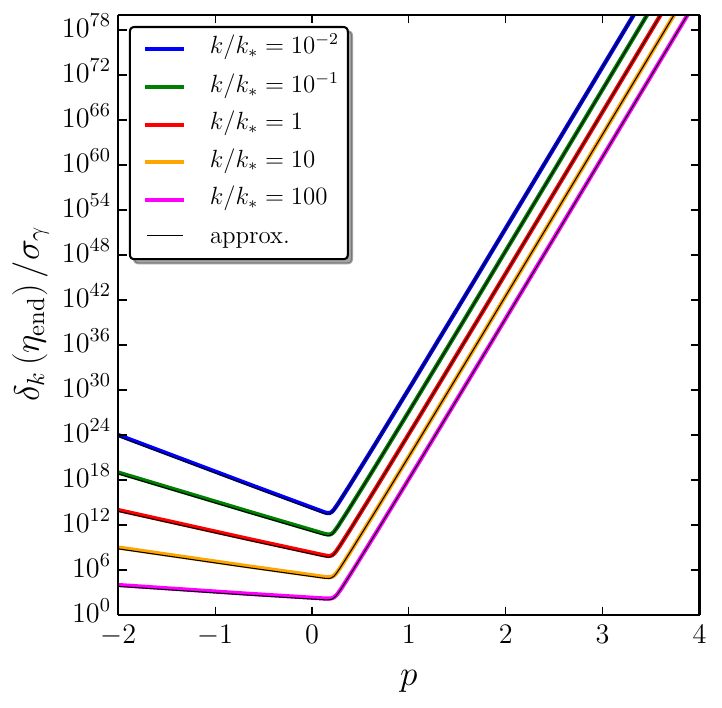}
\caption{Left panel: decoherence parameter $\delta_{\bm{k}}$ (rescaled by $\sigma_\gamma$), computed with \Eq{eq:exactdelta:quad}, as a function of time (labeled with the number of \efolds~since Hubble exit of the pivot scale), for a few values of $p$ and a fixed value of $k=k_*$. Right panel: decoherence parameter at the end of inflation, as a function of $p$ and for a few values of $k$. The coloured lines correspond to the exact formula~(\ref{eq:exactdelta:quad}) while the black lines stand for the analytical approximation~(\ref{eq:deltapprox:quad}) (they are hard to distinguish because of the perfect matching). The values chosen for the parameters are $H_*\lE =10^{-3}$, $\Delta N_*=50$, and $N_{_{\mathrm T}}= N_\uend - N_{{}_\mathrm{IR}} = 10^4$. }
\label{fig:delta:quad}
\end{center}
\end{figure}
Since the calculation is being performed at linear order in $\gamma$, the right-hand side of \Eq{eq:quad:delta:intPvv} must be evaluated in the free theory where $P_{vv} = \vert v_{\bm{k}} \vert^2$, and the mode function $ v_{\bm{k}}$ is given by \Eq{eq:modeFunction:SR}. Making use of \Eq{eq:quad:source:renormalised} for the source function with \Eq{eq:Ck:appr} for the environment correlator, one obtains  
\bea
\label{eq:exactdelta:quad}
\delta_{\bm{k}}\left(\eta\right) = -\frac{\sigma_\gamma}{3 \sin^2\left(\pi\nu\right)}\left(\frac{k_*}{k}\right)^{\alpha_2+1}\left[I_3\left(\nu\right)+I_3\left(-\nu\right)-2\cos\left(\pi\nu\right) I_4\left(\nu\right)\right]\, ,
\eea
where $\sigma_\gamma$ has been defined in \Eq{eq:sigmagamma:def}, $I_3$ and $I_4$ are defined in \App{subsec:solvingthirdquadratic} in \Eq{eq:defI3I4}, and where, neglecting slow-roll corrections for the reason given in \Sec{sec:quad:source}, $\nu=3/2$ and $\alpha_2=2-p$. Let us notice that the structure of \Eq{eq:exactdelta:quad} is very similar to the corresponding expression in the case of linear interactions, namely \Eq{eq:exactdelta}.

The corresponding time evolution of $\delta_{\bm k}$ is displayed on the left panel of \Fig{fig:delta:quad} for different values of $p$ and at the pivot scale $k_*$. The main difference with the case of linear interactions shown in \Fig{fig:deltatime} is that, here, $\delta_{\bm{k}}$ continues to increase at late time for all values of $p$ that are shown. This can be understood analytically by computing $\delta_{\bm{k}}$ close to the end of inflation, when the modes of astrophysical interest today are well outside the Hubble radius and \Eq{eq:exactdelta} can be expanded in the limit $-k\eta \ll 1$. Further assuming that $\lE  \ll H_*^{-1}$, the integrals $I_3$ and $I_4$ can be expanded according to \Eqs{eq:I3:appr} and~(\ref{eq:I4:appr}) respectively, and one obtains
\bea
\label{eq:deltapprox:quad}
\delta_{\bm{k}}\left(N\right)\simeq \frac{2\sigma_\gamma}{3\pi} & \left\lbrace 
\left(\frac{k}{k_*}\right)^{-3}\ee^{p \left(N-N_*\right)}\frac{p\left(N-N_{{}_\mathrm{IR}}\right)-1}{p^2}
\right. \\ & \left.
+\left(\frac{k}{k_*}\right)^{p-3}
\frac{\left(H_*\lE\right)^{p-2}}{2-p}\left[
\frac{1}{2-p}+\ln\left(H_*\lE\right) + \ln\left(\frac{k}{k_*}\right)+N_{{}_\mathrm{T}}-\Delta N_*
\right]
\right\rbrace\, .
\eea
In the right panel of \Fig{fig:delta:quad}, we have represented $\delta_{\bm{k}}$ calculated at the end of inflation using the exact result~(\ref{eq:exactdelta:quad}) (coloured lines) and the analytical approximation~(\ref{eq:deltapprox:quad}) (black lines), and they match very well. 

In \Eq{eq:deltapprox:quad}, one can see that if $p>0$, the first term dominates at late time and $\delta_{\bm{k}}$ grows on large scales, in agreement with what can be seen on the left panel of \Fig{fig:delta:quad}. If, on the contrary, $p<0$, the exponential in the first term becomes very quickly negligible and one is left with the second term, which is constant. Recalling that decoherence at observable scales is characterised by the condition $\delta_{{\bm k}_*}\gg 1$, \Eq{eq:deltapprox:quad} allows us to calculate the minimum interaction strength that is required for decoherence to complete before the end of inflation,
\bea
\label{eq:quad:condgamma:decoherence}
\sigma_{\gamma} \gg
\begin{cases}
\dfrac{\left(H_*\lE\right)^{2-p}}{\ln\left(H_*\lE\right)+N_{{}_\mathrm{T}}-\Delta N_*} \quad&\mathrm{if}\quad p<0\, ,\\
\dfrac{\ee^{-p\Delta N_*}}{N_{{}_\mathrm{T}}}\quad&\mathrm{if}\quad p>0\, .
\end{cases}
\eea
\subsubsection{Combining with observational constraints}
\begin{figure}[t]
\begin{center}
\includegraphics[width=0.95\textwidth]{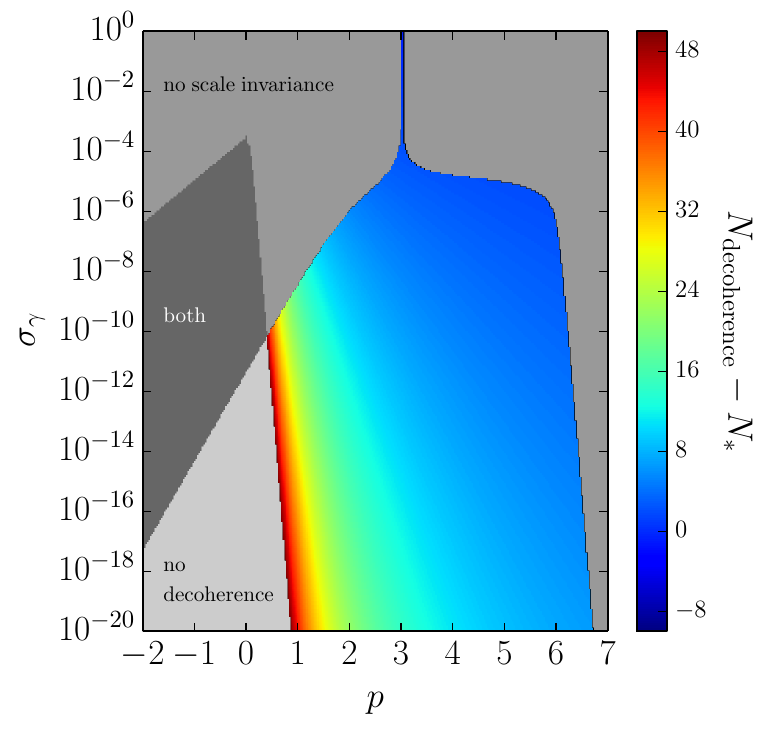}
\caption{Regions in parameter space $(p,\sigma_\gamma)$ where
  decoherence and quasi scale invariance can or cannot be realised. The light grey region corresponds to values of $p$ and $\sigma_\gamma$ where the interaction strength with the environment, parametrised by $\sigma_\gamma$, see \Eq{eq:sigmagamma:def}, is too small to lead to decoherence. The medium grey region is where it is too large to preserve quasi scale invariance, and the dark grey region is where both problems occur (no decoherence and scale invariance breaking). The coloured region corresponds to parameters where perturbations decohere and scale invariance is preserved. The colour code, indicated by the vertical bar, quantifies how many \efolds~since Hubble crossing it takes before complete decoherence is reached. This map is obtained with similar conventions as in \Fig{fig:deltamap}.}
\label{fig:deltamap:quad}
\end{center}
\end{figure}
As for the case of linear interactions, these lower bounds can be combined with the upper bounds derived in \Sec{sec:quad:PSconstraints} from the requirement that the quasi scale invariance of the power spectrum is preserved. One obtains that the range of values for the interaction strength, here parametrised by $\sigma_\gamma$, such that decoherence occurs without spoiling scale invariance is given by
\bea
\label{eq:Constraint:quadratic:finalAppr}
{\ee^{-p \Delta N_*}} \ll N_{{}_\mathrm{T}} \sigma_\gamma\ll 
\begin{cases}
{\left(H_*\lE\right)^{2-p}} \quad &\mathrm{if}\quad 0<p<2\, ,\\
1 \quad &\mathrm{if}\quad 2<p<6\, ,\\
\ee^{\left(6-p\right)\Delta N_*} \quad &\mathrm{if}\quad p>6\, .
\end{cases}
\eea
In particular, one can see that when $p<0$, the correction to the power spectrum given in \Eq{eq:dP3quadratic} and the decoherence parameter given by the first line of \Eq{eq:deltapprox:quad} are directly related, $\left.\Delta {\cal P}_{\bm k}\right \vert _3  = 2 \delta_{\bm{k}}$, which explains why decoherence cannot occur without spoiling the quasi scale invariance of the power spectrum.

The situation is summarised in \Fig{fig:deltamap:quad}, which represents the regions in parameter space where quasi scale invariance and decoherence can or cannot be realised. The conventions are the same as the ones used in \Fig{fig:deltamap} in the case of linear interactions, and all remarks made about that figure apply here too. As for linear interactions, the striking feature of \Fig{fig:deltamap:quad} is the presence of a thin vertical line centred at $p=3$, for which the correction to the power spectrum caused by the Lindblad term is scale invariant, and observations do not constrain the interaction strength. Let us stress again that $p\simeq3$ precisely corresponds to the model described in \App{sec:massivescalarfield} where inflationary perturbations couple to heavy scalar degrees of freedom.
\section{Generalisation to higher-order interactions}
\label{sec:generalisation}
So far, we have shown that the effective inclusion of environmental degrees of freedom, through a Lindblad equation, gives rise to a modified power spectrum and to a decoherence parameter that can be calculated from a source function, see \Eq{eq:exactsolquadratic} and \Eq{eq:quad:delta:intPvv} respectively. In the case of linear interactions, $n=1$, the source function is given by \Eq{eq:source:linear:def}, namely
\bea
\label{eq:source:linear}
S_1\left(\bm{k},\eta\right)=2 (2\pi)^{3/2}\gamma
\tilde{C}_{R}\left(k\right)\, .
\eea
For quadratic interactions, $n=2$, one can rewrite \Eq{eq:source:quadratic:def} as
\bea
\label{eq:source:quadratic}
S_2\left(\bm{k},\eta\right)=\frac{8\gamma}{(2\pi)^{3/2}}
\int \dd   {\bm p}_1\, \tilde{C}_{R}\left({\bm{k} - \bm{p}}_1\right)
P_{vv}\left( \bm{p}_1\right)\, ,
\eea
where the power spectrum has to be evaluated in the free theory at leading order in $\gamma$. 
As explained in \App{subsec:source}, the convolution product between the power spectrum and the environment correlator, which appears in \Eq{eq:source:quadratic}, is a priori dominated by its UV contribution, which is nonetheless removed by adiabatic subtraction.
Since the environment correlator $\tilde{C}_R$ selects out modes that have crossed out the environment correlation length $\lE$, which is much smaller than the Hubble radius, the source function is well approximated by the environment correlator times the integrated power spectrum,
\bea
\label{eq:source:quadratic:simp}
S_2\left(\bm{k},\eta\right)\simeq
\frac{8\gamma}{(2\pi)^{3/2}}
\tilde{C}_{R}\left({ k}\right)
\int_{p_1<aH} \dd   {\bm p}_1P_{vv}\left(p_1\right)\, .
\eea

Before concluding this work, let us try to generalise our approach to higher-order interaction terms. For instance, let  us consider a cubic interaction, $\hat{A}=\hat{v}^3$. In that case, a long but straightforward calculation, similar to the one presented at the beginning of \Sec{subsec:eomquadratic}, gives the same system as in \Eq{eq:vvquadratic:vpquadratic:pvquadratic:ppquadratic}, except that the second line of the last entry now reads
\bea
\label{eq:neq3:Ppp}
&
-\frac{1}{2}\mathrm{Tr}\left\lbrace \int\dd \bm{x} \dd\bm{y} C_R\left(\bm{x}-\bm{y}\right)\left[\left[\hat{p}_{\bm{k}_1}\hat{p}_{\bm{k}_2},\hat{v}^3\left(\bm{x}\right)\right],\hat{v}^3\left(\bm{y}\right)\right] \hat{\rho}_v \right\rbrace
\\ &\qquad \qquad =
\frac{9}{\left(2\pi\right)^{9/2}} \delta\left(\bm{k}_1+\bm{k}_2\right) \int\dd\bm{p}_1\dd\bm{p}_2\dd\bm{p}_3 \tilde{C}_R\left(\bm{p}_1+\bm{p}_2+\bm{k}_1\right) \mathcal{T}_4\left(\bm{p}_1,\bm{p}_2,\bm{p}_3\right)\, .
\eea
In this expression, the trispectrum has been defined according to $\langle \hat{v}_{\bm{p}_1}\hat{v}_{\bm{p}_2}\hat{v}_{\bm{p}_3}  \hat{v}_{\bm{p}_4}\rangle =  \mathcal{T}_4 (\bm{p}_1,\bm{p}_2,\bm{p}_3 ) \delta\left(\bm{p}_1+\bm{p}_2+\bm{p}_3+\bm{p}_4\right) $.  One then obtains the differential equation~(\ref{eq:thirdPvv:source}) for the power spectrum, where the source function is given by $S_3$, which is such that the quantity written in \Eq{eq:neq3:Ppp} equals $S_3(\bm{k}_1) \delta(\bm{k}_1+\bm{k}_2)/2$. At leading order in $\gamma$, it can be evaluated in the free theory where the state is Gaussian and one has $\left\langle \hat{v}_{\bm{p}_1}\hat{v}_{\bm{p}_2}\hat{v}_{\bmk_3}\hat{v}_{\bm{p}_4}  \right\rangle= P_{vv}\left(p_1\right)P_{vv}\left(p_3\right)\delta(\bm{p}_1+\bm{p}_2)\delta(\bm{p}_3+\bm{p}_4) +P_{vv}\left(p_1\right)P_{vv}\left(p_2\right)\delta(\bm{p}_1+\bm{p}_3)\delta(\bm{p}_2+\bm{p}_4)+P_{vv}\left(p_1\right)P_{vv}\left(p_2\right)\delta(\bm{p}_1+\bm{p}_4)\delta(\bm{p}_2+\bm{p}_3)$ according to Wick theorem. This gives rise to
\bea
S_3\left(\bm{k},\eta\right)=\frac{18 \gamma}{\left(2\pi\right)^{9/2}} &
\left[
2 \int\dd\bm{p}_1\dd\bm{p}_2 \tilde{C}_R\left(\bm{k}-\bm{p}_1-\bm{p}_2\right) P_{vv}\left(\bm{p}_1\right) P_{vv}\left(\bm{p}_2\right)
\right. \\  & \left. 
+ \tilde{C}_R\left(\bm{k}\right) \int\dd\bm{p}_1\dd\bm{p}_2 P_{vv}\left(\bm{p}_1\right) P_{vv}\left(\bm{p}_2\right)
\right]\, .
\label{eq:source:cubic}
\eea
As for quadratic interactions, in the limit $\lE\ll H^{-1}$ and keeping the IR component of these integrals only, this reduces to
\bea
\label{eq:source:cubic:simp}
S_3\left(\bm{k},\eta\right)\simeq \frac{54\gamma}{\left(2\pi\right)^{9/2}} \tilde{C}_R\left(k\right) 
\left[\int\dd\bm{p}P_{vv}\left(\bm{p}\right) \right]^2\, .
\eea

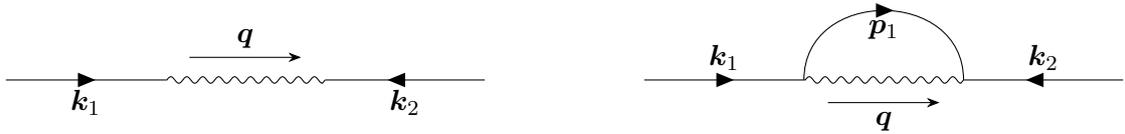
\begin{figure}[t]
\begin{tikzpicture} 
\deflen{mylength}{2.1cm}
\begin{feynman}
\vertex (a) ;
\vertex [right=\mylength of a] (b);
\vertex [right=\mylength of b] (c);
\vertex [right=\mylength of c] (d) ;
\vertex [right=\mylength of d] (e) ;
\vertex [right=\mylength of e] (f) ;
\vertex [right=\mylength of f] (g) ;
\vertex [right=\mylength of g] (h) ;
\vertex [below=2*\mylength of a] (m) ;  
\vertex [right=\mylength of m] (n) ;  
\vertex [right=\mylength of n] (o) ;  
\vertex [right=\mylength of o] (p) ;  
\vertex [below =\mylength of n] (nd) ;  
\vertex [below =\mylength of o] (od) ; 
\vertex [below=2*\mylength of e] (q) ;   
\vertex [right=\mylength of q] (r) ;    
\vertex [right=\mylength of r] (s) ;   
\vertex [right=\mylength of s] (t) ;  
\vertex [below =\mylength of r] (rd) ;  
\vertex [below =\mylength of s] (sd) ; 
\diagram*{
(a) -- [fermion, edge label'=\(\bm{k}_1\)] (b),
(b) -- [photon, momentum=\( \bm{q}\)] (c) ,
(c) -- [anti fermion, edge label'=\(\bm{k}_2\)] (d),
(e) -- [fermion, edge label=\(\bm{k}_1\)] (f),
(f) -- [photon, momentum'=\( \bm{q}\)] (g) ,
(f) -- [fermion, half left, edge label'=\( \bm{p}_1\)] (g) ,
(g) -- [anti fermion, edge label=\(\bm{k}_2\)] (h),
};
\end{feynman}
\end{tikzpicture}
\caption{Feynman diagrams representation of the source function. Straight and wiggly lines stand for propagators of the Mukhanov-Sasaki variable $\hat{v}$, and of the environment operator $\hat{R}$ it couples to, respectively. The first diagram corresponds to linear interactions $\propto \hat{v}\hat{R}$, and the second diagram is for quadratic interactions $\propto \hat{v}^2\hat{R}$.\label{fig:feynman:1}}
\end{figure}
\subsection{Diagrammatic calculation of the source}
\begin{figure}[t]
\begin{tikzpicture} 
\begin{feynman}
\vertex (a1) ;
\diagram*{
(a) -- [fermion, edge label'=\(\bm{k}_1\)] (b),
(b) -- [fermion, half left, looseness=1.5, edge label'=\( \bm{p}_1\)] (c) ,
(b) -- [photon, momentum'=\( \bm{q}\) ] (c) ,
(b) -- [fermion, half right, looseness=1.5, edge label'=\( \bm{p}_2\)] (c) ,
(c) -- [anti fermion, edge label'=\(\bm{k}_2\)] (d),
(m) -- [fermion, edge label'=\(\bm{k}_1\)] (n),
(n) -- [photon, momentum=\( \bm{q}\)] (o) ,
(o) -- [anti fermion, edge label'=\(\bm{k}_2\)] (p),
% home made tadpoles
(n) -- [fermion,half left, looseness=0.8] (nd),
(nd) -- [fermion,half left, looseness=0.8, edge label'=\(\bm{p}_1\)] (n),
(o) -- [fermion,half left, looseness=0.8] (od),
(od) -- [fermion,half left, looseness=0.8, edge label'=\(\bm{p}_2\)] (o),
(e) -- [fermion, edge label'=\(\bm{k}_1\)] (f),
(f) -- [fermion, half left, looseness=2.5, edge label'=\( \bm{p}_1\)] (g) ,
(f) -- [fermion, half left, looseness=1.5, edge label'=\( \bm{p}_2\)] (g) ,
(f) -- [photon, momentum'=\( \bm{q}\) ] (g) ,
(f) -- [fermion, half right, looseness=1.5, edge label'=\( \bm{p}_3\)] (g) ,
(g) -- [anti fermion, edge label'=\(\bm{k}_2\)] (h),
(q) -- [fermion, edge label'=\(\bm{k}_1\)] (r),
(r) -- [photon, momentum=\( \bm{q}\)] (s) ,
(s) -- [anti fermion, edge label'=\(\bm{k}_2\)] (t),
% home made tadpoles
(r) -- [fermion,half left, looseness=0.8] (rd),
(rd) -- [fermion,half left, looseness=0.8, edge label'=\(\bm{p}_1\)] (r),
(r) -- [fermion, half left, looseness=2.5, edge label'=\( \bm{p}_2\)] (s) ,
(s) -- [fermion,half left, looseness=0.8] (sd),
(sd) -- [fermion,half left, looseness=0.8, edge label'=\(\bm{p}_3\)] (s),
};
\end{feynman}
\end{tikzpicture}
\caption{Feynman diagrams representation of the source function, as in \Fig{fig:feynman:1}, for cubic interactions $\propto \hat{v}^3\hat{R}$ (left diagrams), and for quartic interactions $\propto \hat{v}^4\hat{R}$ (right diagrams).\label{fig:feynman:2}}
\end{figure}
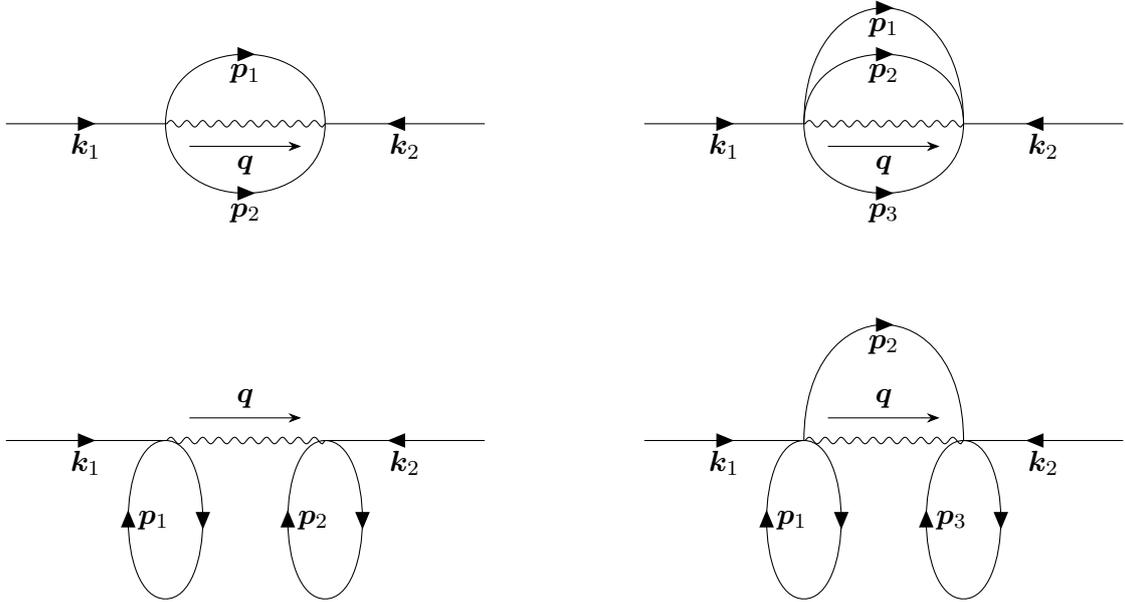

The formulas obtained for the source function for linear interactions in \Eq{eq:source:linear}, for quadratic interactions in \Eq{eq:source:quadratic}, and now for cubic interactions in \Eq{eq:source:cubic}, can be understood with the diagrammatic representation shown in \Figs{fig:feynman:1} and~\ref{fig:feynman:2}. In these Feynman diagrams, straight and wiggly lines represent propagators of the Mukhanov-Sasaki variable $\hat{v}$, and of the environment operator $\hat{R}$ it couples to, respectively. 

The first diagram in \Fig{fig:feynman:1} stands for linear interactions of the form $\hat{v} \hat{R}$. This is why one straight line and one wiggly line are attached to each vertex. Momentum conservation imposes that $\bm{k}_1=-\bm{k}_2=\bm{k}$ and that $\bm{q}=\bm{k}$. This is why the source function is simply proportional to $g^2\tilde{C}_R(\bm{k})\propto \gamma \tilde{C}_R(\bm{k})$, in agreement with \Eq{eq:source:linear}. 

The second diagram in \Fig{fig:feynman:1} stands for quadratic interactions of the form $\hat{v}^2 \hat{R}$, which is why each vertex has two straight lines and one wiggly line. Momentum conservation imposes that $\bm{k}_1=-\bm{k}_2=\bm{k}$ and that $\bm{q}=\bm{k}-\bm{p}_1$. The loop integral then gives rise to $\int \dd\bm{p}_1 P_{vv}(\bm{p}_1) \tilde{C}_R(\bm{k}-\bm{p}_1)$, which indeed corresponds to \Eq{eq:source:quadratic}.

The two left diagrams in \Fig{fig:feynman:2} stand for cubic interactions of the form $\hat{v}^3 \hat{R}$ and correspond to the two ways one can have three straight lines and one wiggly line per vertex, while having a single wiggly line in the diagram. In the left top diagram, momentum conservation imposes that $\bm{k}_1=-\bm{k}_2=\bm{k}$ and that $\bm{q}=\bm{k}-\bm{p}_1-\bm{p}_2$. The loop integral then gives rise to $\int \dd\bm{p}_1 \dd\bm{p}_2 P_{vv}(\bm{p}_1) P_{vv}(\bm{p}_2) \tilde{C}_R(\bm{k}-\bm{p}_1-\bm{p}_2)$, which indeed corresponds to the first term of \Eq{eq:source:cubic}. In the left bottom diagram, momentum conservation imposes that $\bm{k}_1=-\bm{k}_2=\bm{k}$ and that $\bm{q}=\bm{k}$, so the loop integral is given by $\int \dd\bm{p}_1 \dd\bm{p}_2 P_{vv}(\bm{p}_1) P_{vv}(\bm{p}_2) \tilde{C}_R(\bm{k})$, which indeed corresponds to the second term in \Eq{eq:source:cubic}. The multiplicity of the top diagram is $2$ since the lines labeled by $\bm{p}_1$ and $\bm{p}_2$ are indistinguishable, again in agreement with \Eq{eq:source:cubic}.

This diagrammatic representation is useful since it allows one to easily guess the form of the source function if the coupling is of arbitrary order $n$,
\bea
\label{eq:source:n:FeynmanExpanded}
S_n\left(\bm{k},\eta\right)=
\frac{2 n^2}{\left(2\pi\right)^{\frac{3}{2}\left(2n-3\right)}}
 \gamma
\int \prod_{i=1}^{n-1} \dd\bm{p}_i  
P_{vv}\left(\bm{p}_i\right)
\sum_{j=0}^{n-1} a_j
\tilde{C}_R\left(\bm{k}-\sum_{m=1}^j \bm{p}_m\right)\, .
\eea
In this expression, $j$ is the number of system propagators that are involved in the loop with the environment propagator, such that $n-1-j$ is the number of tadpole system propagators, and $a_j$ is the corresponding multiplicity factor. Let us discuss in more detail the above expression using again the case $n=3$, see \Fig{fig:feynman:2} (left diagrams). We have seen that there are two diagrams, one with
no tadpole (top-left diagram) and the other one with two tadpoles (bottom-left diagram). For the
first one, one has $j=2$ and, indeed, one checks that $n-1-j=0$. As already mentioned, the corresponding
multilplicity factor is $a_2=2$. The argument of the correlation
function is given by $\bm{k}-\sum_{m=1}^2 \bm{p}_m=\bm{k}-\bm{p}_1-\bm{p}_2$ in agreement
with \Eq{eq:source:cubic}. For the second diagram, $j=0$ since we see that no
system propagator is involved in a loop with the environment
propagator. As already mentioned, this implies that the number of
tadpoles is $n-1-j=2$. The
multiplicity factor is $a_0=1$. Let us also notice that the
sum appearing in the argument of the correlation function is
such that the upper bound (zero) is less that the lower bound (one).
In that case, this simply means that this sum vanishes by notational convention. There is no diagram with one tadpole (the number of tadpoles
being always even) and, hence, $a_1=0$. The total number
of terms is therefore $\sum_{j=0}^{n-1}=a_0+a_2=3$ which
is indeed the number of terms obtained by expanding a four-point
correlation function by means of the Wick theorem.

This discussion can be repeated for higher-order cases. For instance,
for $n=4$, see \Fig{fig:feynman:2} (right diagrams), one still has two terms, one with no tadpole
and multiplicity $a_3=6$ (right top diagram) and another with two tadpoles and multiplicity
$a_1=9$ (right bottom diagram), the total number of terms being $15$ (using $a_0=a_2=0$),
again in agreement with Wick theorem.

In practice, we do not need to specify the value of $a_j$ for each diagram since each term in the sum over $j$ yields the same contribution. Indeed, as for quadratic interactions, in the limit $\lE\ll H^{-1}$, the IR component of the integrals appearing in \Eq{eq:source:n:FeynmanExpanded} all reduce to the environment correlator evaluated at mode $\bm{k}$ times the integrated power spectrum, and one obtains
\bea
\label{eq:source:n:FeynmanExpanded:simp}
S_n\left(\bm{k},\eta\right)\simeq
\frac{2 n^2 \left(2n-3\right)!!}{\left(2\pi\right)^{\frac{3}{2}\left(2n-3\right)}}
\tilde{C}_R\left(\bm{k}\right) \gamma
\left[\int_{p<aH}\dd\bm{p}P_{vv}\left(\bm{p}\right)\right]^{n-1}\, .
\eea
Here, $(2n-3)!!=\sum_{j=0}^{n-1}a_j$ is, by definition, the number of possibilities to partition a set of $2(n-1)$ elements into $n-1$ pairs and follows from Wick's theorem, and the power spectrum is integrated on super-Hubble scales only. One can check that when $n=1$, \Eqs{eq:source:n:FeynmanExpanded} and~(\ref{eq:source:n:FeynmanExpanded:simp}) reduce to \Eq{eq:source:linear}, when $n=2$, they reduce to \Eqs{eq:source:quadratic} and~(\ref{eq:source:quadratic:simp}) respectively, and when $n=3$, they reduce to \Eqs{eq:source:cubic} and~(\ref{eq:source:cubic:simp}) respectively.

As explained in \Sec{sec:quad:source}, the calculation of the integrated power spectrum cannot be performed generically in the slow-roll approximation since it involves a wide range of modes over which slow-roll expansions may break down. This is why a slow-roll analysis can be carried out only on an inflationary model-by-model basis. Otherwise, neglecting slow-roll corrections, one has
\bea
S_n\left(\bm{k},\eta\right)=
\frac{2 n^2 \left(2n-3\right)!!}{\left(2\pi\right)^{2n-\frac{7}{2}}}
\tilde{C}_R\left(\bm{k}\right)\frac{\gamma}{\eta^{2n-2}}
\ln^{n-1}\left(\frac{\eta_{\mathrm{IR}}}{\eta}\right)\, ,
\eea
where we recall that $\eta_{\mathrm{IR}}$ is an IR cutoff, that \eg corresponds to the time at which inflation started. Making use of \Eqs{eq:gamma} and~(\ref{eq:Ck:appr}), one obtains
\bea
\label{eq:source:generic}
S_n\left(\bm{k},\eta\right)=
\frac{k_\gamma^{4-2n}}{\eta^{2n-2}}
 \left(\frac{\eta_*}{\eta}\right)^{\left(p-3\right)\left(1+\epsilon_{1*}\right)}
\ln^{n-1}\left(\frac{\eta_{\mathrm{IR}}}{\eta}\right)
\Theta\left(\frac{k\lE}{a}\right)\, ,
\eea
where $k_\gamma$ is a comoving scale that characterises the strength of the interaction with the environment and that is defined by 
\bea
\label{eq:kgamma:generic}
k_\gamma^{4-2n}=\frac{4 n^2 \left(2n-3\right)!!}{3\left(2\pi\right)^{2n-3}}\frac{ \bar{C}_R\lE^3\gamma_*}{a_*^3} \, .
\eea
When $n=1$, this is consistent with \Eq{eq:kbreak:def}. When $n=2$, the above expression is singular and the interaction strength is quantified by a dimensionless parameter $\sigma_\gamma$ instead of a comoving scale $k_\gamma$, see \Eq{eq:sigmagamma:def}. 

In \Eq{eq:source:generic}, one can see that the source function varies in time as $S_n\propto \eta^{5-2n-p}$, if one neglects the logarithmic term and slow-roll corrections (which are in fact of the same order). The crucial remark is that if the environment consists of a heavy scalar field, we have shown that $p=7-2n$, see \Eq{eq:gammastar:MassiveEnv}, so $S_n\propto \eta^{-2}$, which does not depend on $n$ anymore, and which is precisely the behaviour that produces a scale-invariant correction to the power spectrum, as we have shown when $n=1$ in \Sec{sec:lin:PowerSpectrumConstraints:MassiveFieldAsEnv} and when $n=2$ in \Sec{sec:quad:PSconstraints}. This implies that the remarkable property found in this work, namely that an environmental heavy test scalar field $\psi$ does not spoil the observed quasi scale invariance of the power spectrum for linear and quadratic interactions, is in fact fully generic and does not depend on the order of the interaction. By linearity, it is true for any coupling of the type $f(\phi) g(\psi)$, as soon as $f$ and $g$ can be Taylor expanded.
\subsection{Power Spectrum}
As shown in \Sec{subsec:psquadratic}, the form~(\ref{eq:exactsolquadratic}) for the power spectrum is valid for any source function. In the case of higher-order interactions, one therefore has
\bea
\label{eq:exactsolquadratic:generic} 
P_{vv} = v_{\bm k}\left(\eta\right)
v^*_{\bm k}\left(\eta\right) +2\int_{-\infty}^\eta S_n\left(\eta'\right)
\mathrm{Im}^2\left[v_{\bm k}\left(\eta'\right)
v^*_{\bm k}\left(\eta\right)\right]
\dd\eta'\, .
\eea
Plugging in the result from the diagrammatic calculation of the source function detailed above, see \Eq{eq:source:generic}, one obtains
\bea
\label{eq:Pvv:modified:generic}
& \kern-1em P_{vv}\left(k\right)  =  v_{\bm k}\left(\eta\right)
v^*_{\bm k}\left(\eta\right) + \frac{\pi^2}{8\sin^2\left(\pi\nu\right)}\frac{-k\eta}{k} \left(\frac{\kgamma}{k_*}\right)^{4-2n} \left(\frac{k}{k_*}\right)^{-1-\alpha_n}\\
& \times \left[
J_{-\nu}^2\left(-k\eta\right) I_{2n-1}\left(\nu\right)
+J_{\nu}^2\left(-k\eta\right) I_{2n-1}\left(-\nu\right)
-2 J_{\nu}\left(-k\eta\right)J_{-\nu}\left(-k\eta\right)I_{2n}\left(\nu\right)
\right]\, ,
\eea
where the integrals $I_{2n-1}(\nu)$ and $I_{2n}(\nu)$ are defined by
\bea
\label{eq:I2n}
I_{2n-1}\left(\nu\right) &= \int_{-k\eta}^{\left(H_*\lE\right)^{-1}}\dd z z^{\alpha_n} \ln^{n-1}\left(\frac{-k\eta_{\mathrm{IR}}}{z}\right)J_{\nu}^2\left(z\right)\\
I_{2n}\left(\nu\right) &= \int_{-k\eta}^{\left(H_*\lE\right)^{-1}}\dd z z^{\alpha_n} \ln^{n-1}\left(\frac{-k\eta_{\mathrm{IR}}}{z}\right)J_{-\nu}\left(z\right)J_{\nu}\left(z\right)
\eea
and $\alpha_n=3-2n-\left(p-3\right)\left(1+\epsilon_{1*}\right)\simeq 6-2n-p$ since for the reasons already explained, we retain the leading contributions in slow roll only. When $n=1$ and $n=2$, one can check that \Eqs{eq:Pvv:modified:generic} and~(\ref{eq:I2n}) match \Eqs{eq:calculJ} and~(\ref{eq:I1:I2}), and \Eqs{eq:exactPvvquadratic} and~(\ref{eq:defI3I4}), respectively. 

If $n=1$ or $n=2$, the integrals $I_{2n-2}$ and $I_{2n-1}$ can be expressed in terms of generalised hypergeometric functions, see \Eqs{eq:I1} and~(\ref{eq:I2}), and \Eqs{eq:primitiveI3} and~(\ref{eq:primitiveI4}), respectively. When $n\geq 3$ this is no longer the case, but these integrals can still be approximated in the following way. Depending on the value of $\alpha_n$, the integrals $I_{2n-2}$ and $I_{2n-1}$ receive their dominant contribution from the neighbourhood of their lower bound $z\sim -k\eta \ll 1$ (case 1 in the language of \Sec{sec:linear:PS:SR}), from the neighbourhood of the intermediate value $z\sim 1$ (case 2 in the language of \Sec{sec:linear:PS:SR}), or from the neighbourhood of their upper bound $z\sim (H_*\lE)^{-1}\gg 1$ (case 3 in the language of \Sec{sec:linear:PS:SR}). In each case, the logarithm term does not vary much over the integration range that provides the main contribution, and so it can be approximated as a constant over that range. This corresponds to a leading-order saddle-point, or steepest-descent, approximation, and in practice, it boils down to taking \Eq{eq:I1:I2:limit} and to multiplying each term by $\ln^{n-1}(-k\eta_{\mathrm{IR}}/z)$, where $z$ needs to be replaced by either $-k\eta$, $1$, or  $(H_*\lE)^{-1}$, depending on where the contribution comes from. This gives rise to
\bea
\label{eq:I2n:appr}
I_{2n-1}(\nu) & \sim
\frac{\left(H_*\lE \right)^{-\alpha_n}}{\pi\alpha_n}
\ln^{n-1} \left[-\frac{k\eta_{_\mathrm{ IR}}}{\left(H_*\lE \right)^{-1}}\right]
\\ & 
+\frac{1}{2\sqrt{\pi}}
\frac{\Gamma(-\alpha_n/2)\Gamma(1/2+\alpha_n/2+\nu)}
{\Gamma(1/2-\alpha_n/2)\Gamma(1/2-\alpha_n/2+\nu)}\ln^{n-1} \left(-k\eta_{_\mathrm{ IR}}\right)
 \\ &
-\frac{2^{-2\nu}}{(1+\alpha_n+2\nu)\Gamma^2(1+\nu)}
(-k\eta)^{1+\alpha_n+2\nu}
\ln^{n-1} \left(\frac{\eta_{_\mathrm{ IR}}}{\eta}\right)\\
I_{2n}(\nu) &\sim
\frac{\left(H_*\lE \right)^{-\alpha_n}}{\pi \alpha_n}
\cos(\pi \nu)
\ln^{n-1} \left[-\frac{k\eta_{_\mathrm{ IR}}}{\left(H_*\lE \right)^{-1}}\right]
 \\ &
+\frac{1}{(1+\alpha_n)\sqrt{\pi}}
\frac{\Gamma(3/2+\alpha_n/2)\Gamma(-\alpha_n/2)}{\Gamma(1/2-\alpha_n/2-\nu)
\Gamma(1/2-\alpha_n/2+\nu)}
\ln^{n-1} \left(-k\eta_{_\mathrm{ IR}}\right)
 \\ &
-\frac{(-k\eta)^{1+\alpha_n}}{(1+\alpha_n)}
\frac{1}{\Gamma(1-\nu)\Gamma(1+\nu)}
\ln^{n-1} \left(\frac{\eta_{_\mathrm{ IR}}}{\eta}\right)
\, .
\eea
Obviously, when $n=1$ this reduces to \Eq{eq:I1:I2:limit}. When $n=2$ however, \Eqs{eq:I3:appr} and~(\ref{eq:I4:appr}) are recovered only at leading order in the logarithms. In principle, one could carry out the saddle-point approximation at higher orders, by Taylor expanding the logarithmic terms, but in practice, \Eq{eq:I2n:appr} already provides reliable estimates. Together with \Eq{eq:Pvv:modified:generic}, it leads to the following relative correction to the power spectrum
\bea
\label{eq:PowerSpectrum:generic:finalAppr}
\left.\Delta\calP_{\bm{k}}\right\vert_{p<6-2n}&\sim \frac{\left(H_*\lE \right)^{p+2n-6}}{2\left(6-2n-p\right)}\left(\frac{\kgamma}{k_*}\right)^{4-2n} \left(\frac{k}{k_*}\right)^{p+2n-7} 
\\ & \times
\left[N_{{}_\mathrm{T}}-\Delta N_*+\ln\left(H_*\lE \frac{k}{k_*}\right)\right]^{n-1}
\\
\left.\Delta\calP_{\bm{k}}\right\vert_{6-2n<p<10-2n}&\sim
\frac{\sqrt{\pi}}{4}\left(\frac{\kgamma}{k_*}\right)^{4-2n} \left(\frac{k}{k_*}\right)^{p+2n-7} 
\frac{\Gamma(n-3+p/2)\Gamma(5-n-p/2)}
{\Gamma(n+p/2-5/2)\Gamma(n+p/2-1)}
\\ & \times
\left[N_{{}_\mathrm{T}}-\Delta N_*+\ln\left(\frac{k}{k_*}\right)\right]^{n-1}
\\
\left.\Delta\calP_{\bm{k}}\right\vert_{p>10-2n}&\sim
\frac{2\ee^{\left(p+2n-10\right)\left(N-N_*\right)}\left(N-N_{\mathrm{IR}}\right)^{n-1}}{\left(p+2n-10\right)\left(p+2n-7\right)\left(p+2n-4\right)}\left(\frac{\kgamma}{k_*}\right)^{4-2n} \left(\frac{k}{k_*}\right)^{3} \, .
\eea
The three cases are similar to the ones discussed in the situation of linear or quadratic interactions. If $p<6-2n$, the power spectrum freezes out on large scales and the amplitude of the correction to the standard result depends on the environment correlation length $\lE$. If $6-2n<p<10-2n$, the power spectrum still freezes out on large scales but its amplitude no longer depends on $\lE$. Since the correction scales as $k^{p+2n-7}$, the specific case $p\simeq 7-2n$ preserves the quasi scale invariance of the power spectrum, and this precisely corresponds to the situation where the environment consists of a heavy test scalar field, as already pointed out. If $p>10-2n$, the power spectrum continues to increase on large scales. Except in the case $p\simeq 7-2n$, the observed quasi scale invariance of primordial cosmological fluctuations thus places an upper bound on the interaction strength with the environment, here parametrised by $k_\gamma/k_*$, that will be given below.
\subsection{Decoherence}
For now let us determine whether decoherence proceeds before the end of inflation or not. In the case of linear and quadratic couplings, we have shown that the decoherence parameter is given in terms of the source function by \Eq{eq:quad:delta:intPvv}. Strictly speaking, we have not shown that this formula generalises to higher orders. One can however easily convince oneself that a calculation similar to the one presented in \Sec{sec:DecoherenceCriterion} yields the same structure for the decoherence parameter at higher values of $n$, namely 
\bea
\label{eq:quad:delta:intPvv:generic}
\delta_{\bm{k}} = \frac{1}{2} \int_{-\infty}^\eta S_n\left(\bm{k},\eta'\right) P_{vv}\left(\bm{k},\eta'\right) \dd \eta'\, .
\eea
The only unknown is the numerical prefactor in \Eq{eq:quad:delta:intPvv:generic}, but since it was found to be the same in the cases $n=1$ and $n=2$, here we assume that it does not depend on $n$ (even if it did, it would not change the observational constraints derived below, since they are based on order-of-magnitude considerations only) and is therefore still given by $1/2$. This leads to
\bea
\label{eq:delta:generic:finalAppr}
\delta_{\bm{k}} \left(N\right) = &  \frac{\pi}{8\sin^2\left(\pi\nu\right)} \left(\frac{\kgamma}{k_*}\right)^{4-2n} \left(\frac{k}{k_*}\right)^{-1-\alpha_n}
%\\ & \times
 \left[I_{2n-1}\left(\nu\right) +I_{2n-1}\left(-\nu\right)-2\cos\left(\pi\nu\right) I_{2n}\left(\nu\right)  \right]\, .
\eea
Making use of the approximations~(\ref{eq:I2n:appr}), one obtains
\bea
\delta_{\bm{k}} \sim &\frac{1}{4} \left(\frac{\kgamma}{k_*}\right)^{4-2n} \left(\frac{k}{k_*}\right)^{p+2n-7} \left\lbrace 
\frac{\left(H_*\lE\right)^{p+2n-6}}{6-2n-p} \left[N_{{}_\mathrm{T}}-\Delta N_*+\ln\left(H_*\lE \frac{k}{k_*}\right)\right]^{n-1}
\right. \\ & \left. \
+\frac{\left(N-N_{\mathrm{IR}}\right)^{n-1}}{p+2n-4}
\left(\frac{k}{k_*}\right)^{4-2n-p}
\ee^{(p+2n-4)\left(N-N_*\right)}
\right\rbrace\, .
\eea
If $p<4-2n$, the first term in the braces dominates at late time, which implies that $\delta_{\bm{k}}$ reaches a constant value that depends explicitly on the environment correlation length $\lE$. If $p>4-2n$, the second term dominates and $\delta_{\bm{k}}$ continues to increases on large scales, leading to more efficient decoherence.
\subsection{Observational constraints}
Requiring that decoherence occurs by the end of inflation at observable scales [$\delta_{\bm{k}_*}(N_\uend)\gg 1$] but that the power spectrum remains unaltered at observable scales [$\Delta \calP_{\bm{k}_*}(N_\uend) \ll 1$] finally leads to the constraint
\bea
\ee^{\left(4-p-2n\right)\Delta N_*} \ll N_{{}_\mathrm{T}}^{n-1}\left(\frac{k_\gamma}{k_*}\right)^{4-2n} \ll
\begin{cases}
\left(H_*\lE\right)^{6-p-2n}\quad&\mathrm{if} \ 4-2n<p<6-2n\, ,\\
1 \quad&\mathrm{if}\ 6-2n<p<10-2n\, ,\\
\ee^{\left(10-p-2n\right)\Delta N_*}\quad&\mathrm{if}\  p>10-2n\, .
\end{cases}
\label{eq:Constraint:generic:finalAppr}
\eea
In particular, one can see that when $p<4-2n$, decoherence can never occur without spoiling the quasi scale invariance of the power spectrum, since from \Eqs{eq:PowerSpectrum:generic:finalAppr} and~(\ref{eq:delta:generic:finalAppr}), one has $\Delta \calP_{\bm{k}} \simeq 2 \delta_{\bm{k}}$ in that case. When $n=1$ and $n=2$, \Eq{eq:Constraint:generic:finalAppr} reduces to \Eq{eq:Constraint:linear:finalAppr} and \Eq{eq:Constraint:quadratic:finalAppr}, respectively. This implies that the structure of \Figs{fig:deltamap} and~\ref{fig:deltamap:quad} generalises to higher-order couplings.
\subsection{Case of a massive scalar field as the environment}
The only exception evading these constraints is the case $p\simeq 7-2n$, for which the correction to the power spectrum is itself quasi scale invariant and only the lower bound on $k_\gamma/k_*$ applies.  More precisely, the quasi scale-invariant correction to the power spectrum coming from the environment may improve or deteriorate how a given model fits the data, but one has to study each model separately, as was done in \Sec{sec:lin:PowerSpectrumConstraints:MassiveFieldAsEnv} for the case of linear interactions, and there is no model-independent conclusion to be drawn. Let us also mention that this requires to incorporate slow-roll corrections into the calculation, which we did for linear interactions, but which otherwise involves the calculation of the source function beyond the de-Sitter limit, that a priori depends on the field dynamics over the entire inflationary period.

As already pointed out, it is remarkable that $p\simeq 7-2n$ precisely corresponds to the model proposed in \App{sec:massivescalarfield} where the environment is made of the degrees of freedom contained in a heavy test scalar field. In this case, combining \Eqs{eq:gammastar:MassiveEnv} and~(\ref{eq:MassivePsi:Cbar:tc:lE:maintext}) with \Eq{eq:kgamma:generic}, one obtains
\bea
\left(\frac{k_\gamma}{k_*}\right)^{4-2n}=&\frac{2^{9-3m} n^2 \left(2n-3\right)!!}{3^{1+2m}\left(2\pi\right)^{2n-3}} 
\left(\frac{37}{7\pi^2}\right)^m
{\frac{\left\lbrace \left(2m-1\right)!!-\sigma\left(m\right)\left[\left(m-1\right)!!\right]^2\right\rbrace^{3}}{\left[m^2\left(2m-3\right)!!\right]^2}}
\\ &
\times \lambda^2
\frac{\mu^{8-2n-2m}H_*^{2n+6m-4}}{M^{4+4m}} \, .
\eea
In terms of the physical parameters of the model, we therefore find that, using \Eq{eq:Constraint:generic:finalAppr}, decoherence occurs if the coupling constant satisfies $\lambda>\lambda_{\mathrm{decoherence}}$, where
\bea
\lambda_{\mathrm{decoherence}} = &
\frac{3^{m+\frac{1}{2}}\left(2\pi\right)^{n-\frac{3}{2}}}{2^{\frac{3}{2}\left(3-m\right)} n \sqrt{\left(2n-3\right)!!}}
\left(\frac{7\pi^2}{37}\right)^{\frac{m}{2}}
\frac{m^2\left(2m-3\right)!!}{\left\lbrace \left(2m-1\right)!!-\sigma\left(m\right)\left[\left(m-1\right)!!\right]^2\right\rbrace^{\frac{3}{2}}}\\
& \times
\frac{\ee^{-\frac{3}{2}\Delta N_*}}{N_{{}_\mathrm{T}}^{\frac{n-1}{2}}} 
\left(\frac{M}{H_*}\right)^{2\left(1+m\right)}
\left(\frac{H_*}{\mu}\right)^{4-n-m}\, .
\eea
For instance, if we take $\Delta N_*=50$, $N_{{}_\mathrm{T}}=10^4$, $M=100\,H_*$ and $\mu=H_*$, we find that $\lambda_{\mathrm{decoherence}}\sim 10^{-25}$, $10^{-20}$ and $10^{-15}$ for $n=1$ and $m=1$, $2$ and $3$ respectively, $\lambda_{\mathrm{decoherence}}\sim 10^{-27}$, $10^{-22}$ and $10^{-17}$ for $n=2$ and $m=1$, $2$ and $3$ respectively, and $\lambda_{\mathrm{decoherence}}\sim 10^{-28}$, $10^{-24}$ and $10^{-19}$ for $n=3$ and $m=1$, $2$ and $3$ respectively. There are however cases where the critical value for the coupling constant $\lambda$ above which decoherence occurs is not small. For instance, with $n=1$ and $m=3$, $\mu$ and $N_{{}_\mathrm{T}}$ are irrelevant and taking $\Delta N_*=50$, one finds that $\lambda_{\mathrm{decoherence}}>1$ as soon as $M/H_*>7030$. This is because if the environmental scalar field is too heavy, its condensate is too suppressed to yield efficient decoherence of the system. One concludes that parametrically small values of the interaction strength $\lambda$ are in general enough to lead to successful decoherence of primordial cosmological perturbations, but only if the heavy test scalar field they couple to has a mass no more than a few orders of magnitude larger than the Hubble scale. Furthermore, if $\lambda>\lambda_{\calP_\zeta}$, where 
\bea
\lambda_{\calP_\zeta} =
\ee^{\frac{3}{2}\Delta N_*}
  \lambda_{\mathrm{decoherence}} \, ,
\eea
the power spectrum is modified, but in a quasi scale-invariant (though model-dependent) way. With $\Delta N_*\simeq 50$, one has $\lambda_{\calP_\zeta} \simeq 10^{32}\lambda_{\mathrm{decoherence}}$, so there always is a wide range of value for $\lambda$ such that decoherence occurs while leaving the power spectrum unchanged.
\section{Conclusion}
\label{sec:conclusions}
Let us now recap our main results and discuss possible extensions. In the early Universe, cosmological density perturbations are amplified from vacuum quantum fluctuations, and subsequently seed the formation of all structures in our Universe. This mechanism occurs in the presence of all degrees of freedom present in the standard model and beyond, to which cosmological fluctuations couple (at least gravitationally). They should therefore be described as an open quantum system (as opposed to an isolated one, as usually done), the evolution of which can be modelled with a Lindblad equation, under some conditions that we have clarified. 

This modified evolution leads to corrections to observable predictions such as the power spectrum of curvature fluctuations. Measurements of the CMB temperature and polarisation anisotropies~\cite{Ade:2015tva} therefore constrain the properties of possible environments and place upper bounds on the interaction strengths, that we have derived. On the other hand, the Lindblad evolution also leads to decoherence of the system in the eigenbasis selected by the form of the interaction with the environment. Since quantum decoherence is thought to play a role in the quantum-to-classical transition of cosmological fluctuations, one may also require that decoherence has completed by the end of inflation, which places lower bounds on the interaction strength. We have then identified the viable scenarios where decoherence occurs without spoiling the quasi scale invariance of the power spectrum, see \Figs{fig:deltamap} and~\ref{fig:deltamap:quad} .

In practice, we have considered local interactions of the form $\hat{H}_{\mathrm{int}}\propto \int \dd\bm{x} \hat{v}^n(\bm{x},t)\otimes \hat{R}(\bm{x},t)$, where $\hat{v}$ is the Mukhanov-Sasaki variable that describes scalar cosmological fluctuations and $\hat{R}$ is the operator in the environment sector to which $\hat{v}$ couples. In the case of linear interactions, $n=1$, we have shown that the Lindblad equation can be solved completely, see \Eqs{eq:finalrhomaintext}-(\ref{eq:defK}). In this case, the state remains Gaussian, and the power spectrum and the decoherence parameter can be calculated from the (modified) density matrix directly. For higher-order coupling, $n\geq 2$, this is no longer the case, but we have shown how the power spectrum and the decoherence parameter can still be calculated exactly, see \Eq{eq:exactsolquadratic:generic} and \Eq{eq:quad:delta:intPvv:generic} respectively, in terms of a source functions that involves the correlator of $\hat{R}$ in the environment sector and correlators of $\hat{v}$ in the (free limit of the) system, see \Eq{eq:source:n:FeynmanExpanded}. Since \Eq{eq:exactsolquadratic:generic} and \Eq{eq:quad:delta:intPvv:generic} linearly depend on the source function, this means that we have entirely solved the problem, for any interaction of the form $\hat{H}_{\mathrm{int}}\propto \int \dd\bm{x} f(\hat{v})(\bm{x},t)\otimes \hat{R}(\bm{x},t)$ as long as $f$ can be Taylor expanded.

As an illustration, we have discussed the situation where the environment is made of a heavy test scalar field $\psi$, see \App{sec:massivescalarfield}, and $\hat{R}\propto \hat{\psi}^m$. In that case, the  time dependence of the effective interaction strength scales as $a^{7-2n}$, where $a$ is the scale factor. We have shown that this precisely corresponds to the very peculiar configuration where the correction to the power spectrum is quasi scale invariant. The observational constraints on the parameters describing the environment (here the mass of the heavy test scalar field in Hubble units) and the interaction strength are therefore less straightforward, and, as was shown around \Fig{fig:nsr} in the case of linear interactions $n=1$, are in fact model dependent. Indeed, there are models for which the correction to the power spectrum coming from the environment is too small to be resolved by current CMB measurements and no constraint can be derived, there are models for which the correction improves the fit to the data and lower bounds on the interaction strength can be obtained, and there are models for which the correction deteriorates the fit to the data and upper bounds on the interaction strength can be derived. For higher-order interactions, $n\geq 2$, such an analysis requires to calculate the source function beyond the de-Sitter limit, which again has to be done on a model-by-model basis, and which depends on the inflaton field dynamics over the entire inflationary period. In principle, this may allow one to probe the inflationary potential beyond the last $\sim 50$ \efolds~of inflation and might extend the range of scales one can access in the early Universe beyond the observable horizon~\cite{Hardwick:2017qcw, Torrado:2017qtr}. We leave such an analysis for future work, and we now mention a few other possible prospects. 

Let us first emphasise that our results assume that the power spectrum remains unaffected during preheating and reheating, as it is the case on large scales in the standard approach. In the presence of interactions with extra degrees of freedom, this should however be verified.
Second, in the case of non-linear interactions, $n\geq 2$, we have shown that the quantum state of cosmological fluctuations, modified by its interaction with the environment, is no longer Gaussian. Since tight upper bounds on the amount of primordial non-Gaussianities have been placed from recent CMB measurements~\cite{Ade:2015ava}, this constitutes another channel through which the environment properties and its interaction strength with cosmological fluctuations can be constrained~\cite{Martin:2018lin}. 
Third, in the case where the effective interaction strength with the environment evolves in time as $a^p$ with $p>7-2n$, the power spectrum is modified on small scales, with a blue tilt [comprised between $\nS=1$ and $\nS=4$ depending on the value of $p$, see \Eq{eq:PowerSpectrum:generic:finalAppr}]. This implies that the amplitude of the power spectrum may reach sizeable values for scales that are smaller than the ones probed in the CMB but that are still of astrophysical interest. Observational bounds on the amount of primordial black holes therefore constitute yet another channel to constraint the environment and its interaction strength with the system. Conversely, this also suggests that quantum decoherence might be a promising candidate to produce such primordial black holes, if their role in the dark matter abundance of our Universe or in providing progenitors to the LIGO black-hole merging events is confirmed (see \eg \Refa{Clesse:2017bsw}). 

Let us also mention that decoherence, per se, does not solve the quantum measurement problem~\cite{Adler:2001us, Schlosshauer:2003zy},\footnote{However, if decoherence is considered together with a non-standard interpretation of quantum mechanics (different from the Copenhaguen one), such as the many-world interpretation, then a solution to the measurement problem can be obtained.} but that there are alternatives to the standard formulation of quantum mechanics that do so, \eg dynamical collapse models  of the wavefunction such as the CSL proposal~\cite{Ghirardi:1985mt, Pearle:1988uh, Ghirardi:1989cn, Bassi:2003gd}. In this model, a non-linear and stochastic correction is added to the Schr\"odinger linear equation, that collapses to wavefunction towards one of the eigenstates of the operator appearing in the extra term. Interestingly, the averaged density matrix (where ``averaged'' here refers to the stochasticity of the theory) precisely satisfies a Lindblad equation. Therefore, the present work also allows one to compute observable corrections in the CSL theory, in a way that is complementary\footnote{In \Refa{Martin:2012pea} it is shown how to compute $\langle \left(\hat{v}   - \langle \hat{v} \rangle \right)^2 \rangle $ (which turns out to be not stochastic), and here we have shown how to calculate $\mathbb{E}(\langle \hat{v}^2 \rangle)$, where $\mathbb{E}$ denotes stochastic average. Combining the two results would allow one to calculate $\mathbb{E}(\langle \hat{v} \rangle^2)$ for instance, which corresponds to the power spectrum when the CSL theory is interpreted as in \eg \Refa{Canate:2012ua}.} to \Refa{Martin:2012pea}. 

Finally, other effective approaches~\cite{Achucarro:2012sm, Gao:2012uq, Noumi:2012vr} have been proposed to incorporate heavy scalar fields and their effects on cosmological fluctuations in the early Universe, and it would interesting to compare them with the one used in this work. Let us however stress that our approach is not limited to the case where the environment consists of heavy test scalar fields but is fully generic, and does not assume anything about the environment apart from the validity conditions for the Lindblad equation. It therefore allows one to carry out a model-independent analysis of environmental influence and quantum decoherence in the early Universe. 
\begin{acknowledgments}
V.V. acknowledges funding from the European Union's Horizon 2020 research and innovation programme under the Marie Sk\l odowska-Curie grant agreement N${}^0$ 750491. We thank Jessie de Kruijf and Nicola Bartolo for pointing out a typo in Appendix D and for very interesting discussions. This typo was corrected after publication.
\end{acknowledgments}
\appendix
\section{Deriving the Lindblad equation}
\label{sec:DerivingLindblad}
In this section, we provide a detailed and generic derivation of the
Lindblad equation. Following the usual text book discussions, see \eg
\Refs{CohenTannoudji:1992, LeBellac:2006, Burgess:2006jn, Pearle:2012,
  Brasil:2012}, we pay special attention to the physical
assumptions the Lindblad formalism relies on, and show how they
concretely enter into the calculation.

Let a system ``S'' be in interaction with some environment ``E''. The
Hilbert space ${\cal E}$ of the full system can be written as the
tensorial product of the Hilbert space of the system,
${\cal E}_\mathrm{S}$, with the Hilbert space of the environment,
${\cal E}_\mathrm{E}$, namely
${\cal E}={\cal E}_\mathrm{S}\otimes {\cal E}_\mathrm{E}$. Then, the
corresponding Hamiltonian reads\footnote{In this section, in order to
 avoid cumbersome expressions, operators dot not carry hats.}
\bea
H=H_0+H_\mathrm{int}=H_{\mathrm S}\otimes \setI_\mathrm{E}+\setI_\mathrm{S}\otimes H_\mathrm{E}+ g H_\mathrm{int}\, .
\eea
Here, $H_\mathrm{S}$ is the Hamiltonian of the system and acts in
${\cal E}_\mathrm{S}$, while $\setI_\mathrm{E}$ is the identity operator acting in
${\cal E}_\mathrm{E}$. In the same manner, $H_\mathrm{E}$ is the Hamiltonian
of the environment and acts in ${\cal E}_\mathrm{E}$, while
$\setI_\mathrm{S}$ is the identity operator acting in ${\cal E}_\mathrm{S}$. They
represent the free Hamiltonian $H_0$ acting in the full space
${\cal E}$, while $H_\mathrm{int}$ is an interaction term. It carries a
(supposedly small) dimensionless coupling parameter $g$ characterising
the strength of the interactions between the system and the
environment.

The density matrix $\rho$ of the full system (acting in the
Hilbert space $\cal E$) obeys the unitary Liouville-von Neumann
equation
\bea
i\frac{\dd  \rho}{\dd  t}=\left[H,\rho\left(t\right)\right].
\eea
In what follows, it will be convenient to factor out the time
dependence of $\rho$ due to the free Hamiltonian $H_0$ by going
to the interaction picture. This is why we introduce
\bea
\tilde{\rho}\left(t\right) &\equiv & U^{\dagger}\left(t\right)\rho\left(t\right)U\left(t\right), \\
\tilde{H}_\mathrm{int}\left(t\right) &\equiv & U^{\dagger}\left(t\right)H_\mathrm{int}U\left(t\right),
\eea
where
$U(t)\equiv {\mathrm e}^{-i\int_0^t H_0(t^\prime) \dd 
  t^\prime}$
is the (unitary) free evolution operator. By definition, it satisfies
\bea
i\frac{\dd  U}{\dd  t}=H_0\left(t\right)U\left(t\right)\, .
\eea
As a consequence, the evolution equation of $\tilde{\rho}(t)$ reads
\bea
\frac{\dd  \tilde{\rho}}{\dd  t} = 
-ig\left[\tilde{H}_\mathrm{int},\tilde{\rho}\left(t\right)\right]\, ,
\label{eq:Lindblad:drhotildadt}
\eea
which can be formally integrated as
\bea
\label{eq:formalsol}
\tilde{\rho}\left(t+\Delta t\right)=\tilde{\rho}\left(t\right)
-ig\int _{t}^{t+\Delta t}{\dd }t'
\left[\tilde{H}_\mathrm{int}\left(t^\prime\right),\tilde{\rho}
\left(t^\prime\right)\right]\, .
\eea
This expression gives rise to a Born expansion in $g$ of the
solution of \Eq{eq:Lindblad:drhotildadt}. Indeed, one can iteratively expand the
integrand of \Eq{eq:formalsol} to obtain
\bea
\label{eq:drhodt}
\tilde{\rho}\left(t+\Delta t\right)-\tilde{\rho}\left(t\right)
&=-ig\int _{t}^{t+\Delta t}{\dd }t'
\left[\tilde{H}_\mathrm{int}\left(t'\right),\tilde{\rho}\left(t\right)\right]
 \\ &
-g^2\int_t^{t+\Delta t}{\dd }t'
\int _t^{t'}{\dd }t''
\left[\tilde{H}_\mathrm{int}\left(t'\right),\left[\tilde{H}_\mathrm{int}\left(t''\right),\tilde{\rho}\left(t\right)
\right]\right]+{\cal O}\left(g^3\right)\, .
\eea
This expression is an explicit solution of \Eq{eq:Lindblad:drhotildadt} at order $g^2$. In the second term, the density matrix is evaluated at the time $t$, but evaluating it at any other time comprised between $t$ and $t+\Delta t$ only gives a correction of order $g^3$. This is why, for future convenience, we chose to evaluate it at the time $t''$ instead and to work with the expression
\bea
\label{eq:drhodt}
\tilde{\rho}\left(t+\Delta t\right)-\tilde{\rho}\left(t\right)
&=-ig\int _{t}^{t+\Delta t}{\dd }t'
\left[\tilde{H}_\mathrm{int}\left(t'\right),\tilde{\rho}\left(t\right)\right]
 \\ &
-g^2\int_t^{t+\Delta t}{\dd }t'
\int _t^{t'}{\dd }t''
\left[\tilde{H}_\mathrm{int}\left(t'\right),\left[\tilde{H}_\mathrm{int}\left(t''\right),\tilde{\rho}\left(t''\right)
\right]\right]+{\cal O}\left(g^3\right)\, .
\eea
From now on, for display convenience, we will drop the ${\cal O}(g^3)$ and remember that the calculation is performed at order $g^2$.
 
Let us now restrict the analysis to the reduced density matrix of the system,
$\tilde{\rho}_\mathrm{S}$. It is obtained from the full density matrix by
tracing out the environment degrees of freedom, \ie
\bea
\label{eq:Lindblad:rhoS:def}
\tilde{\rho}_\mathrm{S}\left(t\right)=\mathrm{Tr}_\mathrm{E}
\left[\tilde{\rho}\left(t\right)\right]\, .
\eea
Let us recall that $\tilde{\rho}$ is an operator acting in ${\cal
  E}_\mathrm{S}\otimes {\cal E}_\mathrm{E}$ and, therefore,
$\tilde{\rho}_\mathrm{S}$ is an operator acting in ${\cal E}_\mathrm{S}$ only. From \Eq{eq:drhodt}, it obeys the equation
\bea
\label{eq:deltarho}
\tilde{\rho}_{\mathrm S}\left(t+\Delta t\right)-\tilde{\rho}_{\mathrm S}\left(t\right)
&=-ig\int _{t}^{t+\Delta t}{\dd }t'
\mathrm{Tr}_\mathrm{E}\left[\tilde{H}_\mathrm{int}\left(t'\right),\tilde{\rho}\left(t\right)\right]
\\ &  - g^2
\int_t^{t+\Delta t}{\dd }t'
\int _t^{t'}{\dd }t''
\mathrm{Tr}_\mathrm{E}
\left[\tilde{H}_\mathrm{int}\left(t'\right),\left[\tilde{H}_\mathrm{int}\left(t''\right),\tilde{\rho}\left(t''\right)
\right]\right]\, .
\eea
Similarly to \Eq{eq:Lindblad:rhoS:def}, we can define the reduced
density operator of the environment, acting in ${\cal E}_\mathrm{E}$, by
$\tilde{\rho}_\mathrm{E}(t)\equiv \mathrm{Tr}_\mathrm{S}
\left[\tilde{\rho}\left(t\right)\right]$.
It is important to stress that, in general,
$\tilde{\rho}(t)\neq \mathrm{Tr}_\mathrm{E}\left[\tilde{\rho}(t)\right] \otimes \mathrm{Tr}_\mathrm{S}[\tilde{\rho}(t)]$, namely 
$\tilde{\rho}(t)\neq \tilde{\rho}_\mathrm{S}(t)\otimes \tilde{\rho}_\mathrm{E}(t)$, and one has instead 
\bea
\label{eq:rhocorr:def}
\tilde{\rho}\left(t\right) =\tilde{\rho}_\mathrm{S}\left(t\right)\otimes \tilde{\rho}_\mathrm{E}\left(t\right) +g^p\tilde{\rho}_\mathrm{correl}\left(t\right)\, .  
\eea 
This relation defines the quantity $\tilde{\rho}_\mathrm{correl}$, which
describes the correlations between the system and the environment at
time $t$. It satisfies\footnote{Indeed, one has
  $\mathrm{Tr}_\mathrm{E}(\tilde{\rho})\equiv\tilde{\rho}_\mathrm{S}
  =\tilde{\rho}_\mathrm{S}\mathrm{Tr}_\mathrm{E}(\tilde{\rho}_\mathrm{E}) +g^p \mathrm{Tr}_\mathrm{E}(\tilde{\rho}_\mathrm{correl}) =\tilde{\rho}_\mathrm{S}\mathrm{Tr}_\mathrm{E} \mathrm{Tr}_\mathrm{S}(\tilde{\rho}) +g^p \mathrm{Tr}_\mathrm{E}(\tilde{\rho}_\mathrm{correl}) =\tilde{\rho}_\mathrm{S} +g^p \mathrm{Tr}_\mathrm{E}(\tilde{\rho}_\mathrm{correl})$,
  which implies that
  $\mathrm{Tr}_\mathrm{E}(\tilde{\rho}_\mathrm{correl})=0$, where we have used that $\mathrm{Tr}_\mathrm{E} \mathrm{Tr}_\mathrm{S}(\tilde{\rho}) = \mathrm{Tr}(\tilde{\rho})=1$. The formula
  $\mathrm{Tr}_\mathrm{S}(\tilde{\rho}_\mathrm{correl})=0$ can be shown
  in the same manner.}
$\mathrm{Tr}_\mathrm{E}(\tilde{\rho}_\mathrm{correl})=0$ and
$\mathrm{Tr}_\mathrm{S}(\tilde{\rho}_\mathrm{correl})=0$. It is clear
that, if we start from a situation where the density operator is
factorised and $\tilde{\rho}_\mathrm{correl}=0$, correlations can only
appear if the interaction term is switched on. Hence, the term
$\tilde{\rho}_\mathrm{correl}$ must carry some $g$-charge, which is
explicitly displayed in \Eq{eq:rhocorr:def} as $g^p$, $p$ being an
unknown natural integer. Plugging \Eq{eq:rhocorr:def} into
\Eq{eq:deltarho}, one then obtains four terms,
\bea
\label{eq:deltarhoexpand}
\tilde{\rho}_\mathrm{S}\left(t+\Delta t\right) -\tilde{\rho}_\mathrm{S}\left(t\right)
& =-ig\int _{t}^{t+\Delta t}{\dd }t' \mathrm{Tr}_\mathrm{E}\left[\tilde{H}_\mathrm{int}\left(t'\right),\tilde{\rho}_\mathrm{S}\left(t\right)\otimes 
\tilde{\rho}_\mathrm{E}\left(t\right)\right]
\\  &
-ig^{p+1}\int _{t}^{t+\Delta t}{\dd }t' \mathrm{Tr}_\mathrm{E}\left[\tilde{H}_\mathrm{int}\left(t'\right),\tilde{\rho}_\mathrm{correl}\left(t\right)\right]
 \\   &
-g^2 \int_t^{t+\Delta t}{\dd }t' \int
_t^{t'}{\dd }t'' \mathrm{Tr}_\mathrm{E} \left[\tilde{H}_\mathrm{int}\left(t'\right),\left[\tilde{H}_\mathrm{int}\left(t''\right),
\tilde{\rho}_\mathrm{S}\left(t''\right)\otimes\tilde{\rho}_\mathrm{E}\left(t''\right)
    \right]\right]
\\   &
-g^{p+2} \int_t^{t+\Delta t}{\dd }t' \int
_t^{t'}{\dd }t'' \mathrm{Tr}_\mathrm{E} \left[\tilde{H}_\mathrm{int}\left(t'\right),\left[\tilde{H}_\mathrm{int}\left(t''\right),
\tilde{\rho}_\mathrm{correl}\left(t''\right)
    \right]\right]\, .
\eea
In order to determine which of these terms dominate, we now need to
specify the interaction Hamiltonian $H_\mathrm{int}$. Let us first assume that
it can be written as
\bea
\label{eq:Lindblad:Hint:decomp}
H_\mathrm{int}\left(t\right)=A\left(t\right)\otimes R\left(t\right)\, ,
\eea
where $A$ acts in ${\cal E}_\mathrm{S}$ and $R$ acts in
${\cal E}_\mathrm{E}$. More generic interaction Hamiltonians will be considered below, see \Eq{eq:Hint:general}. The evolution operator $U$ can be factorised
as $U_{\mathrm S}\otimes U_{\mathrm E}$ because it describes the time evolution in
the case where $H_{\mathrm int}=0$, that is to say when the system and the
environment evolve independently. As a consequence, 
\bea
\tilde{H}_\mathrm{int}\left(t\right)=\left(U^{\dagger}_\mathrm{S}\otimes
  U^{\dagger}_\mathrm{E}\right) \left(A\otimes R\right)\left(U_\mathrm{S}\otimes U_\mathrm{E} \right) =\left(U^{\dagger}_\mathrm{S}AU_\mathrm{S}\right) \otimes \left(U^{\dagger}_\mathrm{E}RU_\mathrm{E}\right)
\equiv \tilde{A}\left(t\right)\otimes \tilde{R}\left(t\right)\, .  
\eea
Let us now evaluate the first term of
\Eq{eq:deltarhoexpand}. Since\footnote{This can be shown as follows. Let us, without loss of generality, write $\tilde{A}$
  and $\tilde{R}$ in terms of their spectra,
\bea
\tilde{A}=\sum_{ij}a_{ij}\vert i\rangle \langle j\vert, \quad 
\tilde{R}=\sum_{\alpha \beta }r_{\alpha \beta}\vert \alpha 
\rangle \langle \beta \vert\, ,
\eea
where latin letters label the eigenvectors of $\tilde{A}$ and greek letters label the eigenvectors of $\tilde{R}$. One then has
\bea
\tilde{A}\otimes \tilde{R}=\sum_{ij}\sum_{\alpha \beta}
a_{ij}r_{\alpha\beta}\vert i\rangle \langle j\vert 
\otimes \vert \alpha \rangle \langle \beta \vert\, .
\eea
Using the definition of the partial trace, $\mathrm{Tr}_\mathrm{E}(\tilde{A}\otimes \tilde{R})=\sum _{\gamma }\langle \gamma
\vert\tilde{A}\otimes \tilde{R} \vert \gamma \rangle$, one obtains
\bea
\mathrm{Tr}_\mathrm{E}\left(\tilde{A}\otimes \tilde{R}\right)
=\sum _{ij}a_{ij}\vert i \rangle \langle j\vert 
\sum _{\gamma}r_{\gamma \gamma},
\eea
namely
$\mathrm{Tr}_\mathrm{E}(\tilde{A}\otimes \tilde{R})=\tilde{A} \mathrm{Tr}_\mathrm{E}(\tilde{R})$.}
$\mathrm{Tr}_\mathrm{E}(\tilde{A}\otimes \tilde{R})=\tilde{A} \mathrm{Tr}_\mathrm{E}(\tilde{R})$, one has
\bea
\mathrm{Tr}_\mathrm{E}\left[\tilde{H}_\mathrm{int}\left(t'\right),\tilde{\rho}_\mathrm{S}\left(t\right)\otimes \tilde{\rho}_\mathrm{E}\left(t\right)\right] 
&= \tilde{A}\left(t'\right)\tilde{\rho}_\mathrm{S}\left(t\right)\otimes \mathrm{Tr}_\mathrm{E} \left[\tilde{R}\left(t'\right)\tilde{\rho}_\mathrm{E}\left(t\right)\right]
\\ &
-\tilde{\rho}_\mathrm{S}\left(t\right)\tilde{A}\left(t'\right)\otimes \mathrm{Tr}_\mathrm{E}
\left[\tilde{\rho}_\mathrm{E}\left(t\right) \tilde{R}\left(t'\right)\right]
\\ &=
\mathrm{Tr}_\mathrm{E} \left[\tilde{R}\left(t'\right)\tilde{\rho}_\mathrm{E}\left(t\right)\right]
\left[\tilde{A}\left(t'\right),\tilde{\rho}_\mathrm{S}\left(t\right)\right] \, .
\label{eq:Lindblad:firstterm}
\eea
In order to proceed further, we need to formulate a few
approximations. Since the system is supposed to be ``small'' compared to
the environment, let us first assume that its influence on the
evolution of the environment is negligible. Under this condition,
$\tilde{\rho}_{\mathrm E}(t)\simeq \tilde{\rho}_{\mathrm E}(0)\equiv
\tilde{\rho}_{\mathrm E}$
is constant in time in the interaction picture.  Notice that this does
not mean that $\rho_{\mathrm E}$ does not depend on time. A second
approximation consists in assuming that the environment is in a
stationary state, namely that the environment Hamiltonian $H_{\mathrm E}$
is not explicitly time dependent, and that
$\left[\tilde{\rho}_{\mathrm E},H_{\mathrm E}\right]=0$. The fact that the
environment Hamiltonian is not explicitly time dependent implies that
its evolution operator can be written as $U_{\mathrm E}={\mathrm e}^{-iH_{\mathrm E}t}$. The
condition $\left[\tilde{\rho}_{\mathrm E},H_{\mathrm E}\right]=0$ can thus be
written as $\left[\tilde{\rho}_{\mathrm E},U_{\mathrm E}\right]=0$. This also
means that
$\rho_\mathrm{E}(t)=\ee^{-iH_\mathrm{E}t}\tilde{\rho}_\mathrm{E}\ee^{i
  H_\mathrm{E}t}$
itself is time independent and, in fact,
$\rho_{\mathrm E}=\tilde{\rho}_{\mathrm E}$. As a consequence,
$[\rho_{\mathrm E},H_{\mathrm E}]=0$, and the environment density operator
can be written as 
\bea
\label{eq:rhoE}
\tilde{\rho}_{\mathrm E}=\sum _n p_n\vert n \rangle \langle n\vert\, ,
\eea
where $\vert n\rangle $ are eigenvectors of $H_{\mathrm E}$ with
eigenvalue $E_n $, \ie $H_{\mathrm E}\vert n\rangle =E_n\vert n\rangle$,
and $p_n$ are constant real coefficients. Finally, a third assumption
is that the mean value of $R(t)$ vanishes, namely\footnote{In
  practice, this condition can be achieved by redefining the
  Hamiltonian of the system and the interaction Hamiltonian according
  to
  $ H_\mathrm{ S} \rightarrow H_\mathrm{ S}+A\mathrm{Tr}_\mathrm{ E}
  \left(\tilde{\rho}_\mathrm{ E}R\right)$
  and
  $H_\mathrm{int} \rightarrow A\otimes R -A\mathrm{ Tr}_\mathrm{ E}
  \left(\tilde{\rho}_\mathrm{ E}R\right)\otimes \setI_\mathrm{ E}$,
  which leaves the total Hamiltonian unchanged but ensures that
  \Eq{eq:Lindbald:meanRvanishes} is satisfied. In \Sec{sec:massivescalarfield}, a concrete example is considered and the procedure discussed here 
is carried out in \Eq{eq:action}.}  
\bea
\label{eq:Lindbald:meanRvanishes}
\langle R\rangle=\mathrm{ Tr}_\mathrm{ E}\left(R \tilde{\rho}_\mathrm{ E}\right)=0\, .
\eea
Here, the trace is taken in ${\cal E}_\mathrm{ E}$ since
$\tilde{\rho}_\mathrm{ E}$ and $R$ act in ${\cal E}_\mathrm{ E}$. Notice that,
using the cyclic property of the trace and the fact that the density
operator $\tilde{\rho}_\mathrm{ E}$ commutes with $U_\mathrm{ E}$, this also
means that
\bea
\mathrm{ Tr}_\mathrm{ E}\left(\tilde{R}\tilde{\rho}_\mathrm{ E}\right)&=
\mathrm{ Tr}_\mathrm{ E}\left(U_\mathrm{ E}^{\dagger}R U_\mathrm{ E}\tilde{\rho}_\mathrm{ E}\right)=
\mathrm{ Tr}_\mathrm{ E}\left(U_\mathrm{ E}^{\dagger}R \tilde{\rho}_\mathrm{ E} U_\mathrm{ E}\right)=
\mathrm{ Tr}_\mathrm{ E}\left( U_\mathrm{ E} U_\mathrm{ E}^{\dagger}R \tilde{\rho}_\mathrm{ E} \right)
 \\ &
=
\mathrm{ Tr}_\mathrm{ E}\left( R \tilde{\rho}_\mathrm{ E} \right)=0\, .
\eea
This implies that the right-hand side of \Eq{eq:Lindblad:firstterm},
hence the first term of \Eq{eq:deltarhoexpand}, vanishes.

This now allows us to determine the value of the positive integer
$p$. Indeed, at leading order in $g$, the left-hand side of
\Eq{eq:deltarhoexpand},
$\tilde{\rho}_\mathrm{ S}(t+\Delta t) -\tilde{\rho}_\mathrm{ S}(t)$,
is directly proportional to $g^p \tilde{\rho}_\mathrm{correl}$, since,
in the absence of the interaction term, $\tilde{\rho}_\mathrm{S}$ does
not evolve in the interaction picture. The left-hand side of
\Eq{eq:deltarhoexpand} is therefore of order $p$, while the right-hand
side contains terms of order $p+1$, $2$ and $p+2$. The only
possibility that allows us to identify the dominant terms in both
sides is that $p=2$. As a consequence, the terms of order $p+1$ and
$p+2$ in the right hand side of \Eq{eq:deltarhoexpand} are
sub-dominant and \bea
\label{eq:deltarhoexpand:oneterm}
\tilde{\rho}_\mathrm{ S}\left(t+\Delta t\right) -\tilde{\rho}_\mathrm{ S}\left(t\right) 
=  -g^2 \int_t^{t+\Delta t}\dd t' \int_t^{t'}\dd t'' 
\mathrm{ Tr}_\mathrm{ E} \left[\tilde{H}_\mathrm{ int}\left(t'\right),
\left[\tilde{H}_\mathrm{ int}\left(t''\right), \tilde{\rho}_\mathrm{ S}\left(t''\right)
\otimes\tilde{\rho}_\mathrm{ E}\right]\right]\, .
\eea
This expression is valid at leading order in $g$, which is why we need
to make a fourth assumption, namely that the interactions modify the
dynamics of the system in the perturbative regime only.

Let us now evaluate the remaining term of \Eq{eq:deltarhoexpand}, that
is to say, the right hand side of
\Eq{eq:deltarhoexpand:oneterm}. Plugging in
\Eq{eq:Lindblad:Hint:decomp} and expanding the double commutator, one
has
\bea
\label{eq:doublec}
\mathrm{ Tr}_\mathrm{ E}& \left[\tilde{H}_\mathrm{
    int}\left(t'\right),\left[\tilde{H}_\mathrm{ int}\left(t''\right),
\tilde{\rho}_\mathrm{ S}\left(t''\right)\otimes\tilde{\rho}_\mathrm{ E}
    \right]\right]
 \\ & =
 \tilde{A}\left(t'\right)\tilde{A}\left(t''\right)\tilde{\rho}_\mathrm{ S}\left(t''\right)
\mathrm{ Tr}_\mathrm{ E}\left[\tilde{R}\left(t'\right)\tilde{R}\left(t''\right)
\tilde{\rho}_\mathrm{ E}\right]
-\tilde{A}\left(t''\right)\tilde{\rho}_\mathrm{ S}\left(t''\right)\tilde{A}\left(t'\right)
\mathrm{ Tr}_\mathrm{ E}\left[\tilde{R}\left(t''\right)
\tilde{\rho}_\mathrm{ E}\tilde{R}\left(t'\right)\right]
 \\ &-
\tilde{A}\left(t'\right)\tilde{\rho}_\mathrm{ S}\left(t''\right)\tilde{A}\left(t''\right)
\mathrm{ Tr}_\mathrm{ E}\left[\tilde{R}\left(t'\right)
\tilde{\rho}_\mathrm{ E}\tilde{R}\left(t''\right)\right]
+
\tilde{\rho}_\mathrm{ S}\left(t''\right)\tilde{A}\left(t''\right)\tilde{A}\left(t'\right)
\mathrm{ Tr}_\mathrm{ E}\left[\tilde{\rho}_\mathrm{ E}
\tilde{R}\left(t''\right)\tilde{R}\left(t'\right)\right]
 \\ & =
\tilde{A}\left(t'\right)\tilde{A}\left(t''\right)\tilde{\rho}_\mathrm{ S}\left(t''\right)
\mathrm{ Tr}_\mathrm{ E}\left[\tilde{\rho}_\mathrm{ E}\tilde{R}\left(t'\right)\tilde{R}\left(t''\right)
\right]
-\tilde{A}\left(t''\right)\tilde{\rho}_\mathrm{ S}\left(t''\right)\tilde{A}\left(t'\right)
\mathrm{ Tr}_\mathrm{ E}\left[
\tilde{\rho}_\mathrm{ E}\tilde{R}\left(t'\right)\tilde{R}\left(t''\right)\right]
 \\  & -
\tilde{A}\left(t'\right)\tilde{\rho}_\mathrm{ S}\left(t''\right)\tilde{A}\left(t''\right)
\mathrm{ Tr}_\mathrm{ E}\left[
\tilde{\rho}_\mathrm{ E}\tilde{R}\left(t''\right)\tilde{R}\left(t'\right)\right]
+
\tilde{\rho}_\mathrm{ S}\left(t''\right)\tilde{A}\left(t''\right)\tilde{A}\left(t'\right)
\mathrm{ Tr}_\mathrm{ E}\left[\tilde{\rho}_\mathrm{ E}
\tilde{R}\left(t''\right)\tilde{R}\left(t'\right)\right]
 \\ & =
 \left[\tilde{A}\left(t'\right)\tilde{A}\left(t''\right)\tilde{\rho}_\mathrm{ S}\left(t''\right)
-\tilde{A}\left(t''\right)\tilde{\rho}_\mathrm{ S}\left(t''\right)\tilde{A}\left(t'\right)
\right]C_R\left(t'-t''\right)
\\  & -
\left[\tilde{A}\left(t'\right)\tilde{\rho}_\mathrm{ S}\left(t''\right)\tilde{A}\left(t''\right)
-
\tilde{\rho}_\mathrm{ S}\left(t''\right)\tilde{A}\left(t''\right)\tilde{A}\left(t'\right)
\right]C_R\left(t''-t'\right)\, ,
\eea
where we have introduced the two-point correlation function
\bea
C_R\left(t,t'\right)\equiv \mathrm{ Tr}_\mathrm{ E}\left[\tilde{\rho}_\mathrm{ E}
\tilde{R}\left(t\right)\tilde{R}\left(t'\right)\right]\, .
\eea
Because the environment is in a stationary state, one can show that
$C_R(t,t')$ is in fact a function of $\tau\equiv t-t'$ only. Indeed, one
has
\bea
C_R\left(t,t'\right) &= \mathrm{ Tr}_\mathrm{ E}\left[\tilde{\rho}_\mathrm{ E}
\ee^{iH_\mathrm{ E}t}\tilde{R}(0)\ee^{-iH_\mathrm{ E}t}
\ee^{iH_\mathrm{ E}t'}
\tilde{R}(0)\ee^{-iH_\mathrm{ E}t'}\right]\\
&= \mathrm{ Tr}_\mathrm{ E}\left[\tilde{\rho}_\mathrm{ E}
\ee^{iH_\mathrm{ E}t}\ee^{-iH_\mathrm{ E}t'}
\ee^{iH_\mathrm{ E}t'}
\tilde{R}(0)\ee^{-iH_\mathrm{ E}\tau}
\tilde{R}(0)\ee^{-iH_\mathrm{ E}t'}\right]\\
&= \mathrm{ Tr}_\mathrm{ E}\left[\tilde{\rho}_\mathrm{ E}
\ee^{iH_\mathrm{ E}t'}\ee^{iH_\mathrm{ E}\tau}
\tilde{R}(0)\ee^{-iH_\mathrm{ E}\tau}
\tilde{R}(0)\ee^{-iH_\mathrm{ E}t'}\right]\\
&= \mathrm{ Tr}_\mathrm{ E}\left[\tilde{\rho}_\mathrm{ E}
\ee^{iH_\mathrm{ E}t'}
\tilde{R}(\tau)
\tilde{R}(0)\ee^{-iH_\mathrm{ E}t'}\right]\\
&= \mathrm{ Tr}_\mathrm{ E}\left[\ee^{-iH_\mathrm{ E}t'}\tilde{\rho}_\mathrm{ E}
\ee^{iH_\mathrm{ E}t'}
\tilde{R}(\tau)
\tilde{R}(0)\right]\\
&= \mathrm{ Tr}_\mathrm{ E}\left[\tilde{\rho}_\mathrm{ E}
\tilde{R}(\tau)
\tilde{R}(0)\right]\equiv C_R(\tau),
\eea
where in the last step of the calculation, we have used the fact that
the reduced density operator of the environment,
$\tilde{\rho}_\mathrm{ E}$, and the Hamiltonian for the environment,
$H_\mathrm{ E}$, commute. A more explicit form of the correlator
$C_R(\tau)$ can be obtained by making use of \Eq{eq:rhoE}. Indeed, one has
\bea
C_R(\tau)&=\sum_{m}\left\langle m\right\vert \left[
\sum _np_n\vert n\rangle \langle n \vert 
\tilde{R}(\tau)\tilde{R}(0)\right]\left\vert m\right\rangle \\
&= \sum_{n}p_n\langle n \vert 
\tilde{R}(\tau)\tilde{R}(0)\vert n\rangle\\
&= \sum_{n}p_n\langle n \vert \ee^{iH_\mathrm{ E}\tau}
\tilde{R}(0)\ee^{-iH_\mathrm{ E}\tau}\tilde{R}(0)\vert n\rangle\\
&= \sum_{n,m,p,q}p_n\langle n \vert \ee^{iH_\mathrm{ E}\tau}\vert m\rangle 
\langle m\vert
\tilde{R}(0)\vert p \rangle \langle p\vert 
\ee^{-iH_\mathrm{ E}\tau}\vert q\rangle \langle q\vert 
\tilde{R}(0)\vert n\rangle\\
&= \sum_{n,p}p_n\ee^{i(E_n-E_p)\tau}
\langle n\vert
\tilde{R}(0)\vert p \rangle 
\langle p\vert 
\tilde{R}(0)\vert n\rangle\\
&= \sum_{n,p}p_n\ee^{i(E_n-E_p)\tau}
\left\vert \langle n\vert
\tilde{R}(0)\vert p \rangle \right \vert^2 .
\eea
In particular, one has $C_R(-\tau)=C_R^*(\tau)$. More specifically, one
can see that $C_R(\tau)$ is a sum of exponentials oscillating at the
Bohr frequencies of the environment. In the limit where the
environment is large and contains an almost continuous set of
energy levels, destructive interference occurs that quickly drives $C_R(\tau)$
to zero with a characteristic time $t_\mathrm{ c}$,
$C_R(\tau)\simeq C_R(0) \ee^{-\vert \tau\vert/t_\mathrm{ c}}$.

Let us now further simplify the right-hand side of
\Eq{eq:deltarhoexpand:oneterm}. Using \Eq{eq:doublec}, one has
\bea
\int_t^{t+\Delta t}\dd t' & \int
_t^{t'}\dd t'' \mathrm{ Tr}_\mathrm{ E} \left[\tilde{H}_\mathrm{
    int}\left(t'\right),\left[\tilde{H}_\mathrm{ int}\left(t''\right),
\tilde{\rho}_\mathrm{ S}\left(t''\right)\otimes\tilde{\rho}_\mathrm{ E}
    \right]\right] \
 \\ = &
\int_t^{t+\Delta t}\dd t' \int
_t^{t'}\dd t''\biggl\{
\left[\tilde{A}\left(t'\right)\tilde{A}\left(t''\right)\tilde{\rho}_\mathrm{ S}\left(t''\right)
-\tilde{A}\left(t''\right)\tilde{\rho}_\mathrm{ S}\left(t''\right)\tilde{A}\left(t'\right)
\right]C_R\left(t'-t''\right)
 \\  & 
- \left[\tilde{A}\left(t'\right)\tilde{\rho}_\mathrm{ S}\left(t''\right)\tilde{A}\left(t''\right)
- \tilde{\rho}_\mathrm{ S}\left(t''\right)\tilde{A}\left(t''\right)\tilde{A}\left(t'\right)
\right]C_R\left(t''-t'\right)\biggr\}\, .
\label{eq:Lindblad:multitint}
\eea
\begin{figure}[t]
\begin{center}
\includegraphics[width=0.5\textwidth,clip=true]{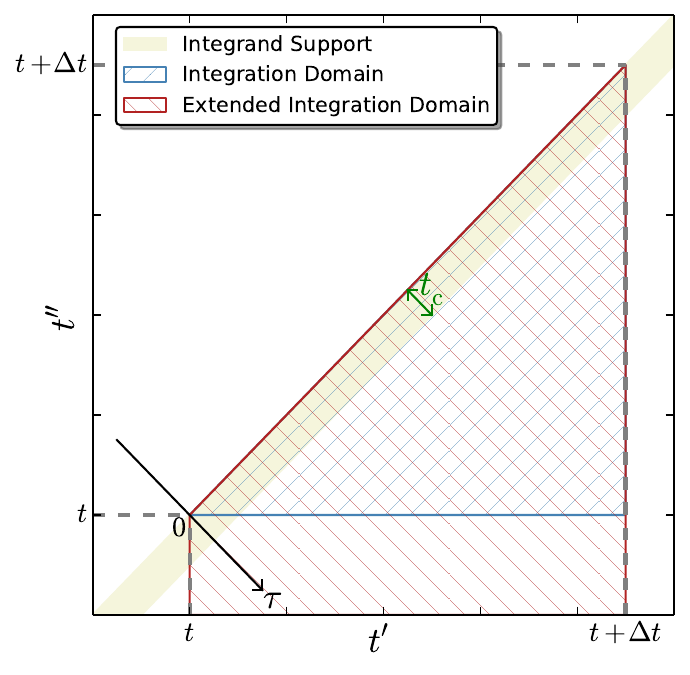}
\caption{Integration domain of \Eq{eq:Lindblad:multitint} (hatched blue surface). In the limit where $t_\uc \ll\Delta t$,
  the extended integration domain (hatched red surface) almost
  coincides with the initial one when restricted to the region where
  the integrand is not vanishingly small (pale green surface).}
\label{fig:integrationdomain}
\end{center}
\end{figure}
The integration domain appearing in this expression is displayed in
\Fig{fig:integrationdomain} as the hatched blue surface. When parametrised in terms of the variables $t'$ and $\tau=t'-t''$, it is given by
\bea
\int_t^{t+\Delta t}\dd t'  \int_t^{t'}\dd t'' = 
\int_0^{\Delta t}\dd \tau \int_{t+\tau}^{t+\Delta t}\dd  t'\, .
\eea
Indeed, from \Fig{fig:integrationdomain}, one can see that $\tau$ is
comprised between $0$ and $\Delta t$ in the integration domain. Once
$\tau$ is fixed, one of the blue lines is described, and it is then
clear that $t'$ varies between $t+\tau$ and $t+\Delta t$. Because of
the terms $C_R(\tau)$ and $C_R(-\tau)$ in \Eq{eq:Lindblad:multitint}, the
integrand vanishes when $\left\vert\tau\right\vert\gg t_\uc $,
hence its support is given by the pale green stripe in
\Fig{fig:integrationdomain}. Let us now consider the extended
integration domain 
\bea 
\int_0^{\infty}\dd \tau \int_{t}^{t+\Delta
  t}\dd  t'\, , 
\eea 
where the upper bound on $\tau$ has been extended
to infinity and the lower bound on $t'$ to $t$. It is displayed as the
hatched red surface in \Fig{fig:integrationdomain}. One can see that
compared to the initial integration domain, two regions are added. The
first one lies outside the integrand support and therefore negligibly
contributes to the overall result. The second one is the small
triangle that lies inside the integrand support. In the limit where
$t_\uc\ll\Delta t$, it corresponds to a small area compared
to the integrand support comprised in the initial integration
domain. As a consequence, its inclusion in the integration domain
negligibly changes the result if one follows the evolution of the
reduced density matrix for the system on time scales $\Delta t$ much
larger than the typical correlation time of the environment, 
\bea
\label{eq:taucllDeltat}
t_\uc \ll\Delta t\, .
\eea
Under this fifth and last assumption, one then has 
\bea
\tilde{\rho}_\mathrm{ S} \left(t+\Delta t\right)-\tilde{\rho}_\mathrm{ S}\left(t\right) 
& \simeq -g^2
\int_0^{\infty}\dd \tau \int_{t}^{t+\Delta t}\dd  t'
\biggl\{ 
\biggl[\tilde{A}\left(t'\right)\tilde{A}\left(t'-\tau\right)\tilde{\rho}_\mathrm{ S}\left(t'-\tau\right)
\\ &
-\tilde{A}\left(t'-\tau\right)\tilde{\rho}_\mathrm{ S}\left(t'-\tau\right)\tilde{A}\left(t'\right)
\biggr]C_R\left(\tau\right)
-
\biggl[\tilde{A}\left(t'\right)\tilde{\rho}_\mathrm{ S}\left(t'-\tau\right)\tilde{A}\left(t'-\tau\right)
 \\ &
-\tilde{\rho}_\mathrm{ S}\left(t'-\tau\right)\tilde{A}\left(t'-\tau\right)\tilde{A}\left(t'\right)
\biggr]C_R\left(-\tau\right)\biggr\}\, .
\eea
The time derivative of $\tilde{\rho}_\mathrm{ S}$ can then be
obtained by dividing the left-hand side and the right-hand side by
$\Delta t$. If $\Delta t$ is much smaller than the typical time scale
by which $A$ varies, $\tilde{A}(t')\simeq \tilde{A}(t)$ and
$\tilde{A}(t'-\tau)\simeq \tilde{A}(t-\tau)$, the difference only
giving rise to $\Delta t$ suppressed quantities. Note that because of
that, the condition~(\ref{eq:taucllDeltat}) actually means that the
interaction operator $A$ should vary on time scales much larger than
the autocorrelation time of $R$ in the environment. Moreover, the
variation of $\tilde{\rho}_\mathrm{ S}$ between times $t$ and
$t+\Delta t$ in the above integrals is of order $g^2$. Since the whole
expression is already proportional to $g^2$, in the integrals, one can
simply write $\tilde{\rho}_\mathrm{ S}(t)$. As a consequence, the
integral over $t'$ becomes trivial and $\Delta t$ can be factorised
out of the right-hand side, so that 
\bea
\frac{\Delta
  \tilde{\rho}_\mathrm{ S}}{\Delta t} = & -g^2\int_0^{\infty}\dd \tau
\biggl\{ \left[\tilde{A}\left(t\right) \tilde{A}\left(t-\tau\right)
  \tilde{\rho}_\mathrm{ S}\left(t\right) -\tilde{A}\left(t-\tau\right)
  \tilde{\rho}_\mathrm{ S}\left(t\right) \tilde{A}\left(t\right)
\right]C_R(\tau)
 \\ &  
- \left[\tilde{A} \left(t\right) \tilde{\rho}_\mathrm{
    S}\left(t\right) \tilde{A}\left(t-\tau\right) - \tilde{\rho}_\mathrm{
    S}\left(t\right) \tilde{A}\left(t-\tau\right)
  \tilde{A}\left(t\right) \right]C_R(-\tau)\biggr\}.  
\eea
This is a Markovian master equation, since the right hand side only
depends on $\tilde{\rho}_\mathrm{ S}$ at time $t$. Defining
\bea
L_1\left(t\right) &\equiv g^2\int_0^{+\infty}\dd \tau
C_R\left(\tau\right)\tilde{A}\left(t-\tau\right),
\\
L_2\left(t\right) &\equiv g^2\int_0^{+\infty}\dd \tau
C_R\left(-\tau\right)\tilde{A}\left(t-\tau\right) = g^2\int_0^{+\infty}\dd \tau
C_R^*\left(\tau\right)\tilde{A}\left(t-\tau\right) =L_1^{\dagger}\left(t\right), 
\eea
where the last equality is valid if $\tilde{A}$ is hermitian, it can
be written in the more compact form 
\bea
\label{eq:Lindblad:Loperators}
\frac{\Delta \tilde{\rho}_\mathrm{ S}}{\Delta t}
=-\tilde{A}\left(t\right)L_1\left(t\right)\tilde{\rho}_\mathrm{ S}\left(t\right)
+L_1\left(t\right)\tilde{\rho}_\mathrm{ S}\left(t\right)\tilde{A}\left(t\right)
+\tilde{A}\left(t\right)\tilde{\rho}_\mathrm{ S}\left(t\right)
L_2\left(t\right)
-\tilde{\rho}_\mathrm{ S}\left(t\right)L_2\left(t\right)\tilde{A}\left(t\right)\, .
\nonumber\\
\eea
The operators $L_1$ and $L_2$ can be further simplified under the
condition $t_\uc \ll \Delta t$. Indeed, because of the
profile $C_R(\tau)=C_R(0)\ee^{-\vert\tau\vert /t_\uc }$ established
above, the integrals $L_1$ and $L_2$ are dominated by their
contribution on the interval
$\tau\in[0,\mathrm{a}\ \mathrm{few}\ t_\uc ]$. In the limit
where $t_\uc \ll \Delta t$, the function $\tilde{A}(t-\tau)$
(which, as already said, varies with a typical time scale much larger
than $\Delta t$) does not evolve much in this interval and one then
has
\bea
L_1\left(t\right) = g^2
\int_0^{+\infty}\dd \tau
C_R\left(\tau\right)\tilde{A}\left(t-\tau\right) \simeq g^2
\int_0^{+\infty}\dd \tau
C_R\left(0\right)\ee^{-\tau/t_\uc }\tilde{A}\left(t\right) = g^2
C_R\left(0\right) t_\uc  \tilde{A}\left(t\right)\, , 
\eea
and the same expression for $L_2$. As a consequence,
\Eq{eq:Lindblad:Loperators} reads
\bea
\label{eq:Lindblad:interactionpicture}
\frac{\dd  \tilde{\rho}_\mathrm{ S}}{\dd  t}
=-g^2C_R\left(0\right) t_\uc 
\left[\tilde{A},\left[\tilde{A},\tilde{\rho}_\mathrm{ S} \right] \right]\, .
\eea
Going back to the standard picture, one finally obtains
\bea
  \frac{\dd  \rho_\mathrm{ S}}{\dd  t}= i \left[\rho_\mathrm{ S}, H_\mathrm{
      S}\right]-g^2 C_R\left(0\right) t_\uc  \left[A,\left[A,
      \rho_\mathrm{ S} \right]\right] \, .
\label{eq:Lindblad:finalform}
\eea
This is the standard form of the Lindblad equation. 

It can be generalised to cases where the interaction term is not simply given
by \Eq{eq:Lindblad:Hint:decomp}, namely by the product of an operator
acting in $\mathcal{E}_\mathrm{ S}$ and an operator acting in
$\mathcal{E}_\mathrm{ E}$. Indeed, if one considers a generic interaction
Hamiltonian
\bea
\label{eq:Hint:general}
H_\mathrm{ int}=\sum _i A_i\left(t\right)\otimes R_i\left(t\right),
\eea
one can define correlators in the environment as
\bea
\label{eq:correlator:multi}
C_{R,ij}\left(t,t'\right)\equiv 
\mathrm{ Tr}_\mathrm{ E}\left[\tilde{\rho}_\mathrm{ E}
\tilde{R}_i\left(t\right)\tilde{R}_j\left(t'\right)\right],
\eea
and all the previous steps can be repeated, leading to
\bea
  \frac{\dd  \rho_\mathrm{ S}}{\dd  t}= i \left[\rho_\mathrm{ S}, H_\mathrm{
      S}\right]-g^2  
\sum _{i,j} t_{\uc,ij}  C_{R,ij}(0)\left[A_i,\left[A_j,
      \rho_\mathrm{ S} \right]\right] \, .
\label{eq:Lindblad:finalformgeneralized}
\eea
Here we have assumed that the assumptions discussed before still
hold and that $C_{R,ij}=C_{R,ji}$. The characteristic times of the correlation functions $C_{R,ij}$ are denoted $t_{\uc,ij} $ and must all be much smaller than the typical time scale over which the system varies. If $i$ and $j$ are replaced by continuous indices $\bm{x}$ and $\bm{y}$, the interaction Hamiltonian~(\ref{eq:Hint:general}) is of the form~(\ref{eq:localinter}), and \Eq{eq:Lindblad:finalformgeneralized} gives rise to the Lindblad equation~(\ref{eq:lindbladgeneral}). In this expression $C_R(\bm{x},\bm{y})$ denotes $C_{R,ij}(0)$, any potential dependence of $t_\uc$ on $\bm{x}$ and $\bm{y}$ is absorbed in the function $C_R(\bm{x},\bm{y})$, and 
\bea
\label{eq:gamma:g}
\gamma =2 g^2 t_\uc\, .
\eea

In conclusion, let us summarise the conditions under which
the Lindblad equation has been obtained. This equation describes the evolution of the
reduced density operator of the system, perturbatively coupled to the
environment through an operator $A\otimes R$ where $A$ acts on the system and $R$ on the environment, on time scales much larger than the correlation time of
$R$. It is valid only when, in the interaction picture, $\tilde{A}$
varies with time scales much larger than this correlation time. The
environment is assumed to be in a stationary state, which technically
means that $\langle R\left(t\right)R(t')\rangle$ must depend on $t-t'$ only, and
sufficiently large not to be affected by its interaction with the
system. Finally, the above equation is valid if the interaction
Hamiltonian has vanishing mean value in the environment,
$\langle R\rangle=0$, but this can always be achieved by properly
redefining the free and interaction Hamiltonians.
\section{Concrete example: a massive scalar field as the environment}
\label{sec:massivescalarfield}
In this section, we present a concrete example where the Lindblad
formalism can be applied. We consider the case where two scalar fields
$\phi$ and $\psi$ are present during inflation, $\phi$ being very
light and $\psi$ very heavy. The Fourier modes of $\phi $ play the
role of the system, coupled to an environment made of the degrees of
freedom contained in $\psi$. This allows us to show how the
assumptions used in \App{sec:DerivingLindblad} in order to derive
\Eq{eq:lindbladgeneral} can be realised in practice, how the
parameters $\gamma$ and $C_R(\bm{x},\bm{y})$ that appear in that
equation can be related to microphysical quantities, and how the
Lindblad operator $A$ can be concretely identified.

Let us consider the following action \bea S=-\int\dd ^4x
\sqrt{-g}\left[
  \frac{1}{2}g^{\mu\nu}\partial_\mu\phi\partial_\nu\phi+V\left(\phi\right)
  +\frac{1}{2}g^{\mu\nu}\partial_\mu\psi\partial_\nu\psi
  +\frac{M^2}{2}\psi^2+\lambda\mu^{4-n-m}\phi^n\psi^m\right]\, , \eea
where $V(\phi)$ is the potential of the field $\phi$ that we leave
unspecified for the moment, $M$ is the mass of the field $\psi$ that
we assume to be much larger than the Hubble scale, $M\gg H$, $\lambda$
is a (supposedly small) dimensionless coupling constant and $\mu$ is a
mass scale parameter that appears in the power-law coupling between
$\phi$ and $\psi$. For now we assume both $\phi$ and $\psi$ to be test
fields (\ie they do not contribute much to the energy budget of the
Universe) in an inflating background, but below we comment on how the
result can be generalised to the case where $\phi$ is the inflaton
field and the system corresponds to the observed cosmological
perturbations.

A first important remark is that the action should be written in such
a way that the quantum mean value of the interacting term vanishes in
the stationary configuration of the environment, as required by
\Eq{eq:Lindbald:meanRvanishes}. Following the procedure described in
\App{sec:DerivingLindblad}, this can be easily done by adding
and subtracting $\lambda\mu^{4-n-m}\left\langle\psi^m
\right\rangle_\mathrm{st}\phi^n$,
\bea
S&=-\int\dd ^4x \sqrt{-g}\left[
\frac{1}{2}g^{\mu\nu}\partial_\mu\phi\partial_\nu\phi+V\left(\phi\right)
+\lambda\mu^{4-n-m}\left\langle\psi^m\right\rangle_\mathrm{st}\phi^n
\right. \\ &  \left.
+\frac{1}{2}g^{\mu\nu}\partial_\mu\psi\partial_\nu\psi
+\frac{M^2}{2}\psi^2
+\lambda\mu^{4-n-m}\phi^n\left(\psi^m
-\left\langle\psi^m\right\rangle_\mathrm{st}\right)\right]\, ,
\label{eq:action}
\eea
where $\left\langle\psi^m\right\rangle_\mathrm{st}$ denotes the stationary quantum mean value of $\psi^m$. This simply modifies the effective potential of the field $\phi $ according to
\bea
\label{eq:pot:effective}
V_\ueff(\phi)=V(\phi)+\lambda
\mu^{4-n-m}\left\langle\psi^m\right\rangle_\mathrm{st}\phi^n\, .
\eea
On the other hand, it is now clear that the mean value of the
interacting term taken in the environment sector,
$\psi^m-\left\langle\psi^m\right\rangle_\mathrm{st}$, vanishes. The action~(\ref{eq:action}) can be decomposed as  $S=S_\phi+S_\psi+S_{\phi\psi}$, and $S_{\phi\psi}$ gives rise to the interaction Hamiltonian 
\bea
H_\mathrm{ int}=\lambda\mu^{4-n-m}a^4\int \dd ^3 {\bm x}\,  \phi^n
\left(\psi^m-\left\langle\psi^m\right\rangle_\mathrm{st}\right)\, .
\eea

Assuming that $V_\ueff(\phi)=m^2\phi^2/2$, in Fourier space, one has
\bea
\label{eq:actionphifourier}
S_\phi &=\frac12 \int \dd \eta \int _{\setR^3}\dd ^3{\bm k}
\left[v_{\bm k}'v_{\bm k}^*{}^\prime-
  \left(k^2-\frac{a''}{a}+m^2a^2\right) v_{\bm k}v_{\bm k}^* \right],
\eea 
where we have defined
$v(\eta,{\bm x})\equiv a(\eta) \phi(\eta,{\bm x})$. If one lets $m=0$
in \Eq{eq:actionphifourier}, then one recovers the action for curvature
perturbations if one ignores metric perturbations (and it is even exactly the same for the particular case of
power-law inflation). This suggests that, by identifying the variable
$v(\eta,{\bm x})$ with the Mukhanov-Sasaki variable (which is the
reason why we have used the same notation) in the uniform-curvature
gauge, we can extend the present analysis to the case where $\phi$ is
the inflaton field and the system corresponds to the curvature
perturbations.

In terms of the Mukhanov-Sasaki variable, the interaction Hamiltonian
is given by \bea H_\mathrm{ int}=\lambda\mu^{4-n-m}a^{4-n}\int \dd ^3
{\bm x}\, v^n\left(\eta,{\bm x}\right) \left[\psi^m\left(\eta,{\bm
      x}\right) -\left\langle\psi^m\right\rangle_\mathrm{st}\right]\,
.  \eea This is of the form~(\ref{eq:localinter}) provided $A=v^n$,
$R=\psi^m-\left\langle\psi^m\right\rangle_\mathrm{st}$, and the
effective coupling constant reads $g=\lambda\mu^{4-n-m}a^{4-n}$ which
is, therefore, a time-dependent quantity. In
\App{sec:DerivingLindblad}, we have showed that
$\gamma =2g^2\tau_\mathrm{{c}}$, see \Eq{eq:gamma:g}. In this
expression, $\tau_\mathrm{{c}}$ is the correlation time of $R$. Given
that, in the main text, the Lindblad equation is written in terms of
conformal time, $\tau_\mathrm{{c}}$ in this context must be
interpreted as a conformal correlation time and expressed as
$t_\uc/a$, $t_\uc$ being the cosmic correlation time. This implies
that the ansatz~(\ref{eq:gamma}) is satisfied,
$\gamma=\gamma_* (a/a_*)^p$, if we make the identification \bea
\label{eq:gammastar:tc:app}
\gamma_* = 2 t_\uc \lambda^2 \mu^{8-2n-2m} a_*^{7-2n}
\eea
and $p=7-2n$. For a linear interaction, this leads to $p=5$ and for a quadratic interaction, one has $p=3$. 

The explicit form of $R$ also allows us to calculate the correlation function $C_{R}$.  Massive fields in de-Sitter space-times have long been
studied~\cite{Chernikov:1968zm,Bunch:1977sq,Bunch:1978yq,Birrell:1982ix}. In the small-separation limit,
$\epsilon^2\equiv[(t_1-t_2)^2-a^2({\bm x}_1-{\bm x}_2)^2]/4\ll
\mathrm{min}(1/H^2, 1/M^2)$,
point-splitting renormalisation yields the two-point correlation
function given by Eq.~(3.14) in \Refa{Bunch:1978yq}. In the
regime where $M\gg H$, expanding this formula in powers of $H/M$ leads to
\bea
\label{eq:massivefield:corr}
\left\langle \psi\left(t_1,{\bm x}_1\right)
\psi\left(t_2,{\bm x}_2\right)\right\rangle_\mathrm{ren}
\simeq \frac{37}{504\pi^2}\frac{H^6}{M^4}
\left(1-\frac{M^2\Sigma\epsilon^2}{2}\right)
\eea
at leading order in $H/M$, where $\Sigma=\pm 1$ depends on whether the separation between the
two points $(t_1,\bm{x}_1)$ and $(t_2,\bm{x}_2)$ is timelike or spacelike. Assuming that $\psi$ has
Gaussian statistics (which is correct at leading order in perturbation
theory), Wick theorem leads to
\bea
\label{eq:massivefield:corr:higher:order:start}
&\left\langle \psi^m\left(t_1,{\bm x}_1\right)
\psi^m\left(t_2,{\bm x}_2\right)\right\rangle_\mathrm{ren}
 \simeq \\ 
& \sum_{0\leq p\leq \floor*{\frac{m}{2}}} a_{p,m} 
\left\langle \psi\left(t_1,{\bm x}_1\right)
\psi\left(t_1,{\bm x}_1\right)\right\rangle_\mathrm{ren}^{p}
\left\langle \psi\left(t_2,{\bm x}_2\right)
\psi\left(t_2,{\bm x}_2\right)\right\rangle_\mathrm{ren}^{p}
\left\langle \psi\left(t_1,{\bm x}_1\right)
\psi\left(t_2,{\bm x}_2\right)\right\rangle_\mathrm{ren}^{m-2p}\, .
\eea
In this expression, $\floor{m/2}$ denotes the integer part of $m/2$, \ie it is $m/2$ is $m$ is even and $(m-1)/2$ if $m$ is odd, and $a_{p,m}$ are combinatory coefficients that can be calculated as follows. In the product of pairs written in the sum of \Eq{eq:massivefield:corr:higher:order:start}, $2p$ is the number of $\psi(t_1,\bm{x_1})$ occurrences that go into auto-correlators and $m-2p$ is the number of $\psi(t_1,\bm{x_1})$ occurrences that go into cross-correlators. The number of ways to split the $m$ occurrences of $\psi(t_1,\bm{x_1})$ between these two types of correlators is given by ${{m}\choose{2p}}=m!/[(2p)!(m-2p)!]$, \ie by the number of ways one can draw $2p$ elements out of $m$ elements (or equivalently $m-2p$ elements out of $m$ elements). The same applies for dispatching the occurrences of $\psi(t_2,\bm{x_2})$. Then, once the $2p$ occurrences of $\psi(t_1,\bm{x_1})$ that appear in auto-correlators are chosen, one has to arrange them into pairs, and the number of such arrangements is given by $(2p-1)!!=(2p)!/(2^p p!)$. The same applies for arranging into pairs the $2p$ occurrences of $\psi(t_2,\bm{x_2})$ that go into auto-correlators. Finally, the number of cross-correlators one can build from the $m-2p$ occurrences of $\psi(t_1,\bm{x_1})$ and $\psi(t_2,\bm{x_2})$ is given by $(m-2p)!$. Putting everything together, one obtains
\bea
\label{eq:apm}
a_{p,m} = \left[{{m}\choose{2p}} (2p-1)!!\right]^2(m-2p)!
=\frac{\left(m!\right)^2}{2^{2p}\left(p!\right)^2\left(m-2p\right)!}\, .
\eea
Combining \Eqs{eq:massivefield:corr} and~(\ref{eq:massivefield:corr:higher:order:start}), one also has
\bea
\label{eq:massivefield:corr:higher:order:2}
\left\langle \psi^m\left(t_1,{\bm x}_1\right)
\psi^m\left(t_2,{\bm x}_2\right)\right\rangle_\mathrm{ren}
& \simeq 
 \left(\frac{37}{504\pi^2}\frac{H^6}{M^4}\right)^{m}
 \sum_{0\leq p\leq \floor*{\frac{m}{2}}} a_{p,m} 
\left(1-\frac{M^2\Sigma\epsilon^2}{2}\right)^{m-2p}
\\ &
\simeq 
 \left(\frac{37}{504\pi^2}\frac{H^6}{M^4}\right)^{m}
 \sum_{0\leq p\leq \floor*{\frac{m}{2}}} a_{p,m} 
\left[1-\left(m-2p\right)\frac{M^2\Sigma\epsilon^2}{2}\right]
\, ,
\eea
where in the second line we have expanded in the small $\epsilon$ limit. This expression requires to calculate two sums involving the combinatory coefficients $a_{p,m}$. The first one is straightforward,
\bea
\label{eq:sum:apm}
 \sum_{0\leq p\leq \floor*{\frac{m}{2}}} a_{p,m} = \left(2m-1\right)!!\, ,
\eea
since it corresponds by definition to the number of arrangements of the $2m$ occurrences of $\psi$ into pairs. Making use of \Eq{eq:apm}, the second sum is given by
\bea
\label{eq:sum:apm:m-2p}
 \sum_{0\leq p\leq \floor*{\frac{m}{2}}} \left(m-2p\right) a_{p,m} &=
  \sum_{0\leq p\leq \floor*{\frac{m}{2}}} \left(m-2p\right) \frac{\left(m!\right)^2}{2^{2p}\left(p!\right)^2\left(m-2p\right)!}
  \\ &=
  \sum_{0\leq p\leq \floor*{\frac{m-1}{2}}} \left(m-2p\right) \frac{\left(m!\right)^2}{2^{2p}\left(p!\right)^2\left(m-2p\right)!}   
  \\ &=
m^2\sum_{0\leq p\leq \floor*{\frac{m-1}{2}}}  \frac{\left[\left(m-1\right)!\right]^2}{2^{2p}\left(p!\right)^2\left(m-1-2p\right)!}
  \\ &=
m^2\sum_{0\leq p\leq \floor*{\frac{m-1}{2}}} a_{p,m-1}
  \\ &=
m^2 \left(2m-3\right)!!\, .
\eea
In the second line of this derivation, we have used the fact that if $m$ is even, the last term of the sum corresponds to $p=m/2$, which vanishes because of the coefficient $m-2p$, so one can stop the sum at $p=\floor{(m-1)/2}$, and in the last line, we have used \Eq{eq:sum:apm}. Then, plugging \Eqs{eq:sum:apm} and~(\ref{eq:sum:apm:m-2p}) into \Eq{eq:massivefield:corr:higher:order:2}, one obtains
\bea
\label{eq:massivefield:corr:higher:order}
\left\langle \psi^m\left(t_1,{\bm x}_1\right)
\psi^m\left(t_2,{\bm x}_2\right)\right\rangle_\mathrm{ren}
 \simeq 
\left(2m-1\right)!! \left(\frac{37}{504\pi^2}\frac{H^6}{M^4}\right)^{m}
\left(1-\frac{m^2}{2m-1}\frac{M^2\Sigma\epsilon^2}{2}\right)\, .
\eea
Wick theorem with \Eq{eq:massivefield:corr} also gives rise to
\bea
\label{eq:massivefield:corr:stat}
\left\langle \psi^m \right\rangle_\mathrm{st} &= \sigma(m)\left(m-1\right)!! \left(\frac{37}{504\pi^2}\frac{H^6}{M^4}\right)^{\frac{m}{2}}\, ,
\eea
where $\sigma(m)=1$ if $m$ is even and $0$ is $m$ is odd. Notice that this equation is consistent with \Eq{eq:massivefield:corr:higher:order}. Recalling that $R=\psi^m-\left\langle\psi^m\right\rangle_\mathrm{st}$, \Eq{eq:correlator:multi} gives rise to
\bea
C_{R}\left(t_1, {\bm x}_1; t_2, {\bm x}_2\right)
=\left\langle \psi^m\left(t_1,{\bm x}_1\right)
\psi^m\left(t_2,{\bm x}_2\right)\right\rangle_\mathrm{ren}
-\left\langle\psi^m\right\rangle_\mathrm{st}^2\, ,
\eea
where we have used that $\langle\psi^m\rangle_\mathrm{st}$ is independent of time (which is in fact required by stationarity) and space. Inserting \Eqs{eq:massivefield:corr:higher:order} and~(\ref{eq:massivefield:corr:stat}) into this expression, one obtains
\bea
C_{R}\left(t_1, {\bm x}_1; t_2, {\bm x}_2\right) = &
\left\lbrace \left(2m-1\right)!!-\sigma\left(m\right)\left[\left(m-1\right)!!\right]^2\right\rbrace
\left(\frac{37}{504\pi^2}\frac{H^6}{M^4}\right)^m
\times \\ &
\left\lbrace 1-\frac{m^2\left(2m-3\right)!!}{\left(2m-1\right)!!-\sigma\left(m\right)\left[\left(m-1\right)!!\right]^2}
\frac{M^2\Sigma\epsilon^2}{2}\right\rbrace\, .
\eea

From this formula, the overall amplitude $\bar{C}_R$, the correlation cosmic time $t_\uc$ and the correlation length $\lE  $ of the two-point function of $R$ in the environment can be read off and are given by
\bea
\bar{C}_R &= \left\lbrace \left(2m-1\right)!!-\sigma\left(m\right)\left[\left(m-1\right)!!\right]^2\right\rbrace
\left(\frac{37}{504\pi^2}\frac{H^6}{M^4}\right)^m\, ,\\
t_\uc & = \lE   = 2\sqrt{2}\sqrt{\frac{\left(2m-1\right)!!-\sigma\left(m\right)\left[\left(m-1\right)!!\right]^2}{m^2\left(2m-3\right)!!}} \frac{1}{M}\, .
\label{eq:MassivePsi:Cbar:tc:lE}
\eea
In particular, with \Eq{eq:gammastar:tc:app} this gives rise to
\bea
\gamma_* = 4\sqrt{2}\sqrt{\frac{\left(2m-1\right)!!-\sigma\left(m\right)\left[\left(m-1\right)!!\right]^2}{m^2\left(2m-3\right)!!}} \frac{\lambda^2}{M} \mu^{8-2n-2m}a_*^{7-2n}\, .
\eea
Let us also note that in \Eq{eq:MassivePsi:Cbar:tc:lE}, $\bar{C}_R$ is proportional to $H^{6m}$. In a Universe that is inflating in the slow-roll regime, $H$ is not strictly constant but scales as $a^{-\epsilon_{1}}$, where $\epsilon_1$ is the first slow-roll parameter. This slow time variation can be absorbed in $p$ by replacing $p\rightarrow p-6m\epsilon_{1*}$, leading to
\bea
p=7-2n-6m\epsilon_{1*} \, ,
\eea
and replacing $H$ with $H_*$ in \Eq{eq:MassivePsi:Cbar:tc:lE}. Let us note that this assumes that \Eq{eq:massivefield:corr} is valid even when $H$ varies with time. In that case however, there is no known analytical solution for the two-point correlation function of the field $\psi$ (however, in the case $H\propto a^{-\epsilon_1}$, see \Refa{Janssen:2009pb}). Nevertheless, since $M\gg H$, the relaxation time of this correlation function, $1/M$, is much smaller than the typical time scale over which $H$ varies, $1/(H\epsilon_{1})$, and one can assume $\psi$ to adiabatically follow \Eq{eq:massivefield:corr}.

In \App{sec:DerivingLindblad}, it was made clear that the
Lindblad equation relies on the validity of a number of
assumptions. Let us now derive the conditions under which those
assumptions are verified in the simple model considered
here.

First, one needs to check that $\psi$ is a test field
(strictly speaking this is not a required condition for the derivation
of the Lindblad equation but is necessary for the consistency of the
model presented here). In order for $\psi$ to play a
negligible role in the energy budget of the Universe,
$M^2\left\langle\psi^2\right\rangle_\mathrm{st}$ has to be small compared to the total energy density of the Universe $\rho$,
\bea
\label{eq:cond:psi:test}
M^2\left\langle\psi^2\right\rangle_\mathrm{st} \ll 3\Mp^2 H^2\, ,
\eea
where the Friedmann equation $\rho=3\Mp^2 H^2$ has been used. Making use of \Eq{eq:massivefield:corr:stat} with $m=2$, one has 
\bea
\frac{M^2\left\langle\psi^2\right\rangle_\mathrm{st}}{3\Mp^2H^2}
=\frac{37}{1512\pi^2}\frac{H^4}{M^2\Mp^2}\,.  
\eea 
Since $M\gg H$, and since $H/\Mp\lesssim 10^{-4}$ due to the current
observational bound~\cite{Ade:2015lrj} on the tensor-to-scalar ratio,
this number is necessarily very small and the condition is satisfied.

Second, if $\phi$ is taken to be a test field, a similar condition must be satisfied, $V_\ueff(\phi) \ll 3\Mp^2 H^2$, where $V_\ueff(\phi)$ is given in \Eq{eq:pot:effective}. Assuming that the initial potential $V(\phi)$ is such that $\phi$ is indeed a test field, $V(\phi)\ll 3\Mp^2 H^2$, the condition $V_\ueff(\phi) \ll 3\Mp^2 H^2$ reduces to
\bea
\lambda\mu^{4-n-m}\left\langle \psi^m\right\rangle_\mathrm{st}
\phi^n\ll 3\Mp^2 H^2\, .
\label{eq:cond:phi}
\eea If, on the other hand, $\phi$ is taken to be the inflaton field,
one must ensure that the correction to its potential arising in
\Eq{eq:pot:effective} does not spoil its flatness. If \Eq{eq:cond:phi}
is satisfied, this is obviously the case since
$3\Mp^2 H^2 \simeq V(\phi)$ in the slow-roll approximation and
\Eq{eq:cond:phi} implies that $V_\ueff (\phi)\simeq V(\phi)$. The
condition~(\ref{eq:cond:phi}) also guarantees that the effect of the
environment on the system can be treated perturbatively. Making use of
\Eq{eq:massivefield:corr:stat}, it can also be expressed as
\bea
\frac{\lambda\mu^{4-n-m}\left\langle \psi^m\right\rangle_\mathrm{st}\phi^n}{3\Mp^2 H^2} = 
\sigma(m)\frac{\left(m-1\right)!!}{3} \left(\frac{37}{504\pi^2}\right)^{\frac{m}{2}}
\lambda\frac{\mu^{4-n-m}\phi^nH^{3m-2}}{\Mp^2 M^{2m}}\ll 1\, .
\eea
This means that, in practice, the coupling constant must be
sufficiently small, but the constraint depends on the value of $\phi$ which can only be specified by choosing an explicit model.

Third, one must check that the interaction term does not affect much the 
behaviour of the environment. This means that
\bea
\lambda \mu^{4-n-m}\phi^n \psi^m\ll 
M^2\psi^2 \, .
\eea
Notice that because of \Eq{eq:cond:psi:test}, this condition ensures that \Eq{eq:cond:phi} is satisfied too. Making use of \Eq{eq:massivefield:corr:stat}, one has
\bea
\frac{\lambda \mu^{4-n-m}\phi^n \left\langle\psi^m\right\rangle_{\mathrm{st}}}{M^2\left\langle \psi^2 \right\rangle_{\mathrm{st}}} = \left(m-1\right)!!
 \left(\frac{37}{504\pi^2}\right)^{\frac{m}{2}-1}
\lambda\frac{ \mu^{4-n-m}  \phi^n }{H^{6-3m} M^{2m-2}} \, .
\eea
Again, this condition is verified if the coupling constant $\lambda$ is sufficiently small, but the precise upper bound on $\lambda$ depends on the value of $\phi$ which has to be specified by choosing a model. 

Fourth, one needs to make sure that, when the environment is in its stationary state,
$R=\psi^m-\left\langle\psi^m\right\rangle_\mathrm{st}$ has
autocorrelation time $t_\uc$ much smaller than the typical time
scale for the evolution of the system in the interaction picture, noted
$\tilde{T}_{A}$. Since the system is a light scalar field, it evolves with a time scale that is
typically of order $H^{-1}$. In the interaction picture, the terms
$\exp ( i\int H_{v})$ must be included as well, but the pulsation
$\omega^2$ given in \Eq{eq:defomega} also gives rise to variations over the Hubble time, such that $\tilde{T}_{A}=H^{-1}$. Given that $t_\uc \sim 1/M$, see \Eq{eq:MassivePsi:Cbar:tc:lE}, one has
\bea
\frac{t_\uc }{\tilde{T}_{A}}\sim \frac{H}{M}\, .
\eea
Since we assumed $\psi$ to be a heavy field, $M\gg H$, the condition
$t_\uc \ll \tilde{T}_{A}$ is always
satisfied.
\section{Density matrix for linear interaction}
\label{sec:dmlinearinteraction}
In this appendix, we show how the equation~(\ref{eq:diffdensitymatrix}) for the elements of the density matrix in the basis $\vert v_{\bm k}^s\rangle $ and in presence of linear interaction can be solved. Let us consider $\left\langle v^{s,(1)}_{\bm k}\right\vert \hat{\rho}_{\bm k}^s
\left\vert v^{s,(2)}_{\bm k}\right\rangle\equiv f(d,D,\eta)$
as a generic function of
\bea
\label{eq:decoherence:dD:def}
d=v^{s,(2)}_{\bm k}-v^{s,(1)}_{\bm k}
\, ,\qquad
 D=\frac{v^{s,(1)}_{\bm k}
+v^{s,(2)}_{\bm k}}{2}
\eea
and conformal time $\eta$, where we have dropped the dependence on $s$ and $\bm k$ of the variables $d$ and $D$ in order not to cluster notations too much (strictly speaking one should write $d^s_{\bm k}$
and $D^s_{\bm k}$, but since the Lindblad equation factorises
into independent equations in each Fourier subspace, no confusion can arise). Written in terms 
of $d$ and $D$, \Eq{eq:diffdensitymatrix} takes the form
\bea
\label{eq:diffdensitymatrix:dD}
\frac{\partial f\left(d,D,\eta\right)}{\partial\eta} 
= \left[-i\frac{\partial}{\partial d} \frac{\partial}{\partial D} 
+i\omega^2\left(k\right)dD
-\frac{\gamma}{2}\left(2\pi\right)^{3/2}\tilde{C}_R(\bm k)d^2
\right]
f\left(d,D,\eta\right)\, .
\eea
This is a linear partial differential equation of order $2$ with respect to the three variables $d$, $D$ and $\eta$. 
\subsection{General solution}
\label{subsec:generaldm}
This equation can be cast as a family of first-order partial differential equations
if we Fourier transform the $D$ coordinate, namely by introducing
\bea
f\left(d,D,\eta\right) = \frac{1}{\sqrt{2\pi}} 
\int\dd  r \ee^{i r D} \tilde{f}\left(d,r,\eta\right)\, .
\eea
Plugging this expansion into \Eq{eq:diffdensitymatrix:dD}, the Fourier component $\tilde{f}(d,r,\eta)$ can be shown to obey the following equation
\bea
\label{eq:fourierfeq}
\frac{\partial \tilde{f}\left(d,r,\eta\right)}{\partial\eta} 
= \left[r\frac{\partial}{\partial d}-\omega^2\left(k\right) d 
\frac{\partial }{\partial r}-\frac{\gamma}{2}
\left(2\pi\right)^{3/2}\tilde{C}_R(\bm k) d^2 \right]\tilde{f}\left(d,r,\eta\right)  \, .
\eea
This is now a linear first-order partial differential equation and, therefore, it can
be solved with the method of characteristics. To this end let us considered a characteristic line $\eta(\tau)$, $r(\tau)$ and $d(\tau)$ in the three dimensional space $(\eta,r,d)$, parametrised by the curvilinear coordinate $\tau$. Along this line, $\tilde{f}$ is a function of $\tau$ only and \Eq{eq:fourierfeq} allows us to write
\bea
\label{eq:fourierfeq:charac}
 \frac{\dd}{\dd\tau}  \tilde{f}\left[d\left(\tau\right),r\left(\tau\right),\eta\left(\tau\right)\right] & =
 \Bigg\lbrace \left[r\left(\tau\right) \eta'\left(\tau\right)+d'\left(\tau\right)\right]\frac{\partial }{\partial d}
+\left[r'\left(\tau\right)-\omega^2\left(k\right)d\left(\tau\right)\eta'\left(\tau\right)\right]\frac{\partial }{\partial r}
 \\  & 
-\frac{\gamma}{2}\left(2\pi\right)^{3/2}\tilde{C}_R\left(\bm k\right) d^2\left(\tau\right)\eta'\left(\tau\right)\Bigg\rbrace
\tilde{f}\left[d\left(\tau\right),r\left(\tau\right),\eta\left(\tau\right)\right]\, ,
\eea
where a prime here denotes derivation with respect to $\tau$. In order to remove the contribution from the partial derivatives along $d$ and $r$ in this equation, let us now choose the characteristic line such that it satisfies
\bea
\label{eq:charaeqs}
\eta'\left(\tau\right)=1\, , \qquad 
d'\left(\tau\right)=-r(\tau)\, , \qquad 
r'\left(\tau\right)=\omega^2\left(k\right)d(\tau),
\eea
\ie in such a way that \Eq{eq:fourierfeq:charac} can be written as an ordinary differential equation of the variable $\tau$ only, namely
\bea
\label{eq:charaf}
\frac{\dd  \tilde{f}}{\dd  \tau}=-\frac{\gamma}{2}(2\pi)^{3/2}
\tilde{C}_R(\bm k)d^2 \tilde{f}.
\eea

Let us now specify the initial conditions. Requiring the initial state
to be in the Bunch-Davies vacuum,\footnote{Since the Lindblad
  correction to the standard dynamics of $\hat{\rho}_{\bm{k}}^s$
  vanishes when $a/k<\lE$, see \Eq{eq:Ck:appr}, it does not spoil the
  ability to set the Bunch-Davies vacuum in the sub-Hubble limit, as
  long as one also is in the sub-$\lE$ limit.} one has\footnote{In the
  sub-Hubble limit, $\omega^2(k)\simeq k^2$ and the ground state is
  given by \bea \Psi \left(v_{\bm k}^s\right)
  \underset{k\eta\rightarrow -\infty} {\longrightarrow}
  \left(\frac{k}{\pi}\right)^{1/4}\ee^{-\frac{k}{2}v_{\bm k}^s{}^2}\,
  .  \eea This implies that 
\bea 
\left\langle v_{\bmk}^{s,(1)}
    \right\vert \hat{\rho}_{\bm k}^s\left\vert
    v_{\bmk}^{s,(2)}\right\rangle \underset{k\eta\rightarrow -\infty}
  {\longrightarrow} \sqrt{\frac{k}{\pi}}
  \ee^{-\frac{k}{2}\left[{v_{\bmk}^{s,(1)}}^2+{v_{\bmk}^{s,(2)}}^2\right]}
  = \sqrt{\frac{k}{\pi}}
  \ee^{-\frac{k}{2}\left(2D^2+\frac{d^2}{2}\right)}\, .  \eea Fourier
  transforming the above expression in the coordinate $D$ leads to
  \Eq{eq:fourierBD}.}  \bea
\label{eq:fourierBD}
 \tilde{f}\left(d_\uin,r_\uin,\eta_\uin\right)\underset{k \eta_\uin \rightarrow -\infty}
{\longrightarrow}\frac{1}{\sqrt{2\pi}}\ee^{-\frac{kd^2_\uin}{4} 
-\frac{r^2_\uin}{4k}}\, .
 \eea
With this initial condition, the solution of \Eq{eq:charaf} can be found and reads
\bea
\label{eq:solfr}
\tilde{f}\left[d\left(\tau\right),r\left(\tau\right),\eta\left(\tau\right)\right]=
\frac{\ee^{-\frac{kd^2_\uin}{4} 
-\frac{r^2_\uin}{4k}}}{\sqrt{2\pi}}\exp\left\lbrace -\frac{(2\pi)^{3/2}}{2}
\int _{\tau_\uin}^\tau \dd  \tau' \gamma\left[\eta\left(\tau'\right)\right]\, 
\tilde{C}_R\left[\bm k, \eta\left(\tau'\right)\right]
\, d^2(\tau')\right\rbrace\, .
\eea
One still needs to solve for the functions $\eta\left(\tau\right)$ and $d(\tau)$ since these functions appear in the above expression. This can be done through the integration of \Eqs{eq:charaeqs}, which leads to
\begin{align}
\label{eq:charaeta}
\eta(\tau)&= \tau, \\
\label{eq:charad}
d\left(\tau\right) &= u_{\bm k}\left(\tau\right) d_\uin
+ w_{\bm k}\left(\tau\right) r_\uin\, ,\\
\label{eq:charar}
r\left(\tau\right) &= -u_{\bm k}'\left(\tau\right) d_\uin
- w_{\bm k}'\left(\tau\right) r_\uin\, .
\end{align}
Here we have chosen the first integration constant such that $\eta_\uin=\tau_\uin$, and since $d(\tau)$ and $r(\tau)$ satisfy two coupled linear first-order differential equations, we have simply written that $d$ is a linear combination of $d_\uin$ and $r_\uin$, and calculated $r=-d'$ accordingly. The two real functions $u_{\bm k}$ and $w_{\bm k}$ (where the dependence on $\bm{k}$ has been reestablished) will be calculated below, but for now, plugging \Eqs{eq:charaeta} and~(\ref{eq:charad}) into \Eq{eq:solfr}, one obtains
\bea
\label{eq:fretaudrini}
& \tilde{f}\left(d,r,\eta\right)=
\\ & \qquad\quad
\frac{\ee^{-\frac{kd^2_\uin}{4} 
-\frac{r^2_\uin}{4k}}}{\sqrt{2\pi}}\exp\left\lbrace -\frac{(2\pi)^{3/2}}{2}
\int _{\eta_\uin}^\eta \dd  \eta' \gamma\left(\eta'\right)\, 
\tilde{C}_R\left(\bm k, \eta'\right)
\, \left[u_{\bm k}\left(\eta'\right) d_\uin
+ w_{\bm k}\left(\eta'\right) r_\uin\right]^2\right\rbrace\, .
\eea
In order to proceed further, one needs to express $d_\uin$ and $r_\uin$ in terms of $\eta$, $d$ and
$r$. To do so, looking at \Eqs{eq:charaeqs}, we notice that $d$
satisfies 
\bea
\label{eq:app:MSequation}
\frac{\dd ^2d}{\dd \eta^2}+\omega^2\left(k\right)d^2=0\, ,
\eea
namely the same
equation as the Mukhanov-Sasaki variable $v_{\bm k}^s$. As a consequence, one can
write $d=v_{\bm k}^s$ and $r=-v_{\bm k}^s{}'$. It
follows that \Eqs{eq:charad} and~(\ref{eq:charar}) can be
reexpressed as
\bea
v_{\bm k}^s\left(\eta\right) &= u_{\bm k}\left(\eta\right) 
v_{\bm k}^s(\eta_\uin)
+w_{\bm k}\left(\eta\right)\left[-v_{\bm k}^s{}'(\eta_\uin)\right] , \\
-v_{\bm k}^s{}'\left(\eta\right) &= -u_{\bm k}'\left(\eta\right) 
v_{\bm k}^s(\eta_\uin)
- w_{\bm k}'\left(\eta\right) \left[-v_{\bm k}^s{}'(\eta_\uin)\right]\, .
\eea
Evaluating these equations at initial time $\eta_\uin$, one can see that $u_{\bm{k}}$ and $w_{\bm{k}}$ must satisfy the initial conditions $u_{\bm k}(\eta_\uin)=1$,
$w_{\bm k}(\eta_\uin)=0$, $u_{\bm k}'(\eta_\uin)=0$
and $w_{\bm k}'(\eta_\uin)=-1$. Furthermore, since the initial condition for the
Mukhanov-Sasaki variable, in the Bunch-Davies vacuum and up to an irrelevant phase, is given by $v_{\bm k}(\eta_\uin)=1/\sqrt{2k}$ and $v_{\bm k}'(\eta_\uin)=i\sqrt{k/2}$, one concludes that
$v_{\bm k}=(u_{\bm k}-ikw_{\bm k})/\sqrt{2k}$. Then, it is easy to
show that the Wronskian
$v_{\bm k}^*v_{\bm k}'-v_{\bm k}v_{\bm k}'{}^*=i$ and using the
relation established before between $v_{\bm k}$ and $u_{\bm k}$ and
$w_{\bm k}$, this implies that
$u_{\bm k}'w_{\bm k}-u_{\bm k}w_{\bm k}'=1$. This has also for
consequence that
\bea
d_\uin &=-w_{\bm k}'(\eta)d(\eta)-w_{\bm k}(\eta)r(\eta)\, ,\\
r_\uin &= u'_{\bm k}(\eta)d(\eta)+u_{\bm k}(\eta)r(\eta)\, .
\eea
Plugging these two formulas into \Eq{eq:fretaudrini}, one obtains
\bea
 \tilde{f}\left(d,r,\eta\right) &= \frac{1}{\sqrt{2\pi}}
\exp\left[-\left(\frac{k w_{\bm k}'{}^2}{4}
+\frac{u_{\bm k}'{}^2}{4k}\right)d^2
-\left(\frac{k w_{\bm k}^2}{4}+\frac{u_{\bm k}^2}{4k}\right)r^2
-2\left(\frac{k w_{\bm k}w_{\bm k}'}{4}
+\frac{u_{\bm k}u_{\bm k}'}{4 k}\right)rd\right]
\\ & \times
\exp\Biggl(-\frac{\left(2\pi\right)^{3/2}}{2}\int_{-\infty}^\eta 
\dd \eta'\, \gamma(\eta') \, \tilde{C}_R({\bm k},\eta')
\bigl\{\left[w_{\bm k}(\eta')u_{\bm k}'(\eta)
-w_{\bm k}'(\eta) u_{\bm k}(\eta')\right]d
\\ &   
+\left[w_{\bm k}(\eta') u_{\bm k}(\eta) 
-w_{\bm k}(\eta) u_{\bm k}(\eta')\right]
r\bigr\}^2\Biggr),
\eea
where we have taken $\eta_\mathrm{ ini}=-\infty$.

Having obtained $ \tilde{f}(d,r,\eta)$, we then need to take
the inverse Fourier transform in the $r$ coordinate to get $f(d,D,\eta)$. This can be easily
done since the above expression is a Gaussian. Expressing $u_{\bm k}$ and $w_{\bm k}$ in terms
of the Mukhanov-Sasaki variable via the relationship $v_{\bm k}=(u_{\bm k}-ikw_{\bm k})/\sqrt{2k}$, and expressing $d$ and $D$ in terms of $v_{\bm{k}}^{s,(1)}$ and $v_{\bm{k}}^{s,(2)}$ via inverting \Eq{eq:decoherence:dD:def}, one obtains
\bea
\label{eq:finalrho}
\left\langle v_{\bmk}^{s,(1)}  \bigl\vert \hat{\rho}_{\bm k}^s\bigr
\vert v_{\bmk}^{s,(2)}\right\rangle & = 
\frac{\left(2\pi\right)^{-1/2}}
{\sqrt{\left\vert v_{\bm k}\right\vert^2+\mathcal{J}_{\bm k}}}
\exp\left\{-\frac{{v^{s,(2)}_{\bm k}}^2+
{v^{s,(1)}_{\bm k}}^2
+i{\left\vert v_{\bm k}\right\vert^2}^\prime
\left[{v^{s,(2)}_{\bm k}}^2-{v^{s,(1)}_{\bm k}}^2\right]}
{4\left(\left\vert v_{\bm k}\right\vert^2
+\mathcal{J}_{\bm k}\right)}\right\}
 \\ & \times
\exp\Biggl\{-\frac{1}{2\left(\left\vert v_{\bm k}\right\vert^2
+\mathcal{J}_{\bm k}\right)}
\biggl(\mathcal{I}_{\bm k}\mathcal{J}_{\bm k}
-\mathcal{K}_{\bm k}^2+\left\vert v_{\bm k}^\prime \right\vert^2
\mathcal{J}_{\bm k}+ \left\vert v_{\bm k}\right\vert^2\mathcal{I}_{\bm k}
 \\ &
-{\left\vert v_{\bm k}\right\vert^2}^\prime \mathcal{K}_{\bm k}\biggr)
\left[v^{s,(2)}_{\bm k}-v^{s,(1)}_{\bm k}\right]^2
-\frac{i\mathcal{K}_{\bm k}}{2\left(\left\vert v_{\bm k}\right\vert^2
+\mathcal{J}_{\bm k}\right)}
\left[{v^{s,(2)}_{\bm k}}^2-{v^{s,(1)}_{\bm k}}^2\right]\Biggr\},
\eea
where ${\cal I}_{\bm k}$, ${\cal J}_{\bm k}$ and
${\cal K}_{\bm k}$ are defined in the main text, see
\Eqs{eq:defI}, (\ref{eq:defJ}) and~(\ref{eq:defK}).
The density matrix defined by \Eq{eq:finalrho} (all other coefficients being zero) is an exact and fully explicit solution of the Lindblad
equation. 

Thanks to the linearity of the interaction term, it still describes a Gaussian state. One can also check that, when $\gamma=0$, namely when the interaction
with the environment is switched off, one recovers the usual two-mode
squeezed state, which is a pure state. Indeed, if $\gamma=0$, then
$\mathcal{I}_{\bm k}=\mathcal{J}_{\bm k}=\mathcal{K}_{\bm k}=0$ and
one has
\bea
\left.\left\langle v_{\bmk}^{s,(1)} \right\vert \hat{\rho}_{\bm k}^s
\left\vert v_{\bmk}^{s,(2)}\right\rangle\right\vert_{\gamma=0} & 
= \Psi\left({v^{s,(1)}_{\bm k}}\right)\Psi^*\left({v^{s,(2)}_{\bm k}}\right) \, , 
\eea
with
\bea 
\Psi\left(v\right)=\left(\frac{1}{2\pi\left\vert
v_{\bm k} \right\vert^2}\right)^{1/4}\exp\left(-\frac{1-i{\left\vert
v_{\bm k}\right\vert^2}^\prime}{4\left\vert
v_{\bm k}\right\vert^2}v^2\right)\, .  
\eea
If we now introduce the quantity
$\Omega_{\bm k}\equiv-iv_{\bm k}^\prime/(2v_{\bm k})$, then 
the above wavefunction can be rewritten as
\bea
\Psi\left(v\right)=
\left[\frac{2\mathrm{Re}\left(\Omega_{\bm k}\right)}{\pi}\right]^{1/4}
\ee^{-\Omega_k v^2}\, , 
\eea
which exactly corresponds to a squeezed state, that is to say the known 
solution in absence of an interaction term.

\subsection{Slow-roll approximation}
\label{subsec:dmsr}

The density matrix given by \Eq{eq:finalrho} is explicitly known if
the three integrals~(\ref{eq:defI}), (\ref{eq:defJ})
and~(\ref{eq:defK}) can be computed exactly. In this section, we show
that this can be done if the slow-roll approximation is used. This
leads to an expression of the integral ${\cal J}_{\bm k}$ that is then
used in the main text to derive the correction to the power
spectrum. This also allows us to establish the expressions of the
integrals ${\cal I}_{\bm k}$ and ${\cal K}_{\bm k}$ that are then used
in the main text to calculate the level of decoherence.

At first order in the slow-roll approximation, one has
$\omega^2 \simeq k^2 - 2
[1+3(2\epsilon_{1*}+\epsilon_{2*})/4]/{\eta^2}$
where $\epsilon_{1*}$ and $\epsilon_{2*}$ are the first and second
slow-roll parameters evaluated at the time when the pivot scale $k_*$
crosses out the Hubble radius. The mode function $v_{\bm{k}}$ is then
the solution of \Eq{eq:app:MSequation} that is normalised to the
Bunch-Davis vacuum in the sub-Hubble limit, \bea
\label{eq:modeFunction:SR}
v_{\bm k}(\eta) = \frac{1}{2}\sqrt{\frac{\pi}{k}}\sqrt{-k\eta} 
\ee^{-i\frac{\pi}{2}\left(\nu+\frac{1}{2}\right)}H_\nu^{(2)}\left(-k\eta\right)\, ,
\eea
where $H^{(2)}_{\nu}(z)$ is the Hankel function of the second kind of
order $\nu$ and where we have defined
$\nu\equiv 3/2+\epsilon_{1*}+\epsilon_{2*}/2$. We also need to specify
the Fourier transform of the correlation function. As explained in the
main text, we work with the ansatz~(\ref{eq:Ck:appr}). Finally, at first order in the slow-roll approximation, the scale factor $a$ scales as $\eta^{-1-\epsilon_{1*}}$ and \Eq{eq:gamma} gives rise to
\bea
\label{eq:gamma:SR}
\gamma &= \gamma_* \left(\frac{\eta_*}{\eta}\right)^{p\left(1+\epsilon_{1*}\right)}\, .
\eea 
Decomposing the Hankel function into real and imaginary parts, $H^{(2)}_\nu(z) = J_\nu(z) - i Y_\nu(z)$, and making use of the relations $Y_\nu\left(z\right) =[J_\nu\left(z\right)\cos\left(\nu\pi\right)-J_{-\nu}\left(z\right)]/[\sin\left(\nu\pi\right)]$ and $H_\nu'(z)=-H_{\nu+1}(z)+\nu/z H_\nu(z)$, the three integrals~(\ref{eq:defI}), (\ref{eq:defJ}) and~(\ref{eq:defK}) can then be
expressed as
\begin{align}
\label{eq:calculI}
\mathcal{I}_{\bm k}(\eta) &= \frac{\pi^2 \kgamma^2}
{8k\sin^2\left(\pi\nu\right)}
\left(\frac{k}{k_*}\right)^{\left(p-3\right)\left(1+\epsilon_{1*}\right)}
\left(-k\eta\right)^{-1}
\Biggl\{
\biggl[\left(\nu+\frac{1}{2}\right){J_{-\nu}\left(-k\eta\right)}
\nonumber \\ &
+\left(-k\eta\right)J_{-\nu-1}\left(-k\eta\right)\biggr]^2
I_1\left(\nu\right)
+ \left[\left(\nu+\frac{1}{2}\right){J_{\nu}\left(-k\eta\right)}
-\left(-k\eta\right)J_{\nu+1}\left(-k\eta\right)\right]^2I_1\left(-\nu\right)
\nonumber \\ &  
-2\left[\left(\nu+\frac{1}{2}\right){J_{-\nu}\left(-k\eta\right)}+\left(-k\eta\right)J_{-\nu-1}\left(-k\eta\right)\right]
\biggl[\left(\nu+\frac{1}{2}\right){J_{\nu}\left(-k\eta\right)}
\nonumber \\ &
-\left(-k\eta\right)J_{\nu+1}\left(-k\eta\right)\biggr]
I_2\left(\nu\right)
\Biggr\}\, ,\\
\label{eq:calculJ}
\mathcal{J}_{\bm k}\left(\eta\right) &= \frac{\pi^2 \kgamma^2}
{8k^3\sin^2\left(\pi\nu\right)}
\left(-k\eta\right)\left(\frac{k}{k_*}
\right)^{\left(p-3\right)\left(1+\epsilon_{1*}\right)}\biggl[
J_{-\nu}^2\left(-k\eta\right)I_1\left(\nu\right)
+J_{\nu}^2\left(-k\eta\right)I_1\left(-\nu\right)
\nonumber \\ &
-2J_{\nu}\left(-k\eta\right)J_{-\nu}\left(-k\eta\right)I_2\left(\nu\right)
\biggr]\, , 
\\ 
\label{eq:calculK}
\mathcal{K}_{\bm k}\left(\eta\right) &= 
-\frac{\pi^2 \kgamma^2}{8k^2\sin^2\left(\pi\nu\right)}
\left(\frac{k}{k_*}\right)^{\left(p-3\right)\left(1+\epsilon_{1*}\right)}
\Biggl\{\biggl[\left(\nu+\frac{1}{2}\right)J_{-\nu}\left(-k\eta\right)
\nonumber \\ &
+\left(-k\eta\right)J_{-\nu-1}\left(-k\eta\right)\biggr]
J_{-\nu}(-k\eta)I_1\left(\nu\right)
+
\biggl[\left(\nu+\frac{1}{2}\right)J_{\nu}\left(-k\eta\right)
-\left(-k\eta\right)J_{\nu+1}\left(-k\eta\right)\biggr]
\nonumber \\ & \times
J_{\nu}(-k\eta)I_1\left(-\nu\right)
+
\biggl[\left(-k\eta\right)J_{-\nu}\left(-k\eta\right)J_{\nu+1}
\left(-k\eta\right)-\left(-k\eta\right)J_{\nu}\left(-k\eta\right)
J_{-\nu-1}\left(-k\eta\right)
\nonumber \\ &
-2\left(\nu+\frac{1}{2}\right)J_{\nu} 
\left(-k\eta\right)J_{-\nu} \left(-k\eta\right)\biggr]
I_2\left(\nu\right)
\Biggr\},
\end{align}
where $\kgamma$ is defined in \Eq{eq:kbreak:def}, and where we have defined the functions $I_1$ and $I_2$ by
\bea
\label{eq:I1:I2}
I_1\left(\nu\right) & \equiv \int_{-k\eta}^{-k\eta_\mathrm{E}}
\dd  z z^{\alpha_1} J_\nu^2\left(z\right), \\
I_2 \left(\nu\right) & \equiv \int_{-k\eta}^{-k\eta_\mathrm{E}}
\dd  z z^{\alpha_1} J_\nu\left(z\right)J_{-\nu}\left(z\right),  
\eea
with $\alpha_1\equiv
1-\left(p-3\right)\left(1+\epsilon_{1*}\right)$. The upper bound in these integrals corresponds to the time when the wavelength $a/k$ of the comoving mode under consideration $k$ crosses the correlation length of the environment $\lE$. At leading order in slow roll, this happens when
\bea
\label{eq:ketaE:def}
-k\eta_\mathrm{E} = \left(1+\epsilon_{1*}\right) \left(H_* \lE\right)^{\epsilon_{1*}-1}\left(\frac{k}{k_*}\right)^{\epsilon_{1*}}\, .
\eea
Notice that keeping the slow-roll corrections in \Eq{eq:ketaE:def} may be problematic given that the environment correlation function has been modelled with a top-hat function, and considering more realistic correlation functions might introduce corrections larger than those. This would however not affect the conclusions drawn in the main text because when the value of $k\eta_\mathrm{E}$ is relevant, the correction to the power spectrum turns out to be strongly scale dependent, regardless of the slow-roll corrections.

These integrals are of the Weber-Schafheitlin type~\cite{Abramovitz:1970aa, nist, Mathematica}, and in
terms of the generalised hypergeometric functions ${}_pF_q$, they are
given by
\begin{align}
\label{eq:I1}
I_1(\nu) & =
\frac{1}{4^\nu\left(1+\alpha_1+2\nu\right)\Gamma^2\left(1+\nu\right)}
\biggl\{
\nonumber \\ & 
\left(-k\eta_\mathrm{E} \right)^{1 + \alpha_1 + 2 \nu}
{}_pF_q\left[ \frac{1}{2}+\nu, \frac{1+\alpha_1}{2}  
+ \nu ;  1 + \nu, 
   \frac{3+\alpha_1}{2} + \nu, 1 + 2 \nu; -(-k\eta_\mathrm{E} )^{2}
\right]  
 \nonumber \\ &  
-(-k\eta)^{1 + \alpha_1  + 2 \nu}
  {}_pF_q\left[\frac{1}{2}+\nu, \frac{1+\alpha_1}{2}  
+ \nu ;  1 + \nu, 
   \frac{3+\alpha_1}{2} + \nu, 1 + 2 \nu ; -(-k\eta)^2\right]
   \biggr\}, \\
\label{eq:I2}
I_2(\nu) & =
\frac{\sin (\pi \nu)}{\pi \nu (1+\alpha_1)}
\biggl\{
\nonumber \\ &
\left(-k\eta_\mathrm{E} \right)^{1 + \alpha_1}
  {}_pF_q\left[ \frac{1}{2}, \frac{1+\alpha_1}{2}  
 ;  \frac{3+\alpha_1}{2}, 1 - \nu, 
 1 +  \nu ; -\left(-k\eta_\mathrm{E} \right)^{2}\right]  
\nonumber \\ &
-(-k\eta)^{1 + \alpha_1}
  {}_pF_q\left( \frac{1}{2}, \frac{1+\alpha_1}{2}  
 ;   \frac{3+\alpha_1}{2}, 1 - \nu, 
 1 +  \nu ; -(-k\eta)^2\right]  
    \biggr\}\, .
\end{align}

The above results are exact but the complexity of the formulae makes
them not very insightful. This is why we now expand these expressions
in two limits. The first one corresponds to the requirement that the
autocorrelation time of the environment $t_\uc$ is much shorter than
the time scale $\Delta t$ over which the system typically evolves, see
\Eq{eq:taucllDeltat}. In \App{sec:DerivingLindblad}, this condition is
shown to be necessary in order for the Lindblad equation to be
valid. As explained in \App{sec:massivescalarfield}, in the present
case the time scale over which the system evolves is the Hubble time,
$\Delta t \sim H^{-1}$. Furthermore, if the environment correlation
time and length are directly related, $\lE\sim t_\uc$, as is the case
in the example discussed in \App{sec:massivescalarfield}, this
condition boils down to $H\lE \ll 1$. Because of \Eq{eq:ketaE:def}, it
implies that $-k\eta_\mathrm{E}\gg 1$. The second limit consists in
evaluating the above expressions when the physical wavelength of the
mode under consideration $a/k$ is well outside the Hubble radius
$H^{-1}$, which is the case at the end of inflation for all modes of
astrophysical interest today. This amounts to taking $-k\eta\ll 1$.
Under these two conditions, the hypergeometric functions of
\Eqs{eq:I1} and~(\ref{eq:I2}) can be expanded, in the large third
argument limit for the first one and in the small third argument limit
for the second one, and one obtains
\begin{align}
\label{eq:I1:I2:limit}
I_1\left(\nu\right) &\simeq
\frac{\left(-k\eta_{\mathrm{E}} \right)^{\alpha_1}}{\alpha_1\pi}
+\frac{1}{2\sqrt{\pi}}
\frac{\Gamma\left(-\alpha_1/2\right)
\Gamma\left(1/2+\alpha_1/2+\nu\right)}{\Gamma\left(1/2-\alpha_1/2\right)
\Gamma\left(1/2-\alpha_1/2+\nu\right)}
-\frac{\left(-k\eta\right)^{1+\alpha_1+2\nu}}{4^\nu\left(1+\alpha_1+2\nu\right)
\Gamma^2\left(1+\nu\right)},
\\
I_2 \left(\nu\right) & \simeq
\frac{\cos\left(\pi\nu\right)}{\alpha_1\pi} 
\left(-k\eta_{\mathrm{E}} \right)^{\alpha_1}
+\frac{1}{2\sqrt{\pi}}
\frac{\Gamma\left(-\alpha_1/2\right)\Gamma\left(1/2+\alpha_1/2\right)}
{\Gamma\left(1/2-\alpha_1/2-\nu\right)\Gamma\left(1/2-\alpha_1/2+\nu\right)}
\nonumber \\ &
- \frac{\sin\left(\pi\nu\right)}{\pi\nu\left(\alpha_1+1\right)}
\left(-k\eta\right)^{\alpha_1+1}.
\end{align}

For the power spectrum, one has to evaluate the integral ${\cal J}_{\bm k}$, see \Eq{eq:solPvv}. Expanding \Eq{eq:calculJ} in the same limits as above, namely $-k\eta_{\mathrm{E}}\gg 1$ and $-k\eta\ll 1$, one has
\bea
\label{eq:approxJ}
\mathcal{J}_{\bm k}\left(\eta\right) 
&\simeq \frac{\pi^2 \kgamma^2}{8k^3\sin^2\left(\pi\nu\right)}
\left(-k\eta\right)^{2-\alpha_1}
\left(\frac{a}{a_*}\right)^{p-3}
\biggl[
\frac{I_1\left(\nu\right)}{\Gamma^2\left(1-\nu\right)} 
\left(-\frac{k\eta}{2}\right)^{-2\nu}
\nonumber \\ &
+\frac{I_1\left(-\nu\right)}{\Gamma^2\left(1+\nu\right)} 
\left(-\frac{k\eta}{2}\right)^{2\nu}
-\frac{2 I_2\left(\nu\right)}
{\Gamma\left(1-\nu\right)\Gamma\left(1+\nu\right)} 
\biggr]\, .
\eea
One can then see which term dominates in the above expressions of
$I_1$ and $I_2$ and, hence, in the power spectrum, depending on the value of $\alpha_1$ (hence $p$). The critical
values for $\alpha_1$ are $-2\nu-1,-2\nu,-1,0,2\nu-1,2\nu$ so we have
seven cases to distinguish a priori, but a more careful study reveals
that, in fact, they can be casted into three cases only. If
$\alpha_1<-2\nu-1$, the dominant terms come from the third ones of
$I_1(\nu)$, $I_1(-\nu)$ and $I_2(\nu)$. If $-2\nu-1<\alpha_1<0$, the
dominant term comes from the second one of $I_1(\nu)$. If $\alpha_1>0$,
the dominant term comes from the first one of $I_1(\nu)$. These
considerations lead to the expressions of $\Delta {\cal P}_i$ in the
main text, see \Eqs{eq:dP1linear}, (\ref{eq:dP2linear})
and~(\ref{eq:dP3linear}).

For decoherence, as explained in the main text, one has to evaluate $\delta_{\bm {k}} = \left\vert v_{\bm k}^\prime \right\vert^2\mathcal{J}_{\bm k}
+ \left\vert v_{\bm k}\right\vert^2\mathcal{I}_{\bm k}
-{\left\vert v_{\bm k}\right\vert^2}^\prime
\mathcal{K}_{\bm k}$, see \Eq{eq:defdeltak:integrals:appr}. Using \Eqs{eq:modeFunction:SR}, (\ref{eq:calculI}), (\ref{eq:calculJ}) and~(\ref{eq:calculK}), one can show that the connection and recurrence relations for Bessel functions~\cite{Abramovitz:1970aa, nist} lead to several cancellations and this expression simplifies to $\delta_{\bm{k}}=I_1\left(\nu\right)+I_1\left(-\nu\right)
-2\cos\left(\pi\nu\right)I_2\left(\nu\right)$, see \Eq{eq:exactdelta}. If $\alpha_1<0$, the dominant contribution comes from the third term of $I_1(-\nu)$, if $0<\alpha_1<2\nu-1$, the dominant contributions come from the third term of $I_1(-\nu)$ and the first terms of $I_1(\nu)$, $I_1(-\nu)$ and $I_2(\nu)$, and if $\alpha_1>2\nu-1$, the dominant contributions come from the first terms of $I_1(\nu)$, $I_1(-\nu)$ and $I_2(\nu)$. This leads to the expression of $\delta_{\bm{k}}$ in the main text, see \Eq{eq:deltapprox}.
\section{Power spectrum for quadratic interaction}
\label{sec:Cr}
In this appendix, we provide additional details for the calculation of the power spectrum in presence of quadratic interactions.
\subsection{Calculation of the source}
\label{subsec:source}
In \Sec{subsec:eomquadratic}, it is shown that the source term in the differential equation satisfied by the power spectrum, \Eq{eq:thirdvquadratic}, involves the convolution product between the power spectrum itself and the Fourier transform of the environment correlator,
\bea
I = \int \dd  ^3 {\bm k}'\, \tilde{C}_{R}\left({ k}'\right)
P_{vv}\left(\left\vert \bm{k}'+\bm{k}\right\vert\right)
=\int\dd^3  \bm p \,  
\tilde{C}_R\left(\left\vert { \bm p - \bm k} \right\vert \right) 
P_{vv}\left( p \right)\, ,
\eea
where we have simply performed the change of integration variable $\bm{k}'=\bm{p}-\bm{k}$. Let us note that $S_2$ is nothing but $8\gamma(2\pi)^{-3/2}I$. If $\theta$ denotes the angle between the vectors $\bm p$ and
$\bm k$, in spherical coordinates where $\theta$ is the polar angle, one has
\bea 
I= 2\pi
\int_0^\infty \dd  p \, p^2 \int_0^\pi \dd \theta \, \sin \theta
\, \tilde{C}_R\left(\sqrt{k^2+p^2-2kp\cos \theta }\right) 
P_{vv}(p)\, ,
\eea
where the integral over the azimuth angle  has been performed. After changing the integration variable $z = k^2+p^2-2kp\cos \theta$, one obtains
\bea
I &= \frac{\pi}{k}\int_0^\infty \dd  p\, p \, P_{vv}(p)
\int_{\left(k-p\right)^2}^{\left(k+p\right)^2} \dd  z
\, \tilde{C}_R\left(\sqrt{z}\right)\, .
\label{eq:integral:Cz}
\eea
This expression is useful since it allows one to perform one-dimensional integrals only. 

In what follows, we perform an explicit calculation of this integral using the ansatz~(\ref{eq:tophatcorrelation}) for the environment correlator. The result can be generalised to any environment correlation function by using the argument presented below \Eq{eq:integral:Cz:maintext}. The Fourier transform of the correlation function~(\ref{eq:tophatcorrelation}) is given by \Eq{eq:Ck:fromstep}, and when plugged into \Eq{eq:integral:Cz}, the second integral can be done exactly, leading to
\bea
I &=
2 \bar{C}_R \frac{\sqrt{2\pi}}{k}\frac{\lE }{a}
\int_0^\infty \dd  p \, p \,
P_{vv}(p)\left[
\frac{\sin\left(\frac{\left\vert k-p\right\vert 
\lE }{a}\right)}
{\frac{\left\vert k-p\right\vert \lE }{a}}
- \frac{\sin\left(\frac{\left\vert k+p\right\vert 
\lE }{a}\right)}
{\frac{\left\vert k+p\right\vert \lE }{a}}
 \right]\, .
\eea
The next step consists in inserting the power spectrum in the above
equation and perform the integral. For simplicity, we neglect slow-roll corrections and work with the piecewise approximation $P_{vv}(p)=(2p)^{-1}$ if
$-p\eta>1$ (sub-Hubble scales) and $P_{vv}(p)=(2p)^{-1}(-p\eta)^{-2}$ if
$-p\eta<1$ (super-Hubble scales). This leads to
$I=I_{_\mathrm{ IR}}+I_{_\mathrm{ UV}}$, with
\begin{align}
\label{eq:Iir}
I_{_\mathrm{ IR}}&= \bar{C}_R \frac{\sqrt{2\pi}}{k \eta^2}
\frac{\lE }{a}
\int_0^{-1/\eta} \frac{\dd  p}{p^2} \left[
\frac{\sin\left(\frac{\left\vert k-p\right\vert \lE }{a}\right)}
{\frac{\left\vert k-p\right\vert \lE }{a}}
-\frac{\sin\left(\frac{\left\vert k+p\right\vert \lE }{a}\right)}
{\frac{\left\vert k+p\right\vert \lE }{a}}
 \right],
 \\ 
\label{eq:Iuv}
I_{_\mathrm{ UV}}&= \bar{C}_R \frac{\sqrt{2\pi}}{k}
\frac{\lE }{a}
\int_{-1/\eta}^\infty \dd  p  \left[
 \frac{\sin\left(\frac{\left\vert k-p\right\vert \lE }{a}\right)}
{\frac{\left\vert k-p\right\vert \lE }{a}}
 - \frac{\sin\left(\frac{\left\vert k+p\right\vert 
\lE }{a}\right)}{\frac{\left\vert k+p\right
\vert \lE }{a}}\right]\, .
\end{align}
As indicated by the notation, the integral $I$ contains an
Ultra-Violet (UV, sub-Hubble scales) part and an Infra-Red (IR, super-Hubble scales) part. 

Let us first discuss the UV integral~(\ref{eq:Iuv}). This contribution
is usually removed through adiabatic subtraction~\cite{Bunch:1980vc, Markkanen:2017rvi} but it is interesting
to notice first that here, it is not divergent and can be calculated exactly
in terms of the Sine integral~\cite{Abramovitz:1970aa, nist} $\mathrm{Si}(x)\equiv \int_0^x \sin(t)\dd t/t$, namely
\bea
 I_{_\mathrm{{UV}}} 
&=\bar{C}_R\frac{\sqrt{2\pi}}{k}\left[\mathrm{Si}\left(\left\vert
k-\frac{1}{\eta}\right\vert
\frac{\lE }{a}\right)-\mathrm{Si}\left(\left\vert
k+\frac{1}{\eta}\right\vert \frac{\lE }{a}\right)\right]\, .
\eea
This formula can then be expanded in the sub-Hubble ($k\gg -1/\eta$)
and super-Hubble ($k\ll -1/\eta$) limits. In the sub-Hubble regime,
one has 
\bea
\left.I_{_\mathrm{UV}}\right\vert_{k\gg -1/\eta} 
\simeq -2\bar{C}_R\frac{\sqrt{2\pi}}{k\eta}
\frac{\lE }{a}
\frac{\sin\left(\frac{k\lE }{a}\right)}
{\frac{k\lE }{a}}. 
\label{eq:IUV:subH}
\eea
The result now depends on whether $k$ is larger or smaller than the
correlation length of the environment. If $k\gg a/\lE $,
that is to say the wavelength of the Fourier mode is smaller than the
correlation length, then the above expression cannot be further
simplified. If, on the contrary, $k\ll a/\lE $, that is to say if
the wavelength is much larger than the correlation length, then
$\left.I_{_\mathrm{UV}}\right\vert_{k\gg -1/\eta} \simeq
-2\bar{C}_R\sqrt{2\pi} \lE /(k\eta a)$.
In the super-Hubble regime, one finds
\bea
\label{eq:IUV:superH}
\left.I_{_\mathrm{UV}}\right\vert_{k\ll -1/\eta} \simeq 
2\bar{C}_R\sqrt{2\pi}\frac{\lE }{a}
\frac{\sin\left(\frac{\lE }{a\eta}\right)}
{\frac{\lE }{a\eta}}\sim 
2\bar{C}_R\sqrt{2\pi}\frac{\lE }{a}
\, ,
\eea
where, in the last equality, we have assumed that the correlation
length of the environment $\lE$ is much smaller than the Hubble radius $H^{-1}$, which is true if $\lE\sim t_\uc$ since the derivation of the Lindblad equation requires $t_\uc\ll H^{-1}$, see \App{sec:DerivingLindblad}.

Let us now consider the IR integral~(\ref{eq:Iir}). Since it does not converge, we
introduce an IR cut-off and replace the lower bound of the integral $0$ with
$-1/\eta_{_\mathrm{ IR}}$, which can be seen as the comoving mode that corresponds to the Hubble radius at the onset of inflation.  We then define the parameter
$K\equiv k\lE /a$ and the new variable $y\equiv kp/a$, which allow us to rewrite the IR integral as
\bea
I_{_\mathrm{ IR}} &=
\bar{C}_R\frac{\sqrt{2\pi}}{k\eta^2}\frac{\lE ^2}{a^2}
\int _{-\lE /(a\eta_{_\mathrm{ IR}})}^{-\lE /(a\eta)} \frac{\dd  y}{y^2}
\left[\frac{\sin\left(K-y\right)}
{K-y}
-
\frac{\sin\left(K+y\right)}
{K+y}
\right]\, .
\eea
This integral can be performed explicitly, and the result reads
\bea
I_{_\mathrm{ IR}}(\eta) &=
\bar{C}_R\frac{\sqrt{2\pi}}{k\eta^2}\frac{\lE ^2}{a^2}
\left[{\cal I}\left(-\frac{\lE}{a\eta}\right)-
{\cal I}\left(-\frac{\lE}{a\eta_{_\mathrm{ IR}}}\right)
\right]\, ,
\eea
where we have introduced the function ${\cal I}$ defined by
\bea
{\cal I}(y)&\equiv 2\frac{\sin K}{K^2}\mathrm{ Ci}(y)
+2\frac{\cos K}{K}\frac{\sin y}{y}
-\frac{\cos K}{K}\mathrm{ Ei}(-iy)
-\frac{\cos K}{K}\mathrm{ Ei}(iy)
\\ &
-\frac{1}{K^2}\mathrm{ Si}(K-y)
-\frac{1}{K^2}\mathrm{ Si}(K+y)\, ,
\eea
where $\mathrm{ Ei}(x)\equiv -\int_{-x}^\infty \ee^{-t} \dd t/t$ is the exponential integral function. For the reason recalled above, we assume the correlation
length of the environment to be much smaller than the Hubble radius, which amounts to
$-\lE /(a\eta)\ll 1$. This also implies that
$-\lE /(a\eta_\mathrm{IR})\ll 1$ since, by definition, $\eta>\eta_{\mathrm{IR}}$ during inflation. As a
consequence, the limit $y\rightarrow 0$ is the relevant one, where
\bea
\lim_{y\rightarrow 0} {\cal I}(y)=\frac{2}{K}\left(\frac{\sin K}{K}-\cos K
\right) \ln \left(\vert y\vert \right)\, .
\eea
Comparing this expression with \Eq{eq:Ck:fromstep}, one can see that this expression is directly related to the Fourier transform of the environment correlator, ${\cal I}(y) \propto \tilde{C}_R(k)$. This should not come as a surprise given the argument presented below \Eq{eq:integral:Cz:maintext}, and allows us to write
\bea
\label{eq:finalIir}
I_{_\mathrm{ IR}} &= \frac{2\pi}{\eta^2} \tilde{C}_R\left(k\right)
\ln\left(\frac{\eta_{\mathrm{IR}}}{\eta}\right)\, .
\eea
We are now in a position to compare the IR and UV contributions. Inside the environment correlation length $\lE$, \ie when $K\gg 1$, one finds $I_{\mathrm{IR}}/I_{\mathrm{UV}}=-H \lE \ln(\eta_{\mathrm{IR}}/\eta)  /\tan(K)$, which is suppressed by $H\lE$ that we assumed was small. Between the correlation length $\lE$ and the Hubble radius $H^{-1}$, \ie when $K\ll 1$ but $-k\eta\gg 1$, one obtains $I_{\mathrm{IR}}/I_{\mathrm{UV}}=k H \lE  \ln(\eta_{\mathrm{IR}}/\eta)  /3 $, which is both suppressed by $K$ and by $H\lE$. At super-Hubble scales, $-k\eta\ll 1$, one finally finds $I_{\mathrm{IR}}/I_{\mathrm{UV}}= (H\lE)^2 \ln(\eta_{\mathrm{IR}}/\eta)$, which is again suppressed by $H\lE$. In all cases, the UV contribution thus exceeds the IR one. However, the inclusion of UV contributions in master equations is a delicate task, since it is ultimately related to the issue of renormalisation in open quantum systems (see for instance \Refa{Agon:2017oia}). In the present work, we adopt the simple viewpoint of adiabatic subtraction, where the UV contribution is removed by hand, and we defer a more detailed analysis of UV effects to future works.
This leads to the source function
\bea
\label{eq:quadratic:source}
S_2=\frac{8\gamma}{\left(2\pi\right)^{3/2}}I
=\frac{8\gamma}{\sqrt{2\pi}\eta^2}
\tilde{C}_R\left(k\right)
\ln\left(\frac{\eta_{\mathrm{IR}}}{\eta}\right)\, .
\eea
\subsection{Calculation of the power spectrum}
\label{subsec:solvingthirdquadratic}
Inserting \Eq{eq:quadratic:source} with the ansatz~(\ref{eq:Ck:appr}) into \Eq{eq:exactsolquadratic} leads to
\bea
P_{vv}(k)=&\vert v_{\bm k}\vert^2
-\frac{32}{3}
 \frac{\bar{C}_R}{2\pi}\frac{\lE ^3}{a_*^3}
\eta_*^{-3}
 \\ & \times
\int_{-\infty}^{\eta}\gamma(\eta')\Theta\left(\frac{k\lE }{a}\right)
\eta'\ln \left(\frac{\eta'}{\eta_{_\mathrm{ IR}}}\right)
\mathrm{ Im}^2\left[v_{\bm k}(\eta')v_{\bm k}^*(\eta)\right]\dd \eta'.
\eea
Then, using the explicit form of the Bunch-Davies normalised mode function given by
\Eq{eq:modeFunction:SR}, one arrives at the following expression
\bea
\label{eq:exactPvvquadratic}
P_{vv}(k)=& \vert v_{\bm k}\vert^2
+\frac{\pi}{3k}\frac{\bar{C}_R}{\sin ^2(\pi \nu)}
\gamma_*\frac{\lE ^3}{a_*^3}(-k\eta)
(-k\eta_*)^{p-3}
 \\ & \times 
\biggl[J_{-\nu}^2(-k \eta)I_3(\nu)
+J_{\nu}^2(-k \eta)I_3(-\nu)
-2J_{\nu}(-k\eta)J_{-\nu}(-k\eta)I_4(\nu)\biggr],
\eea
with $\nu\equiv 3/2$ (since, as explained in \Sec{sec:quad:source}, we neglect all slow-roll corrections) and the integrals $I_3$ and $I_4$ defined by
\bea
\label{eq:defI3I4}
I_3(\nu) &\equiv \int _{-k\eta}^{(H_*\lE )^{-1}}
\dd  z \, z^{\alpha_2}\ln \left(-\frac{z}{k\eta_{_\mathrm{ IR}}}\right)
J_{\nu}^2(z), \\ 
I_4(\nu) &\equiv \int _{-k\eta}^{(H_*\lE )^{-1}}
\dd  z \, z^{\alpha_2}\ln \left(-\frac{z}{k\eta_{_\mathrm{ IR}}}\right)
J_{\nu}(z)J_{-\nu}(z)
\eea
where $\alpha_2\equiv 2-p$.
The goal is now to calculate these two integrals. It turns out that
this can be done exactly in terms of generalised hypergeometric function. For the
integral $I_3(\nu)$, the primitive reads
\bea
\label{eq:primitiveI3}
P& \left[I_3(\nu)\right](z) = -\frac{2^{-2\nu}}{(1+\alpha_2+2\nu)^2}
\frac{z^{1+\alpha_2+2\nu}}{\Gamma^2(1+\nu)}
 \\ & \times 
{}_pF_q\left(\frac12+\nu,\frac12+\frac{\alpha_2}{2}
+\nu,\frac12+\frac{\alpha_2}{2}+\nu ; 1+\nu,\frac{3}{2}+\frac{\alpha_2}{2}+\nu,1+2\nu,
\frac32+\frac{\alpha_2}{2}+\nu ; -z^2\right)
 \\ &
+\frac{2^{-2\nu}}{(1+\alpha_2+2\nu)}
\frac{z^{1+\alpha_2+2\nu}}{\Gamma^2(1+\nu)}\ln \left(-\frac{z}{k\eta_{_\mathrm{ IR}}}\right)
\\ & \times 
{}_pF_q\left(\frac12+\nu,\frac12+\frac{\alpha_2}{2}
+\nu ; 1+\nu,\frac32+\frac{\alpha_2}{2}+\nu,1+2\nu ; 
-z^2\right)\, .
\eea
As can be seen on \Eqs{eq:defI3I4}, and
as it was already the case for the linear interactions, one needs to calculate
this primitive for small values of its argument (since $-k\eta\ll 1$ if the power spectrum is evaluated on super-Hubble scales) and for large
values of its argument [since the environment has a correlation length much smaller than
the Hubble radius, $1/(H\lE )\gg 1$]. In both cases, this allows
us to expand \Eq{eq:primitiveI3} and simplify its expression. For small values
of the argument, one finds that
\bea
\lim_{z \to 0} P\left[I_3(\nu)\right]=\frac{2^{-2\nu}}
{(1+\alpha_2+2\nu)^2\Gamma^2(1+\nu)}
z^{1+\alpha_2+2\nu}
\left[-1+(1+\alpha_2+2\nu)
\ln \left(-\frac{z}{k\eta_{_\mathrm{ IR}}}\right)\right]\, ,
\eea
while, for large values, one obtains that
\bea
\lim_{z \to +\infty}P\left[I_3(\nu)\right] &= \frac{1}{4\sqrt{\pi}}
\frac{\Gamma(-\alpha_2/2)\Gamma(1/2+\alpha_2/2+\nu)}
{\Gamma(1/2-\alpha_2/2)\Gamma(1/2-\alpha_2/2+\nu)}
\Biggl[2
\ln \left(-\frac{1}{k\eta_{_\mathrm{ IR}}}\right)
 \\ &
+\psi\left(\frac{1-\alpha_2}{2}\right)
-\psi\left(-\frac{\alpha_2}{2}\right)
+\psi\left(\frac{1-\alpha_2}{2}+\nu\right)
+\psi\left(\frac{1+\alpha_2}{2}+\nu\right)
\Biggr]
\\ &
-\frac{z^{\alpha_2}}{\pi\alpha_2^2}\left[1-\alpha_2
\ln \left(-\frac{z}{k\eta_{_\mathrm{ IR}}}\right)\right].
\eea
Taking the difference between the two last expressions, the first one being evaluated at $z=-k\eta$ and the second one at $z=(H_*\lE)^{-1}$, we conclude that the integral $I_3(\nu)$ can be expressed as
\bea
\label{eq:I3:appr}
I_3(\nu) & \simeq 
\frac{1}{4\sqrt{\pi}}
\frac{\Gamma(-\alpha_2/2)\Gamma(1/2+\alpha_2/2+\nu)}
{\Gamma(1/2-\alpha_2/2)\Gamma(1/2-\alpha_2/2+\nu)}
\Biggl[2
\ln \left(-\frac{1}{k\eta_{_\mathrm{ IR}}}\right)
+\psi\left(\frac{1-\alpha_2}{2}\right)
-\psi\left(-\frac{\alpha_2}{2}\right)
 \\ &
+\psi\left(\frac{1-\alpha_2}{2}+\nu\right)
+\psi\left(\frac{1+\alpha_2}{2}+\nu\right)
\Biggr]
-\frac{\left(H_*\lE \right)^{-\alpha_2}}{\pi\alpha_2^2}
\left\{1-\alpha_2
\ln \left[-\frac{\left(H_*\lE \right)^{-1}}
{k\eta_{_\mathrm{ IR}}}\right]\right\}
 \\ &
-\frac{2^{-2\nu}}{(1+\alpha_2+2\nu)^2\Gamma^2(1+\nu)}
(-k\eta)^{1+\alpha_2+2\nu}
\left[-1+(1+\alpha_2+2\nu)
\ln \left(\frac{\eta}{\eta_{_\mathrm{ IR}}}\right)\right].
\eea

Let us now calculate the second integral, namely $I_4(\nu)$. It can still be expressed in terms of generalised hypergeometric functions,
\bea
\label{eq:primitiveI4}
P\left[I_4(\nu)\right](z) = & -\frac{z^{1+\alpha_2}}{(1+\alpha_2)^2}
\frac{1}{\Gamma\left(1-\nu\right)
\Gamma \left(1+\nu\right)}
 \\ & \times
{}_pF_q
\left( \frac12,\frac{1+\alpha_2}{2},\frac{1+\alpha_2}{2} ; \frac{3+\alpha_2}{2},1-\nu,1+\nu,\frac{3+\alpha_2}{2};-z^2
\right)
 \\ &
+
\frac{z^{1+\alpha_2}}{1+\alpha_2}
\frac{1}{\Gamma(1-\nu)\Gamma(1+\nu)}
\ln \left(-\frac{z}{k\eta_{_\mathrm{ IR}}}\right)
 \\ & \times 
{}_pF_q\left(\frac12,\frac{1+\alpha_2}{2} ; \frac{3+\alpha_2}{2},1-\nu,1+\nu ; -z^2\right).
\eea

As for $I_3(\nu)$, this primitive needs to be evaluated for small and
large values of its argument $z$. For small values of the argument, one has
\bea
  \lim_{z\to 0}P\left[I_4(\nu)\right]=\frac{z^{1+\alpha_2}}{(1+\alpha_2)^2}
  \frac{1}{\Gamma(1-\nu)\Gamma(1+\nu)}
  \Biggl[-1+(1+\alpha_2)\ln \left(-\frac{z}{k\eta_{_\mathrm{ IR}}}\right)
  \Biggr],
\eea
and for large values of the argument, one finds
\bea
\lim _{z\to +\infty}P\left[I_4(\nu)\right] = & \frac{1}{2(1+\alpha_2)\sqrt{\pi}}
\frac{\Gamma(3/2+\alpha_2/2)\Gamma(-\alpha_2/2)}{\Gamma(1/2-\alpha_2/2-\nu)
\Gamma(1/2-\alpha_2/2+\nu)}
\Biggl[
2\ln \left(-\frac{1}{k\eta_{_\mathrm{ IR}}}\right)
\\ &
-\psi\left(-\frac{\alpha_2}{2}\right)
+\psi\left(\frac{1+\alpha_2}{2}\right)
+\psi\left(\frac12-\frac{\alpha_2}{2}-\nu\right)
+\psi\left(\frac12-\frac{\alpha_2}{2}+\nu\right)\biggr]
 \\ &
+\frac{\cos(\pi \nu)}{\pi \alpha_2^2}z^{\alpha_2}\left[-1+\alpha_2
\ln \left(-\frac{z}{k\eta_{_\mathrm{ IR}}}\right)\right].
\eea
The final expression of $I_4(\nu)$ therefore reads
\bea
\label{eq:I4:appr}
I_4(\nu) &\simeq
\frac{1}{2(1+\alpha_2)\sqrt{\pi}}
\frac{\Gamma(3/2+\alpha_2/2)\Gamma(-\alpha_2/2)}{\Gamma(1/2-\alpha_2/2-\nu)
\Gamma(1/2-\alpha_2/2+\nu)}
\Biggl[
2\ln \left(-\frac{1}{k\eta_{_\mathrm{ IR}}}\right)
 \\ &
-\psi\left(-\frac{\alpha_2}{2}\right)
+\psi\left(\frac{1+\alpha_2}{2}\right)
+\psi\left(\frac12-\frac{\alpha_2}{2}-\nu\right)
+\psi\left(\frac12-\frac{\alpha_2}{2}+\nu\right)\biggr]
 \\ &
+\frac{\cos(\pi \nu)}{\pi \alpha_2^2}
\left(H_*\lE \right)^{-\alpha_2}\left\{-1+\alpha_2
\ln \left[-\frac{\left(H_*\lE \right)^{-1}}
{k\eta_{_\mathrm{ IR}}}\right]\right\}
 \\ &
-\frac{(-k\eta)^{1+\alpha_2}}{(1+\alpha_2)^2}
\frac{1}{\Gamma(1-\nu)\Gamma(1+\nu)}
\Biggl[-1+(1+\alpha_2)\ln \left(\frac{\eta}{\eta_{_\mathrm{ IR}}}\right)
\Biggr].
\eea

Plugging \Eqs{eq:I3:appr} and~(\ref{eq:I4:appr}) into \Eq{eq:exactPvvquadratic}, and expanding the Bessel functions in the same limit $-k\eta \ll 1$ as before, one obtains for the power spectrum at the end of inflation
\bea
\label{eq:finalPvvquadratic}
{\cal P}_\zeta &={\cal P}_{\zeta}\vert_\mathrm{standard}
\Biggl(1-
\frac43 \sigma_\gamma
\left(\frac{k}{k_*}\right)^{p-3}
\Biggl\{
\frac{1}{2\sqrt{\pi}}
\frac{\Gamma(-\alpha_2/2)\Gamma(1/2+\alpha_2/2+\nu)}
{\Gamma(1/2-\alpha_2/2)\Gamma(1/2-\alpha_2/2+\nu)}
 \\ & \times
\Biggl[
\frac12 \psi\left(\frac{1-\alpha_2}{2}\right)
-\frac12 \psi\left(-\frac{\alpha_2}{2}\right)
+\frac12 \psi\left(\frac{1-\alpha_2}{2}+\nu\right)
+\frac12 \psi\left(\frac{1+\alpha_2}{2}+\nu\right)
 \\ &
-\ln \left(\frac{k}{k_*}\right)-N_{_\mathrm{ T}}+\Delta
  N_*
\Biggr]
 % \\ &
-\frac{\left(H_*\lE \right)^{-\alpha_2}}
{\pi\alpha_2^2}
\left[1+\alpha_2
\ln \left(H_*\lE\right)+\alpha_2\ln\left( \frac{k}{k_*}\right)
+\alpha_2\left(N_{_\mathrm{ T}}-\Delta
  N_*
\right)\right]
  \\ & 
+\frac{2^{3-2\nu}}{\Gamma^2(\nu)}
\sin^2\left(\pi\nu\right)
\left(\frac{k}{k_*}\right)^{1+\alpha_2+2\nu}
\ee^{-\left(1+\alpha_2+2\nu\right)\left(N-N_*\right)}
 \\ & \times
\frac{3\left(1+\alpha_2\right)^2-4\nu^2+\left(1+\alpha_2\right)\left[\left(1+\alpha_2\right)^2-4\nu^2\right]\left(N-N_{\mathrm{IR}}\right)}{\left(1+\alpha_2\right)^2\left[\left(1+\alpha_2\right)^2-4\nu^2\right]^2}
\Biggr\}\Biggr),
\eea
where $\sigma_\gamma\equiv \bar{C}_R\lE ^3\gamma_*/a_*^3$. This expression is used in the main text, where the dominant contribution is identified depending on the value of $p$, see \Eqs{eq:dP1quadratic}-(\ref{eq:dP:singular:quadratic}).

\bibliographystyle{JHEP}
\bibliography{Lindblad_PS}
\end{document}